\documentclass[12pt]{article}
\usepackage{amsmath}
\usepackage{mathtools} 
\usepackage[T1]{fontenc}
\usepackage{amssymb}
\usepackage{amsfonts}
\usepackage{amsthm}
\usepackage{mathrsfs}
\usepackage[all]{xy}
\usepackage{tensor}
\usepackage{comment}
\usepackage{bm}
\usepackage{accents} 
\usepackage{bbm}
\usepackage[normalem]{ulem}
\usepackage[usenames,dvipsnames]{xcolor}
\usepackage[colorlinks=true, citecolor=Blue, linkcolor=Blue]{hyperref}
\usepackage{ragged2e}
\usepackage{tikz}
    \usetikzlibrary{decorations.markings}
    \usetikzlibrary{decorations.pathmorphing}
    \usetikzlibrary{decorations.pathreplacing, calligraphy}
    \usetikzlibrary{patterns}
    \usetikzlibrary{patterns.meta}
    \usetikzlibrary{cd} 
    \usetikzlibrary{positioning}
    \usetikzlibrary{shapes.geometric}
    \usepgflibrary{shadings}  

\usepackage{setspace}
\usepackage[labelfont={small, bf, sf}, textfont={small,sf}]{caption}
\usepackage{subcaption}
\numberwithin{equation}{section} 
\usepackage{authblk}
\usepackage[in]{fullpage}
\usepackage{cite}

\newcommand{\op}{\mathcal{O}}
\newcommand{\vep}{\varepsilon} 
\newcommand{\lie}{\pounds}

\newcommand{\tone}{\text{I}}
\newcommand{\ttwo}{\text{II}}
\newcommand{\tthr}{\text{III}}

\newcommand{\alg}{\mathcal{A}}

\newcommand{\vev}[1]{\langle #1\rangle}

\newcommand{\vphi}{\varphi}

\newcommand{\beom}{\mathcal{E}}
\newcommand{\gc}{\varkappa}
\newcommand{\scri}{\mathscr{I}}
\newcommand{\ns}{\mathscr{N}}
\newcommand{\hs}{\mathcal{H}}

\newcommand{\mc}{\mathcal}

\newcommand{\msf}{\mathsf}
\newcommand{\wh}{\widehat}
\newcommand{\tr}{\operatorname{Tr}}
\newcommand{\trinfty}{\tr_\infty}

\newcommand{\ho}[1]{\wh{ \msf{#1}} }

\newcommand{\hqft}{\mathcal{H}_\text{QFT}}

\newcommand{\abif}{\mathcal{B}}
\newcommand{\acut}{\mathcal{C}}

\newcommand{\inc}{\iota}
\newcommand{\aqft}{\alg_{\text{QFT}}}
\newcommand{\sr}{R}

\newcommand{\hpsi}{h_\Omega}
\newcommand{\hor}{\mathscr{H}}

\newcommand{\qasy}{\mathcal{Q}}
\newcommand{\constr}{\mathfrak{C}}

\newcommand{\Hajicek}{H\'{a}j\'{i}\v{c}ek }
\newcommand{\hfut}{\mathcal{H}_{\text{asy}}}
\newcommand{\hcr}{\wh{\mathcal{H}}}
\newcommand{\ainfty}{\mathcal{B}_\infty}
\newcommand{\Kmod}[1][\lambda]{K_{#1}}
\newcommand{\aminfty}{\mathcal{B}_{-\infty'}}

\newcommand{\atwoinfty}{\mathcal{A}_\infty}
\newcommand{\ahor}{\mathcal{A}_\hor}

\newcommand{\defeq}{\coloneqq}

\newcommand{\sigasy}{\Sigma^\text{asy}}
\newcommand{\RT}{R_2}
\newcommand{\Hinf}{H_\infty}

\newcommand{\beq}{\begin{equation}}
\newcommand{\eeq}{\end{equation}}
\newcommand{\bes}{\begin{subequations}}
\newcommand{\ees}{\end{subequations}}
\newcommand{\bea}{\begin{eqnarray}}
\newcommand{\eea}{\end{eqnarray}}

\title{Gravitational algebras and the generalized second
law
}
\author[1]{Thomas Faulkner\thanks{tomf@illinois.edu}}
\author[1]{Antony J. Speranza\thanks{asperanz@gmail.com}}

\affil[1]{\small \it Department of Physics, University of Illinois, Urbana-Champaign, Urbana IL 61801, USA}
\date{ May 1, 2024}

\begin{document}

\maketitle
\begin{abstract}

We derive the generalized second law (GSL)
for arbitrary cuts of Killing horizons from the perspective 
of crossed-product gravitational algebras,  making 
use of a recent proposal by one of us for the construction
of local gravitational algebras.  This construction
relies on the existence of a state whose modular flow is geometric on the horizon. In both free and interacting quantum field theories, such states are guaranteed to exist by the properties of
half-sided translations on the horizon.
Using  geometric 
identities derived from the canonical analysis of
general relativity on null surfaces, we show that the 
crossed product entropy agrees  
with the generalized entropy of the horizon cut
in a semiclassical limit, and further reproduce Wall's 
result relating the GSL to monotonicity of relative entropy
of the quantum field algebras.  We also give a novel generalization
of the GSL for interacting theories in asymptotically flat 
spacetimes involving the concept of an algebra at infinity for 
a half-sided translation, which accounts for triviality of 
the algebra of fields smeared only on the horizon.  
Going beyond the semiclassical limit, we compute 
subleading corrections to the crossed product entropy, but 
are unable to determine if the GSL continues to hold
after accounting for these.  We speculate that an improved  
GSL could follow from a hidden subalgebra structure 
of the crossed products, assuming the existence of an 
operator-valued weight between horizon cut algebras.

\end{abstract}

\flushbottom

\pagebreak

\tableofcontents

\section{Introduction}
Generalized entropy
is a fundamental quantity in semiclassical
quantum gravity providing a link between 
dynamical geometry, thermodynamics, and entanglement. 
It originally appeared in the context of black hole
thermodynamics, where Bekenstein recognized
that one could violate the second law of thermodynamics
by depositing entropy into a black hole 
unless the black hole possessed an intrinsic entropy 
proportional to its horizon area 
\cite{Bekenstein1972, Bekenstein1973a}.  
The resulting generalized entropy consists of a sum
\beq
S_\text{gen} = \frac{A}{4\hbar G_N} + S_\text{out},
\eeq
with the Bekenstein-Hawking area term representing the 
intrinsic black hole entropy, and $S_\text{out}$ denoting
the matter entropy outside of the black hole.  
The generalized second law is then the statement that 
$S_\text{gen}$ cannot decrease under time evolution,
despite the fact that the matter entropy $S_\text{out}$ may decrease
as matter falls across the horizon.  

The connection to entanglement comes from interpreting $S_\text{out}$ 
as the fine-grained entanglement entropy of quantum fields restricted to the 
black hole exterior \cite{Sorkin:2014kta, Bombelli1986, Srednicki1993a,
Frolov1993a}.  
Not only does this give a concrete definition of $S_\text{out}$,
it also leads to a UV-finite notion of entropy in quantum gravity,
owing to an argument of Susskind and Uglum \cite{Susskind:1994sm}
 that the divergences 
in the quantum field entanglement entropy cancel against the 
renormalization of $G_N$ in $S_\text{gen}$.  This cancellation
of divergences has been verified in a numerous examples 
\cite{Jacobson:1994iw, 
Larsen:1995ax, Solodukhin:2011gn, Cooperman:2013iqr, 
Bousso:2015mna}, 
and ultimately motivates interpreting $S_\text{gen}$ as a true count of the 
microscopic quantum gravitational degrees of freedom of the 
black hole.  

Since entanglement entropy can be associated with generic causally
complete subregions in 
quantum field theory, it is natural to consider generalized entropy
for arbitrary subregions in semiclassical quantum gravity 
\cite{Bianchi:2012ev, Myers:2013lva, Bousso:2015mna}.  
In this more general context, 
the area of the codimension-2 boundary of a Cauchy surface 
for the subregion takes the place of the horizon area
in the Bekenstein-Hawking term in $S_\text{gen}$.  
By viewing the generalized entropy as the quantum
analog of the area in semiclassical gravity, one may
upgrade various classical theorems in general relativity
to quantum versions.  Bekenstein's generalized second law is 
one such example, being the quantum version of Hawking's classical
area theorem \cite{Hawking1971, Hawking1972}.  
The quantum focusing conjecture
as a semiclassical version of the classical focusing theorem
is another prominent example \cite{Bousso:2015mna}, 
which has far-reaching 
applications including quantum entropy bounds
(see \cite{Wall:2018ydq} for a review) and 
field theoretic inequalities such as the quantum null energy condition
\cite{Bousso:2015wca,
Balakrishnan:2017bjg, Ceyhan:2018zfg}.  Generalized entropy also features in holographic entanglement
entropy, where the Ryu-Takayanagi and quantum extremal surface formulas
relate the generalized entropy of entanglement wedges in the bulk
to entanglement entropies of subregions in the boundary CFT dual
\cite{Ryu:2006ef, Ryu:2006bv, Hubeny:2007xt, Lewkowycz:2013nqa,
Faulkner:2013ana, 
Engelhardt:2014gca}.  

One obstruction to 
obtaining  rigorous proofs of semiclassical entropy
relations such as the quantum focusing conjecture is the 
occurrence of UV divergences 
in entanglement entropies in quantum field theory. These can be 
eliminated by employing regulators and constructing renormalized
entropies, but this procedure can often obscure important 
quantum information theoretic relations such as strong subadditivity.  
On the other hand, the expected UV-finiteness of the generalized entropy
suggests that coupling the theory to gravity 
results in a universal renormalized notion of entropy in semiclassical
quantum gravity.  If this is the case, there ought to be a continuum
description of this renormalized entropy that is manifestly 
regulator-independent.

Recent progress in this direction comes from considerations
of von Neumann algebras in quantum gravity.  
These began with the work of Leutheusseur and Liu 
\cite{Leutheusser:2021qhd, Leutheusser:2021frk},
who considered the strict large $N$ limit of a holographic
CFT above the Hawking-Page transition, and found
an emergent type $\tthr_1$ von Neumann algebra,
which they took to be a signature of the emergence
of a black hole horizon in the bulk dual.  Subsequently,
Witten argued that $\frac1N$ corrections implement
a crossed product on the algebra, resulting in a 
type $\ttwo_\infty$ von Neumann algebra, which notably 
possesses a well-defined notion of renormalized 
entropy \cite{Witten2021}.  
This entropy agrees with the generalized 
entropy of the horizon in semiclassical states of the 
algebra up to an overall constant, which can 
be interpreted as a universal large background value of the 
entropy \cite{Chandrasekaran2022b}.  
The agreement between algebraic entropy and generalized
entropy is closely tied to the implementation
of the gravitational
constraints, and therefore holds directly in the 
semiclassical bulk description.  This allows for the 
construction of type $\ttwo$ algebras and 
associated entropies in configurations with 
no currently known holographic dual description,
such as the static patch of de Sitter, or general
black holes in asymptotically flat space 
\cite{Chandrasekaran2022a, Kudler-Flam2023}.

The bulk semiclassical gravity picture further suggests
that generic subregions give rise to type $\ttwo$ 
gravitational algebras with a well-defined renormalized
entropy \cite{Jensen2023}.  These algebras each contain
a type $\tthr_1$ 
subalgebra describing quantum fields within
the subregion in a fixed background geometry,
as one  expects from 
from the $G_N\rightarrow 0$ limit, which 
suppresses gravitational backreaction. The emergence 
of such type $\tthr_1$
subregion algebras in the large $N$ limit of holography
has been explored in a number of recent works
\cite{Leutheusser:2022bgi, Bahiru:2022oas, Bahiru:2023zlc, 
Faulkner:2022ada}.  
The local gravitational
algebras also contain 
additional
 operators representing
the asymptotic ADM energy or the energy of a localized
observer degree of freedom, which serve as 
anchors to dress the local operators in the 
gravitational descriptions.  Imposing the gravitational
constraints on both sets of operators results in 
a type $\ttwo$ crossed product algebra that describes 
the fields and semiclassical geometry 
of the subregion.  As in the case of Killing horizons,
the renormalized entropy computed for semiclassical
states in this algebra agrees with the generalized 
entropy up to a state-independent constant.  
Hence, crossed product algebras are a viable framework
for a continuum, regulator-independent
description of semiclassical geometry and entropies.  

Since crossed-product gravitational algebras have manifestly
UV-finite renormalized entropies, they potentially
could provide a new set of tools for analyzing
semiclassical entropy relations such as 
the quantum focusing conjecture \cite{Bousso:2015mna}.  The goal 
of the present work is to initiate such investigations
in a more restricted context by revisiting the generalized
second law (GSL) for Killing horizons.  
In addition to their global boost symmetry, 
Killing horizons possess an enhanced near-horizon 
symmetry that enables a detailed analysis of the 
vacuum entanglement
properties of quantum fields in the spacetime containing the 
horizon.  Due to this symmetry, one can obtain an explicit
expression for the vacuum modular Hamiltonian of 
any subregion whose entangling surface lives on 
an arbitrary cut of the horizon \cite{Wall2011, Casini2017,
WittenLecture19}.  These subregions therefore provide 
nontrivial examples of the {\it geometric modular flow
conjecture} that is necessary in the construction
of gravitational algebras for arbitrary subregions 
proposed by Jensen, Sorce, and Speranza (JSS) \cite{Jensen2023}.
Hence, a second motivation for investigating the 
GSL is to verify the general construction proposed by JSS
in situations where the modular Hamiltonian is not 
determined by global spacetime symmetries.

Wall gave a proof of the GSL for rapidly evolving 
perturbed Killing horizons by 
relating the generalized entropy of cuts of the 
horizon to the relative entropy of quantum fields 
in the region defined by the cut \cite{Wall2011}.  After this,
the GSL becomes equivalent to the monotonicity
of relative entropy under algebra inclusions.  
The first result of the present paper is to reproduce this 
proof using crossed product gravitational algebras.  
The entropy of certain
semiclassical states for the gravitational algebra
manifestly takes the form of a relative entropy 
for the quantum fields, up to a correction involving
the asymptotic gravitational charge.  
Hence, by performing the crossed product separately
at two different cuts of a horizon, we immediately
conclude in section \ref{sec:gslsemi} 
a monotonicity result for the entropies 
of these states, thereby showing that crossed
product algebras can reproduce Wall's proof
of the generalized second law.

This  raises the question of whether
crossed product entropies can lead to stronger
monotonicity results beyond the strict semiclassical
limit for the quantum state.  
The entropy formula for type $\ttwo$ gravitational
algebras receives corrections to the semiclassical
expression due to entanglement between the
asymptotic gravitational charges
and quantum field degrees of freedom.  
One can ask whether the crossed product entropy
continues to satisfy monotonicity after these 
corrections are included.  We explore this 
question by computing the corrections to the 
semiclassical entropy formula in section 
\ref{sec:pertent}.  We demonstrate a number of properties
of these corrections including explicit 
bounds on their size, but are unable to determine if the 
crossed product entropy satisfies a monotonicity
result after accounting for these corrections.  
We also explore the possibility in section
\ref{sec:ovw} that the crossed product algebras 
could exhibit a subalgebra structure that is obscured
in their standard presentation.  This would follow 
from the existence of an operator-valued weight 
between horizon cut algebras, and assuming its existence,
we derive a modified monotonicity result for crossed 
product entropies.  

The overall picture that emerges from these investigations
is that crossed products for subregions indeed capture 
a useful notion of gravitational entropy, and they 
provide powerful tools for investigating 
entropy in semiclassical quantum gravity.  They raise 
the possibility of finding strengthened data-processing
inequalities associated with the change in relative entropy
between horizon cuts, or otherwise 
characterizing temporary violations
of the second law by gravitational effects.  
Ultimately, they could lead to a better understanding
entropy inequalities such as the quantum focusing conjecture,
consequences of causality in quantum field theory,
and low energy constraints on quantum theories of gravity.

\subsection{Summary of results}

This paper contains a number of new results concerning 
crossed product algebras and semiclassical generalized entropy
which we briefly summarize here.

Section \ref{sec:crprod} reviews the definition of the modular
crossed product algebra, and goes over the modular 
theory for the classical-quantum states that are used 
in the remainder of the paper.  We present novel expressions 
for the Tomita operators for these states, the details of 
which are given in appendix \ref{app:modth}, which 
lead to a simplified derivation of the density
matrices for these states originally obtained in \cite{Jensen2023}.
This also produces  expressions for the modular
conjugations for these states which have not appeared previously.  

We then turn to the discussion of gravitational algebras and the 
GSL for Killing horizons in section \ref{sec:semigsl}.  
We begin in section \ref{sec:hstr} reviewing the argument
for the form of the vacuum modular Hamiltonian for 
arbitrary cuts of a Killing horizon.  Versions of this
argument have appeared in various forms across several 
works \cite{Wall2011, Casini2017, WittenLecture19}, 
and this section serves as a summary that presents
the derivation in a unified manner.  We emphasize the connection
between the form of this modular Hamiltonian and 
properties of half-sided translations and half-sided
modular inclusions,
introduced respectively by Borchers and Wiesbrock \cite{Borchers1992, 
Borchers2000, Wiesbrock1992}.  Focusing on half-sided translations
leads to a derivation of the modular 
Hamiltonian that holds even in interacting 
field theories, which specifically does not rely on the existence
of bounded operators localized strictly on the black hole horizon.  

Following this, we present in section \ref{sec:geometric}
various geometric identities associated 
with the gravitational constraint equations 
on the black hole horizon.  These identities 
are derived using the canonical analysis of general relativity
on null surfaces, the details of which are given in appendix
\ref{app:canon}.  This leads to the identification of the 
global gravitational constraint (\ref{eqn:Czetalambda}) that 
relates the asymptotic gravitational charges to the bulk
generator of modular flow, and which is responsible for producing
the crossed product when imposed on the algebra of quantum
fields and asymptotic charges.  We also note that a semilocal
gravitational constraint  leads to a geometric 
identity (\ref{eqn:horcutid})
which relates the area of a horizon cut
to the one-sided modular Hamiltonian and the late-time 
area.  This identity was originally derived by Wall
by directly integrating the Raychaudhuri equation
\cite{Wall2011}; our derivation instead shows how it arises
from the canonical analysis, which directly ties the identity
to the underlying diffeomorphism invariance of the gravitational 
theory.  This has the added advantage of transparently 
identifying the gravitational contribution
to the average null energy on the horizon,
which in the $G_N\rightarrow0$ limit reduces to the 
square of the perturbative horizon shear,
equation (\ref{eqn:tgvv2}).  This identity also corresponds to
the local Smarr relation discussed by JSS
\cite{Jensen2023}, specialized to the 
present case of a subregion bounded by a cut 
of a Killing horizon.

We then turn to the construction of gravitational algebras and 
 analysis of the GSL in section \ref{sec:gslsemi}.  
We show that the gravitational algebra resulting from
imposing the global constraint on the horizon cut algebras 
takes the form of a modular crossed product,
using the form of the vacuum modular Hamiltonian derived in
section \ref{sec:hstr}.  We can then directly compute
the entropy of classical-quantum states on these algebras, and 
show using the semiclassical expansion of this entropy 
(\ref{eqn:SrhoSrel})
that the GSL reduces to monotonicity of relative entropy
for the underlying quantum field theory algebra inclusion.  
This resulting GSL relies on choosing an appropriate normalization
for the traces defined on the gravitational algebras for 
different cuts, which in general is not canonically determined.  
We resolve this by requiring that the trace agree on functions 
of the asymptotic charge $\hat{q}$, which is contained in 
all horizon cut algebras.  This implies that we can use 
the same formula for the trace, (\ref{eqn:Tr}), for any
horizon cut, and it is important that the same quantum 
field vacuum state $|\Omega\rangle$ can be used in this 
formula for different horizon cuts.  

To simplify the discussion, the analysis of section \ref{sec:semigsl}
was restricted to AdS black holes, for which the horizon provides a 
complete Cauchy surface.  Section \ref{sec:ainfty}
generalizes the arguments to spacetimes with other asymptotics,
such as asymptotically flat or de Sitter black holes.  
For free fields, this generalization simply consists of tensoring
in an additional factor to the algebra associated with 
fields that pass through $\scri^+$ or the cosmological horizon.
However, this description fails for interacting theories, 
since there is no  
algebra consisting of fields strictly localized 
to codimension-1 black hole horizon, and hence naively
there is no algebra with which to take the tensor product.  
In section \ref{sec:ainftysemi},
we show that one can handle the case of interacting theories
using the concept of an algebra at infinity for the 
half-sided translation.  This leads to a natural definition
of a subalgebra associated with fields that never 
enter the black hole, but with a trivial relative commutant,
corresponding to triviality of the horizon algebra.  This captures 
the notion that any operator with finite fluctuations must 
be smeared in a region near the horizon, and cannot be 
strictly localized to the horizon.  Using the existence
of a conditional expectation to this algebra at infinity,
we show that one again recovers the semiclassical GSL, 
generalizing the argument of Wall that used the existence 
of tensor-factorized states \cite{Wall2011}.

Having established the structure of this algebra at infinity,
we describe in section \ref{sec:flatgravalg} the construction
of the gravitational algebra.  Adapting the arguments 
of Kudler-Flam, Leutheusser, and Satishchandran
(KFLS) \cite{Kudler-Flam2023}, we show that the gravitational
constraints can separately be imposed on the algebra at 
infinity and on the horizon.  Once this is done, there is 
a final matching constraint associated with the junction
where these null surfaces meet on the Penrose diagram.  
This condition takes the form of a constraint, equation
(\ref{eqn:match}),
matching the two time variables arising in the separate crossed
products for each horizon.  Along with the matching 
conditions for the asymptotic charges, this eliminates the 
asymptotic horizon area in favor of the asymptotic ADM
mass as the extra gravitational charge in the algebra. 
We discuss conditions for which the final algebra
remains semifinite and possesses a trace, and show that these
are satisfied in the example of the Schwarzschild black hole 
in the Hartle-Hawking vacuum.  
Owing to the existence of the conditional expectation
to the algebra at infinity, we are able to conclude
that the crossed product
algebras satisfy a global second law (\ref{eqn:globalgsl}) which
generalizes a similar result obtained in \cite{Chandrasekaran2022b},
and holds for all states, without any semiclassical assumption.  
We also note that the matching 
constraint can be used to eliminate the observer
in the discussion of the algebra for the Schwarzschild-de Sitter
black hole considered by KFLS \cite{Kudler-Flam2023}.
In section \ref{sec:rotUnruh}, we point out some subtleties 
that may arise in obtaining a semifinite algebra 
at infinity when working with asymptotically flat rotating 
black holes or black holes formed from collapse.  

The versions of the GSL obtained in sections 
\ref{sec:gslsemi} and \ref{sec:flatgravalg} rely 
on the semiclassical approximation of the quantum states.
This naturally leads to the question of whether crossed product 
algebras satisfy a GSL beyond the semiclassical limit.  To
begin addressing this question, we compute in section 
\ref{sec:pertent} 
the subleading corrections to the semiclassical entropy
formula (\ref{eqn:SrhoSrel}).  Noting that the classical-quantum
state $|\Phi,f\rangle$ can be viewed as the image under a 
Petz-recovery channel from the quantum field theory 
state $|\Phi\rangle$ to a state on the crossed product,
we show that monotonicity of relative entropy for this 
channel implies that the corrections to the semiclassical
crossed product entropy are strictly nonnegative.  
We then obtain an explicit
expression for this correction, given by 
equation (\ref{eqn:DelSexact}).  We can directly evaluate
the first nontrivial contribution to this correction
in the semiclassical expansion, which we find is second 
order in the expansion parameter, 
and we verify that this correction
is positive.  Unfortunately, we are unable to determine 
if the crossed product entropy, including these corrections,
satisfies a monotonicity result, but note that 
if one existed, it would imply an improved 
data-processing inequality (\ref{eqn:newdpi}).  

Finally, in section \ref{sec:ovw}, we explore the possibility 
that the crossed-product GSL could follow from a hidden subalgebra
structure for the gravitational algebras.  We show that 
if there is an operator-valued weight
\cite{Haagerup1979I, Haagerup1979II} between the horizon 
cut quantum field theory algebras, there is a canonical
way to realize the later horizon-cut gravitational algebra
as a subalgebra of the earlier gravitational algebra. 
Although we do not at 
present have a proof for the existence
of this operator-valued weight, we conjecture  
one should exist by analogy with the 
case of a split inclusions, for which the operator-valued
weight is guaranteed.
Assuming its existence, the result crossed product 
subalgebra structure
should imply a version of the GSL stemming 
from monotonicity of relative entropy for this algebra inclusion.
In order to leverage this, we consider a free energy
quantity (\ref{eqn:crfreeen})  
defined as a relative entropy between the 
state of interest and the dual weight that naturally arises 
in the crossed product construction.  This leads to 
the monotonicity results (\ref{eqn:entineq}) and 
(\ref{eqn:xgslsrel}),
which are nontrivial, but not quite the GSL we were seeking.
However, we note that in the semiclassical limit, these 
inequalities reduce to the GSL, and we leave open the 
possibility that they may lead to the desired result
with a more careful analysis.  

We conclude in section \ref{sec:discussion} by outlining
some questions raised by the present work,
and describing ideas 
for future applications 
to problems in semiclassical quantum gravity.

\paragraph{Note added:} During the final stages of writing this
paper, \cite{Ali:2024jkx} appeared which contains 
some overlap with the 
results presented here.  

\section{Crossed product and modular theory} \label{sec:crprod}

We begin with a brief overview of the crossed product construction
in semiclassical gravity, before describing the application of 
this construction to black hole horizons in section \ref{sec:semigsl}.  

Several recent investigations into semiclassical
quantum gravity have noted the relevance of crossed products
in the algebraic formulation of the theory \cite{Witten2021, 
Chandrasekaran2022a, Chandrasekaran2022b, 
Jensen2023, Kudler-Flam2023}.  The $G_N\rightarrow 0$ limit of quantum
gravity admits a description in terms of quantum field theory
in curved spacetime, 
in which local field algebras
are assigned to arbitrary causally complete subregions.
These algebras are generically von Neumann algebra
factors of type $\tthr_1$, and this structure 
is expected to persist to all orders in the $\sqrt{G_N}$
expansion \cite{Leutheusser:2021qhd, Leutheusser:2021frk, Witten2021}.  However, there exist certain nonlocal
observables such as the global ADM Hamiltonian that interact
nontrivially with these local algebras when going beyond the 
strict $G_N\rightarrow 0$ limit.  
In particular, the gauge constraints arising from diffeomorphism
invariance yield relations between the nonlocal observables
and local algebras which affect the structure of the algebras
even at $G_N = 0$.  Crossed product algebras are the result of 
imposing the constraints on the combined system of local algebras
and global observables.

The construction of gravitational crossed product 
algebras begins with the algebra
$\aqft$ of quantum fields restricted to a 
causally complete subregion
$\sr$ in spacetime.  The subregions of interest
in the present work are the exterior of a bifurcation
surface of an eternal 
black hole, or more generally the exterior
of a cut of the future horizon. $\aqft$  acts on a Hilbert space 
$\hqft$, and we take it to be a factor of type $\tthr_1$
in accord with arguments for the universal
type of subregion algebras in quantum field theory 
\cite{fredenhagen1985modular, buchholz1995scaling, Haag1992}.  
The additional nonlocal degree of freedom is 
an asymptotic gravitational charge, which will generally
be proportional to the perturbation of late-time area 
of a black hole horizon when
working with Killing horizons, and can be related to 
other asymptotic charges such as the ADM Hamiltonian of the spacetime.  
We represent this asymptotic charge
as minus the position operator acting 
on the Hilbert space $\hfut = L^2(\mathbb{R})$,
which reflects the semiclassical assumption that the 
spectrum of $\hat{q}$ is assumed to be continuous and 
unbounded above and below.  The crossed product
algebra $\wh{\alg}$ is the subalgebra of 
$\aqft \otimes \mathcal{B}(\hfut)$ that commutes 
with the complementary asymptotic charge associated, for example, 
with the 
left asymptotic area of a two-sided black hole.  
On the Hilbert space $\wh{\hs} = \hqft\otimes \hfut$, 
this left asymptotic charge is represented by the operator 
\beq
\hat{q}_L = \hat{q} + h_\Omega,
\eeq
where $h_\Omega = -\log \Delta_\Omega$ is the modular
Hamiltonian of a specific vacuum state $|\Omega\rangle \in
\hqft$.  The interpretation of $\hat{q}_L$ as a left 
asymptotic charge follows from a
diffeomorphism constraint in semiclassical 
quantum gravity, and additionally relies on the modular flow generated by 
$h_\Omega$ agreeing with the diffeomorphism
flow along a vector $\xi^a$ on a Cauchy surface for 
the subregion $\sr$.  
As we review in section
\ref{sec:hstr}, vacuum states of the horizon average null energy 
operators satisfy this property for algebras associated to
cuts of the Killing horizon.
We will refer to $|\Omega\rangle$ as the vacuum state
for the QFT algebra since it is a vacuum for the average null energy
operators, although it may not coincide with a global
minimal energy state.

$\wh{\alg}$ can be expressed as 
\beq
\wh{\alg} = \left\langle e^{i\hat{p} \hpsi} \msf{a} e^{-i\hat{p}\hpsi},
\hat{q}\right\rangle,\qquad \msf{a}\in\aqft,
\eeq
where the notation $\langle\cdot\rangle$ means the 
von Neumann algebra generated by the displayed operators by 
taking a double commutant.   The commutant algebra
$\wh{\alg}'$ naturally describes operators
in the causal complement $\sr'$ dressed to the left asymptotic
charge $\hat{q}_L$.  It consists of the operators
\beq
\wh{\alg}' = \left\langle \msf{a}', \hat{q}+\hpsi\right\rangle, \qquad \msf{a}'
\in\aqft'.
\eeq

We are interested in characterizing entropies of 
states on the algebra $\wh{\alg}$ and relating them
to generalized entropies for the Killing horizon.  
The specific states we will focus on
are the classical-quantum states $|\wh{\Phi}\rangle
= |\Phi,f\rangle\defeq |\Phi\rangle_\text{QFT}\otimes f(q)$.
The density matrix for this state 
can be computed by first determining the Tomita operator 
$S_{\wh{\Phi}}$, defined as the unbounded, antilinear
operator which acts on states of the form $\ho a|\wh\Phi\rangle$,
$\ho a\in\wh\alg$ as
\beq
S_{\wh\Phi} \ho a|\wh\Phi\rangle = \ho{a}^\dagger|\wh\Phi\rangle
\eeq
By solving this relation for a spanning 
set of operators of the form $\ho a = e^{i\hat{p}\hpsi} \msf{a}
e^{-i\hat{p}\hpsi} e^{iu\hat{q}}$ (see appendix \ref{app:modth}), 
one finds that 
\begin{align}
S_{\wh{\Phi}} &= J_{\Phi|\Omega} e^{-i\hat{p} \hpsi}\,
\ho s_{\wh{\Phi}} \, \ho s_{\wh{\Phi}}' 
\\
\ho s_{\wh{\Phi}} 
&= e^{i\hat{p}\hpsi}\Delta_{\Phi|\Omega}^{\frac12}
f^*(\hat{q}-\hpsi) e^{\frac{\hat{q}}{2}} e^{-i\hat{p}\hpsi}
\\
\ho s_{\wh{\Phi}}' 
&=\frac{1}{e^{\frac{\hat{q}}{2}}f(\hat{q}+\hpsi)} 
J_\Omega J_{\Omega|\Phi}\Delta_{\Omega|\Phi}^{\frac12}
\end{align}
where $\ho s_{\wh{\Phi}}$ is a linear operator
affiliated with $\wh{\alg}$
and $\ho s_{\wh{\Phi}}'$ with $\wh{\alg}'$.

In the special case of a vacuum classical-quantum state 
$|\wh\Omega\rangle = |\Omega,f\rangle$ with real wavefunction
$f^*(\hat{q}) = f(\hat{q})$, the expression simplifies 
dramatically such that the polar decomposition $S_{\wh{\Omega}}
=J_{\wh\Omega} \Delta_{\wh\Omega}^{\frac12}$ becomes manifest, with
\beq \label{eqn:DelhatPsi}
J_{\wh{\Omega}} = J_\Omega e^{-i\hat{p}\hpsi}, \qquad 
\Delta_{\wh\Omega} = \left[|f(\hat{q})|^2 e^{\hat{q}} \right]
\cdot\left[
\frac{e^{-\hat{q}}\Delta_\Omega}{|f(\hat{q}+\hpsi)|^2}\right]
=\rho_{\wh{\Omega}} \cdot (\rho_{\wh{\Omega}}')^{-1}.
\eeq
The density matrix
$\rho_{\wh{\Omega}}$ is an operator affiliated with $\wh{\alg}$,
and can be used to define a trace on the algebra according to
\cite{Witten2021}
\beq\label{eqn:Tr}
\tr(\ho a) = \langle \Omega,f|\rho_{\wh{\Omega}}^{-1} \ho a|\Omega,f\rangle
= 2\pi \langle \Omega, 0_p| e^{-\hat{q}}\, \ho a|\Omega,0_p\rangle,
\eeq
where $|0_p\rangle$ is the $\delta$-function normalized zero 
$\hat{p}$ eigenstate.  Note that the density matrix 
used to define this trace depends only on the operator
$\hat{q}$, and not on the modular operator $\Delta_\Omega$.  
This will be important  in section \ref{sec:gslsemi} when comparing traces
for crossed product algebras associated with different subregions.
These algebras do not obviously form subalgebras for nested
subregions since the modular Hamiltonian $\hpsi$ depends 
on the choice of subregion.  However, all the algebras share 
the operator $\hat{q}$, and  the density
matrix $\rho_{\wh{\Omega}}$ remains the same in each algebra,
leading to a standard choice for its normalization
and the normalization of the trace
in different crossed product algebras.  

Returning to the more general classical-quantum state 
$|\wh{\Phi}\rangle$, we obtain the density
matrix for $\wh{\alg}$ 
from the relation 
$\rho_{\wh{\Phi}} = \ho s_{\wh{\Phi}}^\dagger \ho s_{\wh{\Phi}}$.
This reproduces the expression derived in
\cite{Jensen2023},
\beq
\rho_{\wh{\Phi}} = e^{i\hat{p}\hpsi}f(\hat{q}-\hpsi) e^{\hat{q}}
\Delta_{\Phi|\Omega}f^*(\hat{q}-\hpsi) e^{-i\hat{p}\hpsi}.
\eeq
For computations of the entropy, it is convenient to rewrite
this density matrix as 
\begin{align}
\rho_{\wh{\Phi}} &= f(\hat{q})e^{\frac{\hat{q}}{2}}
e^{i\hat{p}\hpsi}\Delta_\Omega^{-\frac12}
\Delta_{\Phi|\Omega} \Delta_{\Omega}^{-\frac12}
e^{-i\hat{p}\hpsi} e^{\frac{\hat{q}}{2}}f^*(\hat{q})
\nonumber \\
&=
 f(\hat{q})e^{\frac{\hat{q}}{2}}
e^{i\hat{p}h_{\Omega|\Phi}}\Delta_{\Omega|\Phi}^{-\frac12}
\Delta_{\Phi} \Delta_{\Omega|\Phi}^{-\frac12}
e^{-i\hat{p}h_{\Omega|\Phi}} e^{\frac{\hat{q}}{2}}f^*(\hat{q})
\nonumber \\
&=
e^{i\hat{p}h_{\Omega|\Phi}}f(\hat{q}-h_{\Omega|\Phi}) e^{\hat{q}}
\Delta_\Phi f^*(\hat{q}-h_{\Omega|\Phi}) e^{-i\hat{p}h_{\Omega|\Phi}}
\end{align}
where in the second line we have used the cocycle relations
$\Delta_{\Omega}^{-\frac12}\Delta_{\Phi|\Omega}^{\frac12} = 
\Delta_{\Omega|\Phi}^{-\frac12}\Delta_\Phi^{\frac12}$ and the fact that 
these operators are affiliated with $\aqft$, as well as the 
fact that $h_\Omega$ and $h_{\Omega|\Phi}$ generate the same
action on operators in $\aqft$ (see e.g.\ \cite[Appendix C]{Jensen2023}).  
The logarithm
is then given by
\beq
-\log \rho_{\wh{\Phi}} = 
-h_{\Omega|\Phi} -\hat{q} -\log\left(f(\hat{q}) 
e^{i\hat{p}h_{\Omega|\Phi}}\Delta_\Phi e^{-i\hat{p}h_{\Omega|\Phi}}
f^*(\hat{q})\right). \label{eqn:logrho}
\eeq
Because the remaining operators inside the logarithm
do not commute, the exact expression for the entropy,
described in section
\ref{sec:pertent}, is somewhat complicated. 
The reason for the complications is that
despite appearances, the state $|\Phi,f\rangle$ describes
a state with entanglement between the quantum field degrees
of freedom and the asymptotic charge $\hat{q}$, coming from
the dressing factors $e^{i\hat{p}\hpsi} (\cdot) e^{-i\hat{p}\hpsi}$
for quantum field theory operators in the crossed product algebra.

However, when comparing to standard treatments of the generalized
entropy, there is a useful further restriction 
on the class of states.  This  involves taking the 
wavefunction $f(q)$ to be slowly varying, or, alternatively,
taking its Fourier transform $\tilde{f}(p)$ to be sharply
peaked at low momentum.  Since $\hat{p}$ has the interpretation
of the time shift generated by the asymptotic charge, 
these states describe semiclassical geometries 
in which this global time variable is well-localized.
Away from this limit, the quantum field operators experience
significant smearing in  time due to the factors 
of $e^{i\hat{p}\hpsi}$ acting on the wavefunction $f(q)$.
In keeping with the notation of previous works
\cite{Chandrasekaran2022a, Chandrasekaran2022b, Jensen2023}, 
we will refer to states
with $f(q)$ slowly varying as {\it semiclassical states}.

For semiclassical states, it is valid to treat the operator
$\hat{p}$ as a small parameter, and expand the logarithm in
(\ref{eqn:logrho})
in powers of $\hat{p}$.  Since we will be taking 
an expectation value in the state $|f\rangle$, 
each factor of $\hat{p}$ in the expansion will eventually
act on a wavefunction, resulting in a suppression by derivatives
of $f$ (see section \ref{sec:pertent} for a precise derivation of this 
expansion).  The strict semiclassical entropy comes from
keeping only the leading $\op(\hat{p}^0)$ term in the expansion,
for which the term inside the logarithm in (\ref{eqn:logrho}) becomes
$|f(\hat{q})|^2 \Delta_\Phi$.  The entropy formula in the semiclassical
limit then becomes \cite{Chandrasekaran2022a, Chandrasekaran2022b,
Jensen2023}
\begin{align}
S(\rho_{\wh{\Phi}}) = \langle \wh{\Phi}|-
\log\rho_{\wh\Phi}|\wh{\Phi}\rangle
&\approx 
\langle \wh{\Phi}| -h_{\Omega|\Phi} + h_\Phi -\hat{q}
-\log |f(\hat{q})|^2 |\wh\Phi\rangle
\\
&=
-S_\text{rel}(\Phi||\Omega) - \vev{\hat{q}}_{\wh{\Phi}} +
S_f^q
\label{eqn:SrhoSrel}
\end{align}
This entropy formula will be key for arriving at a second law 
for gravitational crossed product algebras in section
\ref{sec:gslsemi}.

\section{Semiclassical generalized second law}
\label{sec:semigsl}

Having introduced crossed product algebras and 
derived the entropy formula for semiclassical
states, the next task is to relate this formula to 
the generalized second law for Killing horizons.  This 
involves comparing the entropies computed 
for a state on two different crossed product algebras
associated with different cuts of the future Killing
horizon, and demonstrating that this entropy
monotonically increases.  Doing so invokes two
key insights from Wall's original proof of the GSL
\cite{Wall2011}.  
The first is the observation that for 
vacuum states of the horizon algebra, the modular
Hamiltonian for an arbitrary horizon cut 
generates a geometric flow along the horizon.  
The algebraic explanation for this property
is that nested cuts of the horizon define
half-sided modular inclusions \cite{Wiesbrock1992, Borchers2000}, 
which follows, as we review below,
from the average null energy condition.  
The second key insight is the relation between the 
relative entropy of states of the quantum
fields outside a horizon cut and the generalized 
entropy of the same region.  We give a novel 
derivation
of the geometric formulas leading to this identity 
in section \ref{sec:geometric}, utilizing
recent analyses of the phase space of general relativity
on null Cauchy surfaces.  This derivation
directly relates the identity to similar relations
employed by JSS in the construction of 
gravitational algebras for general subregions \cite{Jensen2023}.
Applied to semiclassical states for the crossed product
algebras, these two insights reduce the GSL to the monotonicity
of relative entropy under algebra inclusions, thereby
reproducing Wall's proof.

\begin{figure}
\centering
\tikz [scale = 0.8
        ]
        {
            \def \bhwidth {4};
            \def \bhcut {1.5};
            \def \Clabelarrlength {1};
            \def \Hbmlabel {-\bhwidth/2};

            \tikzdeclarepattern{
                name=mylines,
                parameters={
                    \pgfkeysvalueof{/pgf/pattern keys/size},
                    \pgfkeysvalueof{/pgf/pattern keys/angle},
                    \pgfkeysvalueof{/pgf/pattern keys/line width},
                },
                bounding box={
                    (0,-0.5*\pgfkeysvalueof{/pgf/pattern keys/line width}) and
                    (\pgfkeysvalueof{/pgf/pattern keys/size},
                    0.5*\pgfkeysvalueof{/pgf/pattern keys/line width})},
                tile size={(\pgfkeysvalueof{/pgf/pattern keys/size},
                \pgfkeysvalueof{/pgf/pattern keys/size})},
                tile transformation={rotate=\pgfkeysvalueof{/pgf/pattern keys/angle}},
                defaults={
                    size/.initial=5pt,
                    angle/.initial=45,
                    line width/.initial=.4pt,
                },
                code={
                    \draw [line width=\pgfkeysvalueof{/pgf/pattern keys/line width}]
                    (0,0) -- (\pgfkeysvalueof{/pgf/pattern keys/size},0);
                },
            }
            \tikzset{
                Cutfill/.style = {
                    pattern={mylines[angle=135, size = 5pt, line width=0.5pt]}, pattern color = red
                }
            }
            \tikzset{
                CCutfill/.style = {
                    pattern={mylines[angle=135, size=5pt, line width=0.5pt]},
                    pattern color = orange
                    }
                }
            
            \coordinate (ll) at (-\bhwidth,-\bhwidth);
            \coordinate (ul) at (-\bhwidth,\bhwidth);
            \coordinate (lr) at (\bhwidth,-\bhwidth);
            \coordinate (ur) at (\bhwidth,\bhwidth);
            \coordinate (bif) at (0,0);
            \coordinate (horcut) at (\bhcut,\bhcut);
            \coordinate (cutbdy) at (\bhwidth, 2*\bhcut-\bhwidth);
            \coordinate (cutsing) at (2*\bhcut - \bhwidth, \bhwidth);
            \coordinate (Blabel) at (\bhwidth*2/3, \bhcut-\bhwidth*2/3);
            \coordinate (Clabel) at (\bhwidth +0.5, \bhcut);
            \coordinate (Cplabel) at (-\bhwidth*1/4,\bhcut +\bhwidth*1/4);

            \fill[fill=blue!20] (bif) -- (ur) -- (lr) -- cycle;

            \fill[fill=green!90!black!30] (bif) -- (ll) -- (ul) -- cycle;

            \fill[Cutfill] 
                (horcut) -- (ur) -- (cutbdy) -- cycle;

            \filldraw [CCutfill]
                (ul) decorate [decoration=zigzag] {-- (cutsing)} 
                -- (horcut) -- (ll) -- cycle;
            \draw[decorate, decoration = zigzag] (cutsing) -- (ur);

            \draw (ll) -- (ul);
            \draw (lr) -- (ur);

            \draw [color = blue, line width = 4pt, line cap = round] (bif) -- (ur);
            \draw [color = red, line width = 1.5pt, line cap = round] (horcut) -- (ur);
            \draw (bif) -- (horcut);
            \draw (ul) -- (lr);

            \draw (horcut)--(cutbdy);
            \draw (horcut) -- (cutsing);

            \draw[decorate, decoration=zigzag] (ll) -- (lr);

            \filldraw [fill=blue] (bif) circle (3pt) node[below=3pt, color=blue]{${b}$};
            \filldraw [fill=red] (horcut) circle (3pt) node[left=6pt, color=red, circle, inner sep = 0pt, fill=white]{$c_\lambda$};
            \node (B) at (Blabel) [color=blue]{$R_b$};
            \draw [thick, color=black, ->] (Clabel) node[right=-3pt,color=red]{$R_\lambda$} -- ++(-\Clabelarrlength,0);
            \draw [thick, color=black, ->] (-\bhwidth-0.5,0) node[left=-3pt, color=green!60!black]{$R_b'$} -- ++(\Clabelarrlength,0);
            \node (Cp) at (Cplabel) [color=orange!90!black, circle, inner sep=0pt,
            fill=white]{$R_\lambda'$};


            \draw [pen colour = {black}, ultra thick, decorate, decoration = {calligraphic brace, raise=3pt, amplitude = 10pt, aspect=0.75}] (ur) -- (bif)
            node[pos=0.75,below right=3pt,blue]{$\mathscr{H}_{b}^+$};
            \draw [pen colour = {black}, ultra thick, decorate, decoration = {calligraphic brace, raise=3pt, amplitude = 10pt}] (horcut) -- (ur)
            node[pos=0.5,above left=5pt,red]{$\mathscr{H}_{\lambda}^+$};


        }

    \caption{The two-sided AdS black hole is shown in the Penrose
    diagram above.  The bifurcation surface is labeled as the point $b$, 
    and $\hor_b^+$ denotes the half of the event horizon to the future 
    of $b$.  $\hor_b^+$ serves as a Cauchy surface 
    for the right exterior region $R_b$, shown in blue.  The causal
    complement $R_b'$ involving the left exterior
    of the black hole is shown in green.  The red dot labeled $c_\lambda$ denotes a cut 
    of the future horizon, and the portion of the horizon to the future 
    of this cut, $\hor_\lambda^+$, is a Cauchy surface for the subregion
    $R_\lambda$ in the right exterior, depicted by the  red lines.  
    The causal complement of this region is $R_\lambda'$, labeled by 
    the  orange lines, and consists of the entire left exterior as 
    well as a portion of the black hole interior.  }
    \label{fig:AdSBH}
\end{figure}

\subsection{Half-sided modular inclusions for horizon cuts}
\label{sec:hstr}

The first step in formulating the GSL in  terms of crossed 
product algebras is to demonstrate that quantum field theory 
algebras 
associated with nested horizon cuts define half-sided modular
inclusions.  This fact was originally established 
for free and superrenormalizable theories by 
Wall \cite{Wall2011} (although not using the terminology of 
half-sided modular inclusions), and arguments extending this 
result to arbitrary interacting quantum field theories have been
given by Casini, Teste, and Torroba \cite{Casini2017}, and Witten
\cite{WittenLecture19}. This section serves as a review of these 
results, presented in a way that is directly amenable to application
to the generalized second law for black holes.  

For the initial discussion, we consider the case of a 
two-sided 
black hole in asymptotically anti-de Sitter space,
shown in figure \ref{fig:AdSBH}; 
generalizations including asymptotically flat and de Sitter
 black holes 
will be discussed 
in section \ref{sec:ainfty}.  AdS asymptotics 
have the advantage that boundary conditions 
prevent  radiation from escaping through
infinity, which implies that the black hole event horizon
$\hor$ for
the right asymptotic region defines a complete Cauchy surface 
for the spacetime.  
$\hor$ forms one component of a bifurcate Killing horizon,
and we denote the associated Killing vector as $\xi^a$
and the bifurcation surface at which $\xi^a$ vanishes 
as $b$. The bifurcation surface divides $\hor$ into two 
pieces, $\hor_{b}^+$ to the future and $\hor_{b}^-$ to the 
past.  $\hor_{b}^+$ is a Cauchy surface for the right exterior
region of the black hole, denoted $R_b$, and we define
$\abif = \alg(R_b)$ to be the type $\tthr_1$ von Neumann
algebra of quantum fields restricted to this region.  

To describe the horizon cut algebra, it is convenient to 
employ Gaussian null coordinates $(u,v,y^A)$ near $\hor$ (see
appendix \ref{app:gnuc}).  In these coordinates, $\hor$ 
is located at $u=0$, and $v$ is the affine parameter 
for the null horizon generator $l^a$, which on $\hor$
is related to the Killing vector according to $l^a = \frac{1}{\kappa v}
\xi^a$, where $\kappa$ is the horizon surface gravity. The 
bifurcation surface lies at $ u = v = 0$, and all surfaces 
of constant $v$ are null hypersurfaces associated with 
ingoing light rays that intersect $\hor$.  The coordinate $u$
is an affine parameter for the null tangent vector $n^a$
for these light rays, which is normalized
at $\hor$ to satisfy $n\cdot l = -1$.  A horizon cut $c_\lambda$ 
is specified by a function $\lambda(y^A)$ 
as the surface $v = \lambda(y^A)$ on $\hor$.  The cut 
divides the horizon into a future component $\hor_{\lambda}^+$
and a past component $\hor_{\lambda}^-$, with $\hor_{\lambda}^+$
serving as a Cauchy surface for the region $R_\lambda$ in the 
right exterior that is spacelike separated from the cut.  
The quantum field theory algebra for this region
is denoted $\acut_\lambda = \alg(R_\lambda)$, and the nesting
of the regions $R_\lambda \subset R_b$ implies
the inclusion of the associated algebras $\acut_\lambda \subset
\abif$.

The crucial object in demonstrating that this inclusion is 
half-sided modular is the average null energy
operator, defined in terms of an integral
of the quantum field stress tensor $T_{ab}$
over given null geodesic $\gamma_y$
on the horizon at fixed $y^A$ as
\beq
P(y^A) = \int_{\gamma_y} dv\, T_{ab}l^a l^b 
= \int_{-\infty}^\infty
dv\, T_{vv}(v, y^A).
\eeq
We will make use of  smeared versions of this operator
with respect to a function $\rho(y^A)$,
defined as the stress-energy flux associated with the 
vector $\rho l^a$,
\beq\label{eqn:Psigma}
P_\rho = -\int_{\hor}T\indices{^a_b}\rho l^b\epsilon_{a\ldots}
=\int_{-\infty}^{\infty} dv \int dy^A\sqrt{q}\,T_{vv}
(v,y^A)\,\rho(y^A).
\eeq
Here, the orientation conventions for the volume forms are
$\epsilon = -l\wedge \eta$, where $\epsilon$ is the spacetime
volume form, and $\eta = \sqrt{q} dv\wedge dy^1\wedge \ldots \wedge
dy^{d-2}$ is the induced volume form on the horizon relative the 
affinely parameterized null normal, i.e.\ $\eta = i_l\epsilon$,
and $\sqrt{q}$ determinant of the spatial metric on the horizon.  
$P_\rho$ generates a geometric action on fields localized on the 
horizon, translating a local operator $\op(v,y^A)$ along
the null coordinate to an
operator $e^{i sP_\rho}\op(v,y^A)
e^{-is P_\rho} = \tilde\op(v+s\rho(y^A), y^A)$  at the 
shifted affine coordinate $v+s\rho(y^A)$.  
The precise form of the translated operator
$\tilde\op$ depends on its spin and scaling dimension,
but the important point is that $P_\rho$ sends 
local operators on the horizon to local operators
at the translated coordinate.

Away from $\hor$, $P_\rho$ generates a nonlocal action on
individual operators
since the vector field $\rho l^a$ does not extend
to a Killing vector on the full spacetime.  However,
its geometric action on horizon operators 
implies an overall
geometric action on the algebras determined by horizon cuts.  
Schematically, because $\hor_\lambda^+$ serves as a Cauchy 
surface for the region $R_{\lambda}$ exterior to the cut,
the operators in $R_{\lambda}$ are dynamically determined
by the operators on $\hor_\lambda^+$.  Hence, because 
$e^{is P_\rho}$ translates operators located on $\hor_{\lambda}^+$
to operators located on the translated future horizon $\hor_{{\lambda +s \rho}}^+$,
we should expect it to map the full algebra $\acut_\lambda$ associated with $R_\lambda$
into the algebra $\acut_{\lambda+s \rho}$ associated with the exterior
region $R_{\lambda+s\rho}$ for the translated cut.  
This argument for the geometric action of $P_\rho$ on the algebras $\acut_\lambda$
was first made by Wall in the case of free field theories by
taking advantage of the fact that one can obtain a nontrivial horizon algebra
by smearing certain local operators with functions supported purely on
$\hor$ \cite{Wall2011}.  For interacting field theories, 
the argument is more subtle because local operators must be smeared
over open regions in spacetime in order to produce operators with
finite fluctuations \cite{Bousso2014}.  Here, we will describe an argument
due to Witten \cite{WittenLecture19} 
for the validity of this geometric action for interacting 
field theories; closely related arguments were originally given
by Casini, Teste, and Torroba \cite{Casini2017}.

Although there are no well-defined operators that are smeared solely
on the horizon in interacting theories, the unsmeared local operators
$\op(x)$ nevertheless have finite matrix elements 
$\langle \chi|\op(x)|\phi\rangle$ for a sufficiently large number 
of states $|\phi\rangle, |\chi\rangle$ in the QFT Hilbert space $\hqft$.
For example, these states could be obtained by acting on the vacuum
with operators smeared in a region spacelike separated from $\op(x)$.
Such an object is known as a {\it sesquilinear form}, defined as 
a map $\mathfrak{s}(|\chi\rangle,|\phi\rangle)$ from vectors in a domain
$\mathcal{D}(\mathfrak{s})\subset\hqft$ to $\mathbb{C}$ that is conjugate linear in
the first argument and linear in the second \cite{Schmudgen2012}.  Sesquilinear forms
generalize the notion of an unbounded operator to objects such
as $\op(x)$ which make sense inside of certain matrix elements
but tend to produce nonnormalizable states when acting on 
a vector.  
The domain associated to a given local operator $\op(x)$, viewed
as a sesquilinear form, is expected to be dense in $\hqft$, since,
for example, the object $\frac{1}{(1+H)^n} \op(x) \frac{1}{(1+H)^n}$ will be 
a bounded operator for sufficiently high powers of $n$, where $H$ 
is the positive Hamiltonian associated with global future-directed 
time evolution \cite{Haag1963, Fredenhagen1981}; the domain of $\op(x)$ is then
 contained in the domain of $H^n$.
 Hence, the future horizon of a cut $\hor_\lambda^+$
is associated with a set $\mc{S}_\lambda$ of densely defined
sesquilinear forms constructed 
from local operators on $\hor_\lambda^+$, even though none of these define 
true operators with finite fluctuations on $\hqft$.

The full algebra $\acut_\lambda$ can be constructed from $\mc{S}_\lambda$
by taking a double commutant.  A bounded operator $\msf{b}'$ is said
to commute with a sesquilinear form $\mathfrak{s}$ if $\msf{b}'$ preserves
the domain $\mathcal{D}(\mathfrak{s})$ and 
$\mathfrak{s}\big((\msf{b}')^\dagger|\chi\rangle,|\phi\rangle\big)
=\mathfrak{s}\big(|\chi\rangle,\msf{b}'|\phi\rangle\big)$ for all
$|\chi\rangle,|\phi\rangle\in\mathcal{D}(\mathfrak{s})$.  For the forms
defined by local operators, this condition simply states that 
$\langle \chi|\msf{b}'\op(x)|\phi\rangle 
= \langle\chi|\op(x)\msf{b}'|\phi\rangle$ in states for which these
expressions are finite.  The commutant $\mathcal{S}_\lambda'$ is then 
an algebra of bounded operators localized in the region $R_\lambda'$ 
to the left of the cut, shown in figure \ref{fig:AdSBH}.  Taking a second commutant
results in the algebra of operators to the right
of the cut, $\acut_\lambda = \mc{S}_\lambda''$, which is a von
Neumann algebra since it is defined as a commutant of a collection
of bounded operators $\mc{S}_\lambda'$.
Note that $\mc{S}_\lambda'$ itself need not be a von Neumann
algebra, since $\mc{S}_\lambda$ is not an algebra of bounded operators,
and hence is not contained in $\mc{C}_\lambda$.  The full algebra
for the region $R_\lambda'$ can be defined as $\alg(R_\lambda') = \acut_\lambda'
= \mc{S}_\lambda'''$.
The geometric action of $P_\rho$ on local operators on the horizon
implies that $e^{isP_\rho} \mc{S}_\lambda e^{-is P_\rho} 
= \mc{S}_{\lambda+s\rho}$.  It is also clear then that
$e^{isP_\rho}$ will conjugate elements
of $\mc{S}_\lambda'$ into operators that commute with $\mc{S}_{\lambda + s\rho}$,
and hence $e^{is P_\rho} \mc{S}_\lambda' e^{-isP_\rho} = 
\mc{S}_{\lambda + s\rho}'$.  The same argument applies to $\acut_\lambda
=\mc{S}_\lambda''$, and hence we conclude that $e^{isP_\rho}\acut_\lambda
e^{-isP_\rho} = \acut_{\lambda + s\rho}$, showing that 
the geometric action of $P_\rho$ extends to the horizon cut algebras as a whole.

We now apply this result on the geometric action of the null translations
to the algebra $\abif = \acut_0$ associated with the exterior of the horizon
bifurcation surface.  Fixing a nonnegative profile for the smearing function
$\lambda(y^A)$, we define $U_\lambda(s) = e^{isP_\lambda}$ to be the unitary
implementing a finite null translation.  By the above discussion,
we find that for $s\geq 0$,
\beq
U_\lambda(s) \abif U_\lambda(s)^\dagger =\acut_{s\lambda} \subseteq \abif.
\eeq
This is one necessary condition in order to conclude that $U_\lambda(s)$
generates a {\it half-sided translation} for $\abif$, a concept
introduced by Borchers \cite{Borchers1992, Borchers2000}.  The other
conditions are that $U_\lambda(s)$ have a positive generator, and 
that it preserve a vacuum state
$|\Omega\rangle$ that is cyclic and separating for $\abif$.
Positivity of the generator $P_\lambda$ is implied by the 
average null energy condition (ANEC)
\cite{Roman:1986tp, Borde:1987qr, Roman:1988vv}, 
since $P_\lambda$ is an integral
of the average null energy of each horizon generator, weighted by a positive
function $\lambda$.   Rigorous proofs of the ANEC exist for interacting 
field theories in Minkowski space \cite{Faulkner:2016mzt, Hartman:2016lgu,
Kravchuk:2018htv}.  
In curved spacetimes, it is generally expected to hold for achronal null
geodesics for backgrounds satisfying the semiclassical
Einstein equation \cite{Wald:1991xn, Penrose:1993ud, Graham:2007va, 
Witten:2019qhl, Kontou:2020bta}.
In the present context, since the horizon generators are 
achronal, it is reasonable to assume that the ANEC holds for them,
implying that the $P_\lambda$ are positive operators.
The vacuum state $|\Omega\rangle$ is then simply a ground state
for the smeared 
average null energy operator, satisfying $P_\lambda|\Omega\rangle = 0$.
Note that since average null energy operators on different horizon
generators commute, any other choice of smearing $\rho$ yields
an operator satisfying $[P_\rho,P_\lambda]=0$, and hence
$|\Omega\rangle$ can be defined as a simultaneous ground state
of all smeared horizon average null energy operators \cite{Wall2011}.

Together, the conditions (i) $P_\lambda \geq 0$, (ii) $U_\lambda(s)
\abif U_\lambda(s)^\dagger \subseteq \abif$ for $s\geq 0$, and (iii) 
$P_\lambda|\Omega\rangle = 0$ with $|\Omega\rangle$ cyclic and 
separating for $\abif$ imply that $U_\lambda(s)$ defines a 
half-sided translation for $\abif$.
Borchers's theorem for half-sided translations then states
that the modular flow $\Delta_\Omega^{-is}$
for the vector $|\Omega\rangle$ combines with translations
to form a unitary
representation of the one-dimensional affine group,
with the relation
\beq \label{eqn:affine}
\Delta_\Omega^{-is} U_\lambda(t) \Delta_\Omega^{is} = 
U_\lambda(e^{2\pi s} t).
\eeq
This relation holds for any choice of
smearing $\lambda$ since $|\Omega\rangle$ is the simultaneous
ground state of all smeared translation generators $P_\lambda$.

Specializing for the moment to the constant smearing
$\lambda = 1$, we can demonstrate that modular
flow acts geometrically on local operators on the horizon.
We first express a given horizon operator $\op(v)$
(leaving the $y^A$ dependence implicit) as the translation
of an operator at $v=0$, $\op(v) = U_1(v)\op_1(0)U_1(v)^\dagger$.
Because $\op_1(0)$ lies on the bifurcation surface, it should
commute with operators in both $\abif$ and $\abif'$, and hence
modular flow should map it to another local operator at $v=0$,
$\Delta_\Omega^{-is}\op_1(0)\Delta_\Omega^{is} = 
\op_2(0)$.  Then we find that
\begin{align}
\Delta_\Omega^{-is}\op(v)\Delta_\Omega^{is}
&=\Delta_\Omega^{-is}U_1(v)\Delta_\Omega^{is}
\op_2(0)
\Delta_\Omega^{-is}U_1(v)^\dagger\Delta_\Omega^{is}
=
U_1(e^{2\pi s} v)\op_2(0) U_1(e^{2\pi s} v) 
\nonumber \\
&= \op_3(e^{2\pi s} v),
\label{eqn:opvconj}
\end{align}
for some local operator $\op_3$.  This is precisely the form
of a geometric flow along the vector field 
$2\pi v\left(\frac{\partial}{\partial v}\right)^a 
= \frac{2\pi}{\kappa}\xi^a$, whose generator
can be expressed as an integral over the horizon
of the stress tensor weighted by $\xi^a$.  From
this argument, we conclude that the vacuum modular 
Hamiltonian $h_\Omega^{\abif}$ for $\abif$ is given by
\beq \label{eqn:hOmega}
h_\Omega^{\abif} = -\log \Delta_\Omega = 
-\frac{2\pi}{\kappa}\int_{\hor} T\indices{^a_b}\xi^b\epsilon_{a\ldots}
= 2\pi \int_{-\infty}^\infty dv \int dy^A \sqrt{q}
\,T_{vv}\, v.
\eeq

Because we did not specify the form
of the operator $\op_3$ in equation (\ref{eqn:opvconj}),
in principle we only determine $h_\Omega^{\abif}$ up to 
transformations that fix its location such as local Lorentz
or internal symmetry transformations.  However, the fact that 
equation (\ref{eqn:affine}) implies the
commutation relation
\beq \label{eqn:hPbrack}
[h_\Omega^{\abif}, P_1] = -i2\pi  P_1
\eeq
fixes the form of $h_\Omega^{\abif}$ in terms of the 
stress tensor, up to terms that commute with $P_1$. Such 
terms are related to the algebra at infinity, and are
discussed in section \ref{sec:ainfty}. 
Equation (\ref{eqn:hOmega}) is the standard expression
for the modular Hamiltonian of the Hartle-Hawking
state (in situations where it exists) in terms of the generator
of the Killing symmetry.  However, for more general Killing horizons
such as Kerr black holes and Schwarzschild-de Sitter, equation
(\ref{eqn:hOmega}) remains valid even though there is no
global Hartle-Hawking state.

Finally, we obtain the form of the modular Hamiltonian for an
arbitrary horizon cut $\acut_\lambda$.  First we note that 
$\acut_\lambda = U_\lambda(1) \abif U_\lambda(1)^\dagger$,
which, along with the relation (\ref{eqn:affine}) demonstrates that
for $s\geq 0$,
\beq
\Delta_{\Omega,\abif}^{-is}\,\acut_\lambda\, \Delta_{\Omega,\abif}^{is}
= U_\lambda(e^{2\pi s}-1)\,\acut_\lambda\, U_\lambda(e^{2\pi s}-1)^\dagger
\subseteq \acut_\lambda.
\eeq
Assuming $|\Omega\rangle$ is cyclic for $\acut_\lambda$, this
property implies that $\acut_\lambda\subset\abif$ is a
{\it half-sided modular inclusion}, as defined by 
Wiesbrock \cite{Wiesbrock1992}.  Since $U_\lambda(1)|\Omega\rangle
=|\Omega\rangle$, the modular operator for $\acut_\lambda$ 
is simply obtained from $\Delta_{\abif}$ by conjugation:
$\Delta_{\acut_\lambda} = U_\lambda(1)\Delta_{\abif} U_\lambda(-1)$
(leaving the dependence on $|\Omega\rangle$ implicit).
Again applying (\ref{eqn:affine}), this results in 
\beq
\Delta_{\abif}^{-is}\Delta_{\acut_\lambda}^{is} = 
U_\lambda(e^{2\pi s}-1),
\eeq
which leads to the relation for the modular Hamiltonians
\beq
h_{\abif} - h_{\acut_\lambda} = 2\pi P_\lambda.
\eeq
Since both $h_{\abif}$ and $P_\lambda$ have been expressed
in terms of integrals of the stress tensor in
equations (\ref{eqn:hOmega}) and (\ref{eqn:Psigma}),
we see that $h_{\acut_\lambda}$ may also be expressed 
as a stress tensor integral over the horizon according to
\beq \label{eqn:hclambda}
h_{\acut_\lambda} = -\frac{2\pi}{\kappa}\int_{\hor}
T\indices{^a_b}\zeta_\lambda^b \epsilon_{a\ldots}
=2\pi\int_{-\infty}^\infty dv \int dy^A\sqrt{q}\, T_{vv}\cdot(v-\lambda(y^A)),
\eeq
where we have defined $\zeta_\lambda^a = \xi^a - \kappa\lambda l^a$.
This is Wall's expression for the horizon-cut modular
Hamiltonian \cite{Wall2011}.  

The argument described above for deriving this equation
for interacting theories using half-sided translations and modular
inclusions was first given
for Rindler horizons by Casini, Teste, and Torroba \cite{Casini2017}.
The idea to define
the exterior algebras $\acut_\lambda$ 
as a double commutant of a set $\mc{S}_\lambda$ of sesquilinear
forms consisting of  local operators on the horizon appears in
a lecture by Witten \cite{WittenLecture19}, and immediately
allows one to see that the horizon average null energy operators
generate  half-sided translations.  The idea to relate the 
geometric action of modular Hamiltonians on Killing horizons to 
half-sided modular inclusions appears in work by Summers and 
Verch \cite{Summers:1995kp}, and is closely related to 
older arguments by Sewell for the Hawking temperature 
of Killing horizons in the Wightman formulation of axiomatic 
quantum
field theory \cite{Sewell:1982zz}.  These latter two works
attempt to make sense of a net of operators strictly localized
to the Killing horizon, which appears to only be valid in free 
field theory.  In interacting theories, there are arguments
that no bounded operators can be strictly localized to 
a finite affine parameter interval of a 
codimension-1 null surface \cite{Bousso2014}, and in this case
one must use the more sophisticated 
arguments developed by Casini, Teste, and Torroba, 
and Witten to conclude that horizon cuts define
half-sided modular inclusions.

\subsection{Geometric horizon identities}
\label{sec:geometric}

Connecting the results in section \ref{sec:hstr}
on the form of the horizon cut modular Hamiltonian
to the generalized entropy requires
a discussion of how the quantum fields couple to gravity
in the semiclassical limit $G_N\rightarrow 0$.  
This limit suppresses backreaction, 
allowing for a description in terms of quantum fields 
on the fixed black hole geometry.  However, we must also
account for the gravitational constraints when constructing
a theory that consistently embeds into the interacting 
theory away from $G_N=0$.  
These constraints lead to relations between the 
stress-energy on the horizon and the values of asymptotic
gravitational charges evaluated at the future and past 
boundaries of the horizon $c_{\pm\infty}$ located
at $v\rightarrow\pm\infty$.  In addition, semilocal
constraints associated with one side of a horizon
cut lead to formal relations between the area of the cut
and the asymptotic charges, which are necessary for 
establishing the equivalence of the entropy of the gravitational
algebra and the horizon generalized entropy.  In this section,
we review how these constraints arise  and present
a novel derivation of the quasilocal identity relating the 
horizon area and one-sided boost energy.  This identity
is originally due to Wall \cite{Wall2011}; see also 
\cite{Kudler-Flam2023} for a related recent treatment.

The structure of the constraints and asymptotic charges
can be derived from the classical theory at finite $G_N$;
from these one can obtain perturbative expressions
in the $\sqrt{G_N}$ expansion that then apply
in the semiclassical quantum theory.
Due to the choice of AdS asymptotics, the black hole 
horizon $\hor$ serves as a null Cauchy surface for the 
spacetime.  The canonical formulation of general
relativity on null surfaces has been the subject of 
several recent investigations \cite{CFP2018, Chandrasekaran2020, 
Chandrasekaran2021gen, Ciambelli:2023mir, Odak:2023pga, 
Chandrasekaran:2023vzb, Adami:2021nnf}
and can immediately be applied in the present context 
to derive the form of the Hamiltonians and constraints for
the geometric flows considered in section \ref{sec:hstr}.
We will simply quote some of the main results needed from the 
canonical analysis in this section; additional details can be 
found in appendix \ref{app:canon} and in the cited references.  

We denote by $\phi$ the dynamical fields in the theory,
which consist of the metric $g_{ab}$ and any matter 
fields $\psi$. 
On the null Cauchy surface, 
one can define a symplectic potential current
$\beom[\phi; \delta\phi]$ which, assuming 
minimal coupling between matter and gravity,
is expressed as a sum of a gravitational term and a matter term,
\beq
\beom = \beom^g+\beom^\psi.
\eeq
The gravitational piece, after imposing the gauge conditions 
given in (\ref{eqn:CFPgauge}), 
takes the form
\begin{align}\label{eqn:Eg}
\beom^g[g_{ab};\delta{g}_{ab}] = 
\frac{1}{16\pi G_N} \eta\left(
\sigma^{ab}\delta q_{ab} -\frac{d-3}{d-2}
\Theta q^{ab}\delta q_{ab}\right),
\end{align}
where $\eta$ is the volume form on $\hor$, 
$q_{ab}$ is the degenerate induced metric,
$\Theta = \nabla_a l^a$ is the expansion, and
$\sigma_{ab} = \frac12 \lie_l q_{ab} - \frac{1}{d-2}\Theta q_{ab}$
is the shear, which vanishes in the background but is nonvanishing 
once we consider perturbative fluctuations of the geometry.  
The matter symplectic potential $\beom^\psi$ is theory dependent, but
for definiteness one can have in mind a 
scalar field $\varphi$, for which 
\beq
\beom^\varphi = \eta (l^a\nabla_a \varphi)\delta\varphi.
\eeq
The symplectic form is obtained from $\beom$ by integrating 
its antisymmetric variation 
over the Cauchy surface $\hor$,
\beq
\Omega[\delta_1\phi, \delta_2\phi] = 
\int_\hor \Big(\delta_1\beom[\delta_2\phi]
-\delta_2\beom[\delta_1\phi]
\Big).
\eeq

For a vector field $\zeta^a$ tangent to $\hor$, the associated
Hamiltonian $H_\zeta$ can be defined as (see equation (\ref{eqn:Hxi}))
\beq
H_\zeta = \int_\hor \beom[\phi;\lie_\zeta\phi]
= \frac{1}{16\pi G_N}\int_\hor \eta\Big(\sigma^{ab}\lie_\zeta q_{ab}
-\frac{d-3}{d-2}\Theta q^{ab}\lie_\zeta q_{ab}\Big)
- \int_\hor (T_{(\psi)})\indices{^a_b}\zeta^b \epsilon_{a\ldots}.
\eeq
which displays the contributions from both
the gravitational field and the matter stress 
tensor $(T_{(\psi)})\indices{^a_b}$.
This object generates the flow 
of the diffeomorphism $\zeta^a$ on the gravitational
phase space since it satisfies Hamilton's equation
\beq
\delta H_\zeta = \Omega[\delta\phi, \lie_\zeta\phi],
\eeq
which holds off-shell as long as $\zeta^a$ preserves the asymptotic
boundary conditions imposed at $v\rightarrow\pm\infty$.

Due to diffeomorphism invariance, the stress-energy density 
$\beom[\phi;\lie_\zeta\phi]$ can be written as a total 
derivative $dM_\zeta$, with $M_\zeta$ defined in equation
(\ref{eqn:Mxi}), up to a term involving the constraint 
equation $C_\zeta$, 
\beq\label{eqn:EdM}
\beom[\phi;\lie_\zeta\phi] = d M_\zeta + C_\zeta,
\eeq
Integrating this relation over $\hor$ yields a formula
expressing the integrated constraint $\constr[\zeta]
=\int_\hor C_\zeta$ 
in terms of the generator of the diffeomorphism $H_\zeta$ and the 
asymptotic charges $\qasy_\zeta^\pm$, 
\beq \label{eqn:globconstr}
\constr[\zeta] = H_\zeta - \qasy_\zeta^+ + \qasy_\zeta^-,
\eeq
where
\beq
\qasy_\zeta^\pm = \int_{c_{\pm\infty}}M_\zeta.
\eeq

We now specialize to the vector fields that are relevant
to the discussion of half-sided modular inclusions considered
in section \ref{sec:hstr}, all of which are parallel to the null generator
$l^a$ on $\hor$.  They have the property of being  {\it quasi-Killing
vectors},\footnote{This terminology was introduced by Bob
Wald in a talk related to his work with Hollands and Zhang on
dynamical black hole entropy \cite{Hollands:2024vbe}.} 
meaning that they satisfy Killing's equation $\lie_\zeta 
g_{ab} = 0$ only on $\hor$, i.e.\ they are symmetries of the horizon
which may not extend to symmetries of the full spacetime.  
A generic quasi-Killing vector parallel to $l^a$ on $\hor$ 
can be written in terms of a function $\tau(v,y^A)$ as (see appendix
\ref{app:quasi-Killing}) 
\beq
\zeta^a = \tau l^a + u \nabla^a \tau.
\eeq
The choice $\tau = \kappa v$ yields the global
Killing vector $\xi^a = \kappa v l^a-\kappa u n^a$, 
while $\tau = \lambda(y^A)$ results in
the smeared null translation that appears in the average
null energy operator (\ref{eqn:Psigma}).

The boost vector $\zeta_\lambda^a$ occurring in the modular
Hamiltonian for a horizon cut is also quasi-Killing,
arising from the smearing $\tau(v, y^A) = 
\kappa (v - \lambda(y^A))$ which leads to the expression
\beq \label{eqn:zetalambda}
\zeta_\lambda^a = \kappa\Big[( v - \lambda)l^a -  u \lambda n^a 
- u\nabla^a\lambda\Big].
\eeq
This vector vanishes at the horizon cut $c_\lambda$ at which $ v = 
\lambda$, and its covariant derivative there satisfies
\beq \label{eqn:dzeta}
\nabla_a (\zeta_\lambda)_b \overset{c_\lambda}{=} \kappa n^\lambda_{ab},
\eeq
where
\beq \label{eqn:nabl}
n^\lambda_{ab} = 2 l_{[a} (n_{b]} +\nabla_{b]}\lambda) 
\eeq
is the unit binormal to $c_\lambda$. 
To see this, note that the cut is defined by
the relations $u = 0$ and $v-\lambda(y^A) = 0$,
so (\ref{eqn:nabl}) is proportional to the unit binormal
since $n_a = -\nabla_a v$.  To verify it is unit-normalized,
 we compute
 \beq
 n^\lambda_{ab} n_\lambda^{ab} = 2 (l\cdot l) |n + d\lambda|^2
 - 2 \Big(l\cdot(n+d\lambda)\Big)^2 \overset{\hor}{=} -2.
\eeq

It is notable that $\kappa$ in equation (\ref{eqn:dzeta}) is constant, 
since it means that the modular flow for the horizon
cut corresponds to a boost with constant surface gravity.  
This is a nontrivial verification of the geometric modular flow
conjecture from \cite{Jensen2023}, which states that a modular 
flow should look like a constant-surface-gravity boost
near the entangling surface.  Constancy of $\kappa$ can be
viewed as a {\it zeroth law of modular flow}, since it is closely
related to the KMS condition for the modular flow on the 
horizon cut algebra.  It is notable that $\kappa$ is constant
for a generic cut of the horizon, and does not depend on quantities
such as the extrinsic curvature of the codimension-2 cut, which 
is nonzero in the $u$ direction away from the bifurcation surface.

For the translation quasi-Killing vector $\lambda l^a+ u\nabla^a \lambda$, 
the gravitational contribution to the energy density
obtained from (\ref{eqn:Eg}) is 
\begin{align}
\beom^g[\lie_{\lambda l} g_{ab}] &= 
\frac{1}{8\pi G_N}\eta \,\lambda
\Big(
\sigma^{ab}\sigma_{ab} - \frac{d-3}{d-2}\Theta^2
\Big),
\end{align}
which motivates the definition of the gravitational null energy
density as\footnote{Closely related expressions for
the gravitational energy associated with generic
null surfaces were recently explored in \cite{Ciambelli:2023mir}.} 
\beq \label{eqn:tgvv}
t^{(g)}_{vv} = \frac{1}{8\pi G_N}\left(\sigma^{ab}\sigma_{ab} - 
\frac{d-3}{d-2}\Theta^2\right).
\eeq
Note that this definition of  gravitational null energy
makes sense in principle on generic null surfaces, not 
only for perurbations of a Killing horizon.  
On the other hand, the potential $M_{\lambda l}$ evaluates to
(see equation (\ref{eqn:CFPpot}))
\beq
M_{\lambda l} = -\frac{1}{8\pi G_N}\mu \lambda \Theta,
\eeq
with $\mu$ the induced $(d-2)$-form that pulls back to 
the area form on cuts of $\hor$.
From this, we can then integrate equation (\ref{eqn:EdM})
between two horizon cuts $v=\tau_1(y^A)$ and $v=\tau_2(y^A)$
to derive a relation between the expansions at the two cuts,
\beq\label{eqn:intRay}
\frac{1}{8\pi G_N}\left(\int_{\tau_2}\mu \lambda \Theta
-\int_{\tau_1}\mu \lambda\Theta\right)
=
-\int_{\hor_1^2}\eta\lambda\left(
t^{(g)}_{vv}+ T^{(\psi)}_{ab}l^al^b \right)
+\constr_1^2[\lambda l]
\eeq
where $\constr_1^2[\lambda l]$ is the integral of the constraint
$C_{\lambda l}$ over $\hor_1^2$.  

Equation (\ref{eqn:intRay}) can be viewed
as an integrated version of the Raychaudhuri equation,\footnote{An 
alternative way to obtain this formula 
is to write $d(\Theta \mu) = (\dot\Theta + \Theta^2)\eta$ with 
$\dot\Theta = l^a\nabla_a\Theta$.  One then uses
the Raychaudhuri equation to obtain $\dot \Theta + \Theta^2
= \frac{d-3}{d-2}\Theta^2-\sigma^{ab}\sigma_{ab} - T^{(\psi)}_{ab}l^a l^b
+ C_l$, with $C_l=0$ a component of the Einstein equation. }
which highlights the interpretation of the terms in the equation
as contributions from the matter and gravitational null energy.
As the second cut $\tau_2$ approaches future infinity on the 
horizon, the expansion must go to zero as a boundary condition
specifying that $\hor$ is an event horizon \cite{Wall2011, 
Hollands:2012sf}.  This limit results in an expression
for the expansion on a cut $c_\tau$ in terms of the semi-infinite
null energy (once the constraints $C_{\lambda l} = 0$ are imposed)
\beq\label{eqn:semianec}
\frac{1}{8\pi G_N}\int_{c_\tau} \mu\lambda\Theta = P_\tau^+ = 
\int_{\hor_\tau^+}\eta\lambda\left(t^{(g)}_{vv}+ T^{(\psi)}_{ab}l^al^b \right).
\eeq

Another useful relation comes from examining the charges 
for the boost vector $\zeta_\lambda^a$.  From equation 
(\ref{eqn:CFPpot}), the potential $M_{\zeta_\lambda}$
evaluates to
\beq
M_{\zeta_\lambda} = \frac{\kappa}{8\pi G_N}\mu\Big(1-(v-\lambda) \Theta\Big),
\eeq
and after integrating the flux-balance equation (\ref{eqn:EdM})
between the cut $c_\lambda$ at which $\zeta^a_\lambda$ vanishes
and infinity, we obtain
\beq \label{eqn:horcutid}
\frac{\kappa}{8\pi G_N}(A_\infty - A_\lambda) = H_{\zeta_\lambda}^+
= \kappa\int_{\hor_\lambda^+}\eta (v-\lambda)\left(t^{(g)}_{vv}
+T^{(\psi)}_{ab}l^a l^b\right),
\eeq
again using the future boundary condition $\Theta_{+\infty} = 0$
and imposing the constraint $C_{\zeta_\lambda} = 0$.
This is the nonlinear version of Wall's relation \cite{Wall2011} between
the difference of horizon areas $(A_\infty - A_\lambda)$
and the one-sided boost
energy $H_{\zeta_\lambda}^+$.  It also is the identity corresponding 
to the local Smarr relation discussed by JSS 
\cite{Jensen2023}, specialized
to the case of a cut of a Killing horizon.  
Note that the boost energy
contains both a matter contribution from $T^{(\psi)}_{ab}$ 
and a gravitational contribution $t^{(g)}_{vv}$
involving the shear and expansion according to
(\ref{eqn:tgvv}).
This latter  contribution will
become the graviton null energy density when quantizing
around the Killing horizon background.

The crossed product algebra is the result of imposing the global
constraint (\ref{eqn:globconstr}) for $\zeta_\lambda^a$,
which is the quasi-Killing vector that 
preserves the horizon cut.  We already found in equation 
(\ref{eqn:horcutid}) that the future asymptotic charge
is given by the late time area of the event horizon,
\beq
\qasy_{\zeta_\lambda}^+ = \frac{\kappa A_\infty}{8\pi G_N}.
\eeq
The asymptotic gravitational 
charge at past infinity  is given by
\beq \label{eqn:qasy-}
\qasy_{\zeta_\lambda}^- = \lim_{\tau\rightarrow-\infty}\frac{\kappa}{8\pi G_N}
\left(A_\tau - \int_{c_\tau} \mu (v-\lambda)\Theta\right).
\eeq
Unlike the future horizon boundary, we cannot impose
$\Theta\overset{\tau\rightarrow -\infty}{\longrightarrow} 0$
since the asymptotic value of $\Theta$ is determined via (\ref{eqn:semianec}) by the total 
average null energy on the horizon.  This raises 
the possibility that $\qasy_{\zeta_\lambda}^-$ could 
diverge in the limit.  However, at least when working 
perturbatively about a Killing horizon, the expansion $\Theta$
remains small up to late time scales of order 
$v\sim \op(\frac{1}{G_N})$.  
In this perturbative regime, the terms growing linearly 
with $v$ in the area $A_\tau$ cancel against the linear term
in $v$ in the integral involving the expansion in 
(\ref{eqn:qasy-}), resulting in a finite expression if 
we take $\tau\rightarrow-\infty$ and $G_N\rightarrow 0$
with $|\tau|\ll \frac{1}{G_N}$.

Since we will mostly be concerned with the algebra to the future 
of a cut, the precise expression for the left asymptotic 
charge will not enter into the entropy calculations.
It suffices to know that such a left asymptotic 
charge can be defined, after which the constraint
(\ref{eqn:globconstr}) can be written
\beq \label{eqn:Czetalambda}
\constr[\zeta_\lambda] = H_{\zeta_\lambda} -\frac{\kappa}{8\pi G_N} A_\infty +
\qasy_{\zeta_\lambda}^-
\eeq
It is, however, interesting 
to note that the expression $A_\tau - \int_{c_\tau}\mu v\Theta$
has recently appeared in the dynamical entropy proposal 
of Hollands, Wald, and Zhang \cite{Hollands:2024vbe},
who  showed that this 
combination can be interpreted as the area of 
an apparent horizon inside the black hole event horizon
(see also \cite{Visser:2024pwz}).   As $\tau\rightarrow
-\infty$, this apparent horizon approaches the 
white hole event horizon for the left asymptotic region.
In general, when there is nonzero null energy 
flux through $\hor$, the Einstein-Rosen bridge connecting the 
left and right regions becomes longer, causing the 
left white hole event horizon to differ from the right 
black hole event horizon.  Hence, 
we can use the above result to rewrite the left asymptotic
charge in terms of the left white hole event horizon
asymptotic area $A_{-\infty}^{\text{wh}}$ and the 
asymptotic expansion $\Theta_{-\infty}$ as
\beq
\qasy_{\zeta_\lambda}^- = \frac{\kappa}{8\pi G_N}
\left(A_{-\infty}^{\text{wh}} + \int_{c_{-\infty}}\mu \lambda
\Theta_{-\infty}\right).
\eeq
Taking into account the identity (\ref{eqn:semianec}) in the 
limit $\tau\rightarrow-\infty$, we see that the correction 
in $\qasy_{\zeta_\lambda}^-$ to the left event horizon area
involves the total average null energy flux through the 
horizon, weighted by $\lambda(y^A)$.

We now turn to the description of the various relations above in the 
small $G_N$ limit, which allows for perturbative quantization of the matter 
fields and gravitons about the Killing horizon background.  
The metric can be expanded about its background value
$g^{0}_{ab}$ as
\beq \label{eqn:g0gch}
g_{ab} = g^{0}_{ab} + \gc h_{ab}
\eeq
where $\gc = \sqrt{32\pi G_N}$; this choice ensures that $h_{ab}$ has a
canonical kinetic term in the quadratic Lagrangian.  Similarly, the degenerate
metric on $\hor$ is expanded as 
\beq
q_{ab} = q^0_{ab} + \gc \rho_{ab} +\frac{\gc}{d-2} \rho q^0_{ab}, 
\qquad q_0^{ab}\rho_{ab}=0,
\eeq
and because the shear and expansion vanish in the background,  to 
leading order in $\gc$ they are given by
\begin{align}
\sigma_{ab} &= \frac{\gc}{2} \lie_l \rho_{ab} \equiv \frac{\gc}{2}\dot\rho_{ab}\\
\Theta &=\frac{\gc}{2}\lie_l \rho\equiv \frac{\gc}{2} \dot\rho.
\end{align}
This then implies that the graviton null energy density, given by (\ref{eqn:tgvv}),
is finite as $G_N\rightarrow 0$.
Additionally, the expansion is  constrained to satisfy (\ref{eqn:semianec})
for all choices of smearing $\lambda$.  Since the right hand side is $\mathcal{O}(\gc^0)$,
it must be that the expansion can be nonzero only at $\op(\gc^2)$, implying that in
the linear theory, $\dot\rho = 0$, and the graviton null energy density
simplifies to 
\beq\label{eqn:tgvv2}
t^{(g)}_{vv} = \dot\rho^{ab}\dot \rho_{ab}.
\eeq
Note that this energy density differs from the 
Hollands-Wald canonical energy considered in 
\cite{Hollands:2012sf} by a total derivative 
which integrates to zero over the horizon.  

The vanishing of the expansion at $\op(\gc)$ implies the horizon area
is constant at linear order.  Its value is then set by $A_{\infty}^{(1)}$, the $\op(\gc)$
term in the expansion of the late time horizon area, 
which we choose to set 
to zero as a boundary condition; fluctuations in the area at this order
have the interpretation of describing different background spacetimes.\footnote{Fluctuations at this order 
are expected in the canonical ensemble for the black 
hole, but allowing for such large fluctuations leads to
complications in the construction of the gravitational algebras,
and require working in a formal power series in 
$\gc$, as in \cite{Witten2021}.  Restricting 
fluctuations in the area to be $\op(\gc^2)$ corresponds 
to a microcanonical ensemble, as discussed 
in \cite{Chandrasekaran2022b}.}  On the other hand, the
 $\op(\gc^2)$ term in the late time area,
$A_\infty^{(2)}$, does interact nontrivially with the matter fields, and hence 
we include it as a global gravitational charge.  Together with the second order
contribution to the left gravitational charge $(\qasy_{\zeta_\lambda}^-)^{(2)}$, the global
constraint (\ref{eqn:Czetalambda}) 
for the linearized theory becomes 
\begin{align}
\constr^{(2)}[\zeta_\lambda] 
&= 
\kappa \int_\hor\eta\Big(
t^{(g)}_{vv}+T^{(\psi)}_{ab}l^a l^b \Big)(v-\lambda)
- \frac{\kappa}{2\pi} \frac{A_\infty^{(2)}}{4G_N} + (\qasy_{\zeta_\lambda}^-)^{(2)}
\label{eqn:C2Tvv}
\\
&=\frac{\kappa}{2\pi} \left(h_{\acut_\lambda} -\frac{ A^{(2)}_\infty}{4G_N}
\right)
+ (\qasy_{\zeta_\lambda}^-)^{(2)}
\label{eqn:C2hA}
\end{align}
where in the second line we have applied the formula (\ref{eqn:hclambda}) 
to write the constraint in terms of the vacuum
modular Hamiltonian of the horizon cut, which, including
the graviton contribution, takes the form
\beq\label{eqn:hmodcutgrav}
h_{\acut_\lambda} = 2\pi\int_{\hor}\eta \left(t_{vv}^{(g)}
+ T^{(\psi)}_{ab}l^a l^b\right) (v-\lambda)
=
2\pi \int dy^A \int_{-\infty}^\infty dv\sqrt{q}\left(t^{(g)}_{vv}
+ T^{(\psi)}_{vv}\right)(v-\lambda(y^A)).
\eeq

\subsection{Generalized second law for semiclassical states}
\label{sec:gslsemi}

We now have all the components needed to construct
gravitational algebras associated with horizon cuts
and use them to prove a generalized second law.  
The starting point is the algebra $\acut_\lambda$ 
associated with the region $R_\lambda$ spacelike 
separated from the horizon cut defined by the function
$\lambda(y^A)$.  This algebra describes both matter 
and graviton degrees of freedom.  The matter will
generically involve an interacting quantum field theory;
in particular, we will not make use of any 
free field relations in the matter sector.  In this
case, the classical description of the matter in terms
of a Lagrangian and symplectic form may be a poor approximation, 
but the general arguments leading to the form
of the matter modular Hamiltonian, equation
(\ref{eqn:hclambda}) 
only rely on the existence
of a stress tensor as the generator of diffeomorphisms
and positivity of the average null energy.  
Since all matter couples universally to gravity via
its stress tensor, the matter contribution to the 
gravitational constraint in equation (\ref{eqn:C2hA}) 
remains valid even for interacting theories.  

On the other hand, the gravitons are quantized as a decoupled
theory of free, massless, spin-2 particles, obtained 
by promoting the metric perturbation 
$h_{ab}$ in (\ref{eqn:g0gch}) to a quantum operator and 
performing canonical quantization with 
appropriate gauge-fixing conditions for linearized
diffeomorphisms.  This is the 
correct description in the $G_N\rightarrow 0$ limit, which
suppresses self-interactions between gravitons as well as direct
matter-graviton couplings.  The ability to treat gravitons
as free fields is important at a technical level, since
gravitational interactions  are nonrenormalizable, 
causing the theory to become strongly
coupled at the Planck length.  A description
in this regime would require a fully nonperturbative theory of
quantum gravity, and it is widely expected that 
the language of quantum field theory is not applicable
at such energy scales.  However, assuming the low energy
gravitational 
theory admits an effective field theory description
\cite{Burgess2004, Donoghue:2012zc}, 
the free graviton approximation should give an accurate
description at leading order in $G_N$. In this
case, the classical phase space analysis described
in section \ref{sec:geometric} is directly relevant, and the graviton
contribution to the constraint (\ref{eqn:C2Tvv})
is given by
$t^{(g)}_{vv}$ defined in (\ref{eqn:tgvv2}), which is quadratic
in the graviton field. 

The gravitational algebra additionally includes degrees 
of freedom associated with the asymptotic charges 
$\qasy_\lambda^{\pm}$.  As discussed in section
\ref{sec:geometric}, these charges are constructed 
from the $\op(\gc^2)$ contribution to the metric 
perturbation, and hence are not present in the linearized
theory involving only gravitons \cite{Kudler-Flam2023}. 
In the interacting theory away from $G_N=0$, 
these asymptotic charges 
are present and are related to the matter and graviton 
stress energy, which is nontrivial even in the linearized
theory.  Hence, in order to obtain a theory at $G_N=0$
that consistently lifts to the interacting theory,
we must include the asymptotic charges explicitly and 
couple them to the matter fields via the constraint
(\ref{eqn:C2Tvv}).\footnote{In
a complete description, one should also add charges 
associated with other asymptotic symmetries such as 
rotations \cite{Witten2021, Chandrasekaran2022a, Chandrasekaran2022b}.
Including these charges results in  additional crossed products 
 with respect to the symmetries that they generate.  They do not
change the overall type of the algebra, so they
will not affect the results 
involving finiteness of the entropy and 
should not affect the arguments for the semiclassical
generalized second law, so we do not consider
them in detail in this work.  However, rotational
asymptotic charges are important when obtaining the 
correct matching conditions for rotating black hole 
backgrounds \cite{Kudler-Flam2023}, so it would be worth considering 
them in more detail in the future.
} 

The $\op(\gc^2)$ contribution to the asymptotic horizon
area comprises the future asymptotic charge $\qasy_\lambda^+$ 
that must be included.  We represent this operator
on an independent Hilbert space $\hfut = L^2(\mathbb{R})$
as the position operator,
\beq \label{eqn:qArea}
\hat{q} = -\frac{A_\infty^{(2)}}{4G_N}.
\eeq
This choice reflects two semiclassical assumptions about the 
spectrum of the asymptotic area operator, namely that it is 
continuous and unbounded above and below.  The 
continuous spectrum of this operator implies that we are 
working in a description that does not resolve individual
microstates of the black hole, which ultimately
manifests in the constant shift ambiguity in the 
crossed product entropy.  The unboundedness of the spectrum
corresponds to the fact that we are considering perturbations 
of the area around a fixed background, and because the 
perturbation only describes changes in the geometry at 
order $\gc^2$, it can take on  large positive or negative
values in the $\gc\rightarrow 0$ limit
without 
producing a significant change in the 
background value of the area.

The algebra of gravitationally dressed operators is 
denoted $\wh{\acut}_\lambda$, and is
constructed by imposing the constraint (\ref{eqn:C2hA})
to obtain a gauge-invariant Hilbert space and operators,
following the procedure described in 
\cite{Jensen2023, Chandrasekaran2022a}.  
This results in the crossed product algebra discussed in section 
\ref{sec:crprod}.  It is represented on the Hilbert 
space $\hcr = \hqft\otimes\hfut$, and is given by
\beq\label{eqn:hClambda}
\wh{\acut}_\lambda = \left\langle e^{i\hat{p} h_\lambda} \msf{a}
e^{-i\hat{p}h_\lambda}, \hat{q}\right\rangle, \qquad \msf{a}\in
\acut_\lambda,
\eeq
where $\hat{p} = -i\frac{d}{dq}$, and $h_\lambda \equiv
h_{\acut_\lambda}$ is the modular Hamiltonian for the 
horizon cut in the average null energy 
vacuum $|\Omega\rangle$, given in terms of the matter and graviton
stress tensors by equation (\ref{eqn:hmodcutgrav}).\footnote{Another
way to interpret this algebra is to view it as the commutant
of all operators in $\acut_\lambda'$ as well as a left asymptotic
charge $\qasy_{\zeta_\lambda}^- 
= -\frac{\kappa}{2\pi}(\hat{q} + h_\lambda) $. Because
$\qasy_{\zeta_\lambda}^-$ is not the left event horizon area,
which in the present setup matches to the left ADM Hamiltonian,
most of the operators in the algebra $\wh{\acut}_\lambda$ 
do not commute with the left 
ADM Hamiltonian, with functions of $\hat{q}$ being the exception.  
This is a somewhat peculiar feature of this construction, but 
appears necessary in order to obtain a nontrivial algebra that 
includes the right asymptotic charge $\hat{q}$ as well as a subalgebra 
isomorphic to the algebra of quantum fields in the region
exterior to the cut.  The fact that we can derive
a semiclassical second law also supports this prescription
for defining the subregion gravitational algebra, but this subtlety
deserves closer consideration in the future.  
}

To examine the generalized second law, 
we would like to compute entropies of states
of the form $|\wh{\Phi}\rangle
= |\Phi\rangle\otimes|f\rangle$, where $|\Phi\rangle$
is a state in $\hqft$ and $f$ is a wavefunction 
on $L^2(\mathbb{R})$ that is slowly varying in the position
basis, and hence whose derivatives are suppressed by a 
small parameter $\vep$.  Using the formula for the density
matrix for this state given in (\ref{eqn:logrho}), we can 
apply the semiclassical expression for the entropy
(\ref{eqn:SrhoSrel}) to arrive at
\beq\label{eqn:Srhocut}
S_\lambda(\rho_{\wh\Phi}) = -S_{\text{rel}}^{\acut_\lambda}(\Phi||\Omega)
+ \Big\langle \frac{A^{(2)}_\infty}{4G_N}\Big\rangle_{\wh\Phi}
+ S_f^{A_\infty} + \op(\vep^2).
\eeq
The $\op(\vep^2)$ corrections to this semiclassical entropy
formula are discussed in section \ref{sec:pertent}.

We can convert this expression into something involving
the generalized entropy by employing the horizon-cut
identity (\ref{eqn:horcutid}).  The term on the right
hand side is proportional to the one-sided vacuum
modular Hamiltonian 
$\Kmod$, defined by restricting the range of 
integration of $v$ in (\ref{eqn:hmodcutgrav}) to the future
of the horizon cut, $v>\lambda(y^A)$.  This defines the 
formal split of the modular Hamiltonian $h_\lambda = \Kmod
- \Kmod'$ into an object $\Kmod$ affiliated
with $\acut_\lambda$ and $\Kmod'$ affiliated 
with the commutant $\acut_{\lambda}'$.
The one-sided modular Hamiltonians $\Kmod, \Kmod'$ make
sense as densely-defined sesquilinear forms (but not as unbounded
operators), similar to the local operators discussed 
in section \ref{sec:hstr}.  $\Kmod$ therefore has well-defined 
expectation values for a dense set 
of states.  When coupling to gravity,
the constraint (\ref{eqn:horcutid}) then implies that 
the area of the horizon cut, expanded to second order in 
$\gc$, must also define a sesquilinear form affiliated 
with the crossed product algebra,
\beq\label{eqn:Alambda2}
\frac{A_\lambda^{(2)}}{4G_N} = \frac{A^{(2)}_\infty}{4G_N} - \Kmod
= -\hat{q} - \Kmod.
\eeq

The relative entropy in (\ref{eqn:Srhocut}) also has an expression
in terms of one-sided quantities.  The starting
point is the 
standard expression for the relative entropy between two
density matrices $S_\text{rel}(\rho_{\Phi}||\rho_{\Omega})
= \tr(\rho_\Phi\log\rho_\Phi - \rho_\Phi\log\rho_\Omega)
=-S(\rho_\Phi) +\langle\Phi|-\log\rho_\Omega|\Phi\rangle$.  
Assuming a regulator to make the one-sided QFT entanglement
entropies finite, we can apply this formula to the relative entropy
appearing in (\ref{eqn:Srhocut}).  The density matrix for the 
horizon vacuum state $|\Omega\rangle$ is thermal with respect to the 
one-sided modular Hamiltonian, and hence is formally
expressed as
\beq
\rho_{\Omega} = \frac{1}{Z} e^{-\Kmod},
\eeq
with the normalization set by the requirement that $\Kmod$ 
have zero expectation value in the vacuum, $\langle \Omega|\Kmod
|\Omega\rangle=0$.  
The vacuum entanglement entropy is then simply given
by $\log Z$:
\beq\label{eqn:SPsi}
S_\Omega^{\text{QFT}} = \langle \Omega|-\log\rho_\Omega|\Omega\rangle = \log Z.
\eeq
From these relations, the relative entropy takes the form
\beq
-S_\text{rel}^{\acut_\lambda}(\Phi||\Omega) = 
S_{\Phi,\lambda}^\text{QFT}-\log Z
-\langle \Kmod\rangle_\Phi.
\eeq

Applying this to (\ref{eqn:Srhocut}) and employing the 
identity (\ref{eqn:Alambda2}), we arrive at the semiclassical
expression for the crossed product entropy
\beq
S_\lambda(\rho_{\wh\Phi}) = 
\Big\langle \frac{A^{(2)}_\lambda}{4G_N}\Big\rangle_{\wh\Phi}
+S_{\Phi,\lambda}^\text{QFT} + S_f^{A_\infty} -\log Z +\op(\vep^2).
\eeq
This agrees with the horizon generalized entropy up to a constant,
which can be formally determined by writing the area perturbation
in terms of the background value $A^{(0)}_\lambda$ 
as $A^{(2)}_\lambda = A_\lambda - A_\lambda^{(0)}$.
The resulting entropy formula is then
\beq \label{eqn:Slambdagen}
S_\lambda(\rho_{\wh\Phi}) = 
\Big\langle\frac{A_\lambda}{4G_N}\Big\rangle_{\wh\Phi}
+S_{\Phi,\lambda}^\text{QFT} + S_f^{A_\infty}
-\left( \frac{A^{(0)}_\lambda}{4G_N} +\log Z \right) +\op(\vep^2)
\eeq
The term $\log Z$ should be interpreted as quantum field
vacuum entanglement entropy, and hence the above constant in parentheses
can be viewed as the generalized entropy in the vacuum
state $|\Omega\rangle$ (assuming no contribution from the entropy
associated with the asymptotic area).  Crucially, this constant
is independent of the horizon cut $\lambda$, since the 
vacuum $|\Omega\rangle$
is a stationary state.  
This constant will therefore
drop out when considering differences in entropies between
horizon cuts.

We now examine the algebra associated with a later cut defined 
by $\tilde{\lambda}(y^A)\geq \lambda(y^A)$.  The quantum
field theory algebra $\acut_{\tilde\lambda}$ acts on the same
Hilbert space as before $\hqft$, and we once again adjoin
the same asymptotic charge $\hat{q} = -\frac{A_\infty^{(2)}}{4G_N}$ 
acting on $\hfut=L^2(\mathbb{R})$
when constructing the gravitational algebra.  It is significant
at this point that the asymptotic charge does not depend on the 
horizon cut $\lambda$, which follows from the form of the 
constraint (\ref{eqn:C2hA}) for a generic boost 
quasi-Killing vector $\zeta_\lambda^a$.  The gravitational
algebra again takes the form of a crossed product,
\beq\label{eqn:hClambdatil}
\wh{\acut}_{\tilde\lambda} = \left\langle e^{i\hat{p}h_{\tilde\lambda}}
\msf{a} e^{-i\hat{p}h_{\tilde\lambda}}, \hat{q}\right\rangle,
\qquad \msf{a}\in \acut_{\tilde\lambda}.
\eeq
Even though $\acut_{\tilde\lambda}\subset\acut_\lambda$,
the crossed product algebra $\wh{\acut}_{\tilde\lambda}$ is 
not a subalgebra of $\wh{\acut}_{\lambda}$ in this representation,
since the modular Hamiltonians $h_{\tilde\lambda}$ and $h_{\lambda}$
associated with different cuts are not the same.

An issue that arises when comparing entropies between two type $\ttwo$ 
algebras is that there may not be 
an obvious choice for the relative normalizations 
of the traces.  This results in a potential
ambiguity when comparing entropies between the two
algebras.  In the present context, we can take 
advantage of the fact that the algebras share a common
operator, the asymptotic charge $\hat{q}$,  to find
a canonical relative normalization for the traces.  For both
algebras, the state $|\wh\Omega\rangle =
|\Omega\rangle\otimes|f\rangle$ has 
a density matrix determined by equation (\ref{eqn:DelhatPsi}),
and is given by 
\beq\label{eqn:rhohatPsi}
\rho_{\wh\Omega} = |f(\hat q)|^2 e^{\hat{q}},
\eeq
 which is solely
a function of $\hat{q}$.  Normalizing the density matrices 
for both algebras 
to agree precisely with (\ref{eqn:rhohatPsi}) then 
determines the relative normalization for the traces
on the algebras, and in both cases is given by the 
formula (\ref{eqn:Tr}).  This choice ensures that 
operators in the subalgebra generated by $\hat{q}$ 
have the same trace in both $\wh{\acut}_{\tilde\lambda}$ 
and $\wh{\acut}_\lambda$.  

Considering the same state $|\wh\Phi\rangle = |\Phi\rangle
\otimes |f\rangle$ as before, the computation
of the entropy for the algebra $\wh\acut_{\tilde\lambda}$
now proceeds analogously, giving
\beq\label{eqn:Srhotildecut}
S_{\tilde\lambda}(\rho_{\wh\Phi}) = -S_\text{rel}^{\acut_{\tilde\lambda}}
(\Phi||\Omega) +
\Big\langle \frac{A^{(2)}_\infty}{4G_N}\Big\rangle_{\wh\Phi}
+ S_f^{A_\infty} + \op(\vep^2)
\eeq
Since $\big\langle\frac{A^{(2)}_\infty}{4G_N}\big\rangle_{\wh\Phi}$
and $S_f^{A_\infty}$ depend only on the wavefunction
$f$ and the operator $\hat{q}$, these quantities
agree when computed in either algebra.  They will
therefore drop out when considering the difference 
in entropies $S_{\tilde\lambda} - S_{\lambda}$.  
Hence, in the semiclassical
limit in which we can ignore the $\op(\vep^2)$ contributions
to the entropy, we have that
\beq \label{eqn:GSLII}
S_{\tilde\lambda} - S_\lambda = 
-S_\text{rel}^{\acut_{\tilde\lambda}}
(\Phi||\Omega) + S_{\text{rel}}^{\acut_\lambda}(\Phi||\Omega) \geq 0
\eeq
where the last inequality follows from the monotonicity of 
relative entropy under the inclusion $\acut_{\tilde\lambda}\subset
\acut_\lambda$ \cite{Araki1975, Uhlmann1977, Petz1993}. 
Hence, we have succeeded 
in showing that in the semiclassical limit $\vep\ll 1$,
the crossed product entropies  increase upon restricting to 
horizon cuts that are further to the future.  This is the 
statement of the second law for the crossed product algebras.  

An attractive feature of the second law formulated in terms
of crossed product algebras is that it directly relates
the increase in entropy to the monotonicity
of relative entropy, without appealing to relations
involving involving regulated quantum field 
theory entanglement entropies.  Of course, we argued above
that the crossed product entropy agrees semiclassically
with the generalized entropy by invoking the local
gravitational constraint (\ref{eqn:horcutid}),
so this monotonicity result can instead be stated 
directly in terms of generalized entropy.   
The formula (\ref{eqn:Slambdagen}) is applicable
for the $\wh{\acut}_{\tilde\lambda}$ algebra as well,
and since the constant in parentheses is independent 
of the horizon cut, it drops out of the entropy difference.
Hence, equation (\ref{eqn:GSLII}) also implies 
the standard statement of the generalized second law
in terms of generalized entropies,
\beq
\Big\langle\frac{A_{\tilde\lambda}}{4G_N}\Big\rangle_{\wh\Phi} 
+ S_{\tilde\lambda}^\text{out}
-\Big\langle\frac{A_{\lambda}}{4G_N}\Big\rangle_{\wh\Phi} 
-S_{\lambda}^\text{out} \geq 0,
\eeq
where $S_{\tilde\lambda}^\text{out}\defeq 
S_{\Phi,\tilde\lambda}^{\text{QFT}} + S_f^{A_\infty}$,
and similarly for $S_\lambda^\text{out}$.  This shows
that the crossed product entropies reproduce Wall's proof
of the generalized second law, but they have the added
benefit of giving an algebraic interpretation to the 
generalized entropies appearing in it.

One important aspect of this derivation is that it is 
necessary to work in the representation where the 
crossed product algebras take the form (\ref{eqn:hClambda}) and 
(\ref{eqn:hClambdatil}).  These representations
have the property that the algebraic intersection
consists of operators in the asymptotic algebra,
i.e.\ $\wh{\acut}_\lambda\wedge \wh{\acut}_{\tilde\lambda} = 
\left\langle\hat{q}\right\rangle$.  In some recent investigations 
of gravitational algebras \cite{Witten2021,
Chandrasekaran2022a, Chandrasekaran2022b}, a unitarily equivalent
representation was employed where the algebras take the form
\begin{align}
\tilde{\acut}_\lambda &= e^{-i\hat{p} h_\lambda}\wh{\acut}_\lambda
e^{i\hat{p}h_\lambda} = \left\langle
\msf{a}, \hat{q}-h_\lambda\right\rangle, \quad
\msf{a}\in \acut_\lambda, \\
\tilde{\acut}_{\tilde\lambda} &= e^{-i\hat{p} 
h_{\tilde\lambda} }\wh{\acut}_{\tilde\lambda}
e^{i\hat{p}h_{\tilde\lambda}} = \left\langle\msf{b}, \hat{q}-
h_{\tilde\lambda}\right\rangle, \quad
\msf{b}\in \acut_{\tilde\lambda}.
\end{align}
The procedure described above of considering the same global
state $|\wh{\Phi}\rangle = |\Phi\rangle\otimes|f\rangle$ for the 
two algebras and then comparing the entropies leads to the conclusion
that the entropy {\it decreases} 
in the semiclassical limit.\footnote{This entropy
is equal to the entropy of the twirled
state $|\tilde\Phi\rangle =e^{i\hat{p}h}|\Phi,f\rangle$ 
on the original crossed product algebra $\wh\acut_{\lambda}$.
The density matrix for this state is given by 
equation (\ref{eqn:twirlrho}), and assuming a 
slowly varying wavefunction, the entropy can be 
approximated by 
\beq
S(\rho_{\tilde\Phi})\approx\langle\Phi|h_{\Phi|\Omega}|\Phi\rangle
-\langle f|\hat q + \log g(\hat{q})|f\rangle
= -S_\text{rel}^{\acut_\lambda'}(\Phi||\Omega)
-\langle f|\hat q + \log g(\hat{q})|f\rangle,
\eeq
which now involves the relative entropy for the commutant
quantum field theory algebra $\acut_\lambda'$.  Since the 
inclusion for the commutant algebras is reversed, 
$\acut_\lambda' \subset \acut_{\tilde\lambda}'$,
monotonicity of relative entropy now causes the 
crossed product entropy to decrease between cuts.
}
A possibly related pathology with this parameterization
is that the asymptotic charge is now 
given by $\frac{A_\infty^{(2)}}{4G_N} = -\hat{q} + 
h_{\lambda}$, which now depends on the horizon cut 
$\lambda$.  The expectation value of this asymptotic charge
then evolves as we change the cut, despite fixing the global
state $|\wh{\Phi}\rangle$.
These issues indicate there we should 
not use above parameterization while also fixing the 
state $|\wh\Phi\rangle$. 

As the formulas (\ref{eqn:Srhocut}) and (\ref{eqn:Srhotildecut})
indicate, there are corrections to the entropy in the gravitational
algebras starting at second order in the semiclassical
expansion parameter $\vep$.  This immediately raises the 
question of whether the generalized second law continues
to hold even after including these corrections.  This 
questions is nontrivial because there is not an obvious
way in which the algebra $\wh{\acut}_{\tilde\lambda}$ 
embeds as a subalgebra of $\wh{\acut}_\lambda$, making 
it difficult to appeal to monotonicity results
of entropies under algebra inclusions.  In order to begin
addressing these questions, in section \ref{sec:pertent} we compute
the corrections to the semiclassical entropy expression, and 
discuss various bounds on these corrections.  Unfortunately,
these corrections  do not ostensibly 
obey any monotonicity
properties, but we leave open the possibility that 
a more careful analysis would produce such a result.  
We also explore in section \ref{sec:ovw} another approach
in which we identify a subalgebra of $\wh{\acut}_\lambda$
unitarily equivalent to $\wh{\acut}_{\tilde\lambda}$,
assuming the existence of an operator-valued weight 
between the underlying quantum field theory algebra.
This may ultimately lead to improved monotonicity results
for the subleading corrections to the crossed 
product entropies.

\section{Algebra at infinity}  \label{sec:ainfty}

The discussion of the GSL to this point has focused for simplicity
on asymptotically AdS black holes.  
The Killing horizon is a Cauchy
surface for these spacetimes since 
the Dirichlet boundary conditions imposed at the
AdS boundary cause all outgoing modes to be reflected 
back through the black hole horizon.
In this case, the quantum field algebra associated with the horizon
is complete, and the gravitational algebra enlarges
this only by a 
single additional asymptotic charge, the late time 
area perturbation $\frac{A^{(2)}_\infty}{4G_N}$.
However,
for black holes with alternative asymptotics, 
there are outgoing
quantum field modes that never enter the black hole 
horizon.  For asympotically flat black holes,
these include modes that propagate out to future timelike
infinity $i^+$
or null infinity
$\scri^+$, while for asymptotically de Sitter black holes,
these include degrees of freedom beyond the cosmological horizon
(see figure \ref{fig:SdS}).  
The analysis must therefore be generalized to allow for 
the possibility of these nontrivial algebras, which 
we refer to as the algebra at infinity.  

In this section, we describe the treatment of the generalized
second law with nontrivial algebras at infinity.  In section 
\ref{sec:ainftysemi}, we describe how this algebra at infinity is naturally
incorporated into algebraic structures associated with half-sided translations,
which allow a generalization of Wall's discussion of the GSL
with an algebra at infinity that applies to interacting field theories.
Following this, section \ref{sec:flatgravalg} describes the 
general construction of the gravitational algebra
in these contexts.  Finally, section \ref{sec:rotUnruh}
discusses some subtleties that can arise when analyzing rotating 
black holes and backgrounds involving the Unruh vacuum.

\subsection{Semiclassical GSL}
\label{sec:ainftysemi}

Intuitively, the existence of additional degrees of freedom
localized far from the black hole should not affect the 
derivation of the GSL, since any entropy associated with these 
degrees of freedom would appear in the algebra associated with 
any horizon cut; it should therefore cancel out in when considering
entropy differences between two cuts.  Wall gave a precise
version of this argument assuming the existence of a factorization
between the algebra associated with the horizon and the algebra
associated with $\scri^+$ \cite{Wall2011}.  Such a factorization
is expected to hold for free field theories 
\cite{Dappiaggi2011}, and was recently
utilized for the construction of gravitational algebras 
associated with arbitrary Killing horizons \cite{Kudler-Flam2023}.

\begin{figure}
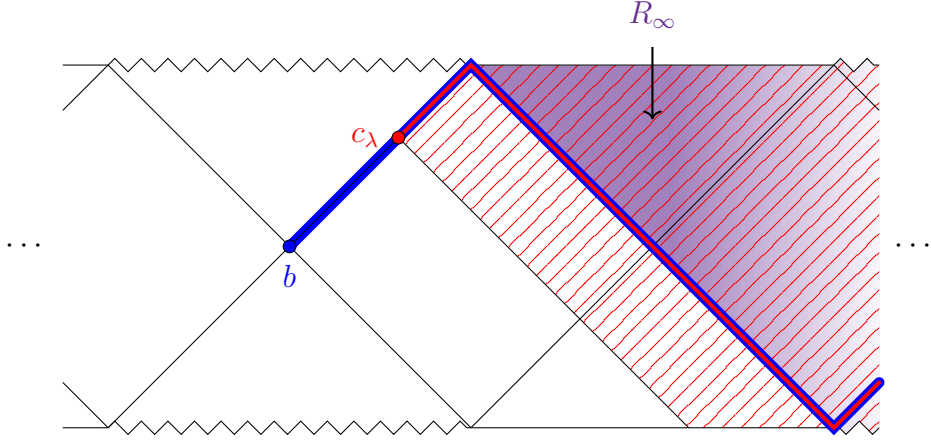

\centering
    \include{sds-figure.tex}
    \caption{The extended Schwarzschild-de Sitter spacetime provides
    an example where one finds an algebra at infinity for the horizon
    translation.  This spacetime is described as an infinite chain 
    of black hole and cosmological horizons, indicated by the ellipses
    ``$\cdots$'' in the Penrose diagram above.  
    The algebra for a cut $c_\lambda$
    is defined as all quantum fields supported in the region spacelike
    to the right of the cut, shown by the red lines above.  The algebra 
    at infinity is associated with the region $R_\infty$ 
    shown in purple, consisting of fields localized beyond
    the cosmological horizon.  }
    \label{fig:SdS}
\end{figure}

However, there are reasons to believe that this factorization
property fails for interacting theories.  
Correlation functions of local operators 
in interacting field theories
have stronger divergences than in free theories, 
which necessitates smearing the fields 
in timelike directions in order to obtain operators
with finite fluctuations \cite{Borchers1964, Wall2011, Witten:2023qsv}.
Smearing only on the horizon is not 
sufficient, implying that there is no nontrivial 
algebra of bounded operators localized solely 
on a horizon cut $\hor_\lambda^+$ \cite{Bousso2014}.  
This prevents a decomposition of the horizon cut algebra 
$\acut_\lambda$ into a tensor product $\alg(\hor_\lambda^+)
\otimes \alg(\infty)$, since $\alg(\hor_\lambda^+)$ is  trivial. 
Fortunately, the existence of a tensor factorization 
is not strictly necessary in order to derive the GSL.  Instead,
we will make use of a weaker condition on the structure of the 
horizon cut algebra $\acut_\lambda$ that follows from the properties
of half-sided translations.  

The argument from section \ref{sec:hstr} that the smeared 
average null energy operator $P_\lambda$ generates a half-sided
translation continues to hold for 
spacetimes with non-AdS asymptotics, 
provided that the horizon cut algebra $\acut_\lambda$ is 
defined as the double commutant of local operators on 
a Cauchy slice consisting of the future horizon cut $\hor_\lambda^+$
and a component $\sigasy$ 
spacelike separated from the horizon. For example,
in asymptotically flat spacetimes, $\sigasy$ consists of $\scri^+$
and a contribution from $i^+$ 
if massive fields are present.  
Since $P_\lambda$ is an integral of the stress tensor 
only over the horizon $\hor$, it acts trivially
on operators localized to $\sigasy$, which are associated 
purely with outgoing modes that do not interact with the horizon
$\hor$.  This allows us to characterize the algebra $\ainfty$ 
associated 
with $\sigasy$ as the collection of operators in $\abif$
(the algebra associated with the exterior of the 
bifurcation surface) that are 
invariant under the translations,
\beq\label{eqn:A+inv}
\ainfty\defeq \{\msf{a}\in \abif| e^{itP_\lambda}\msf{a}
e^{-itP_\lambda} = \msf{a}\}
\eeq

The properties of such a fixed point algebra under the action
of a half-sided translation were studied 
by Borchers \cite{Borchers1997}, 
who showed that the algebra can equivalently
be defined as the operators contained in the translated
algebras $\acut_{t\lambda} = 
e^{itP_\lambda}\abif e^{-itP_\lambda}$ 
for all $t\geq 0$,
\beq \label{eqn:A+int}
\ainfty = \bigwedge_{t\geq 0} \acut_{t\lambda}.
\eeq
Since $\acut_{t\lambda}$ defines an algebra associated 
with a cut further to the future for larger values of $t$, 
the limiting algebra $\ainfty$ 
consists of modes that never fall across
the horizon, and hence escape to the region associated 
with $\sigasy$.  
The fact that the definitions (\ref{eqn:A+inv}) and 
(\ref{eqn:A+int}) agree justifies the statement above that 
the operators localized to the asymptotic regions are fixed 
by the action of $P_\lambda$.

One can further show that the algebra $\ainfty$ is preserved by the 
modular flow of $\abif$ in a vacuum state $|\Omega\rangle$
for $P_\lambda$, $\Delta_{\abif}^{-is}\ainfty\Delta_{\abif}^{is}
= \ainfty$.  By Takesaki's theorem
\cite{Takesaki1972}, this implies 
the existence of a conditional expectation from
$\abif$ to $\ainfty$.  Recall that a conditional
expectation is a completely positive linear map
$E:\abif\rightarrow \ainfty$ such that $E(\msf{b}) = 
\msf{b}$ for all $\msf{b}\in\ainfty$ and  
satisfies the bimodule property $E(\msf{b_1ab_2}) = \msf{b_1}
E(\msf{a})\msf{b_2}$ for $\msf{b_1, b_2}\in \ainfty$, 
$\msf{a}\in\abif$
\cite{Petz1993, TakesakiII}.  Takesaki's theorem further
guarantees that the conditional expectation
preserves the state $\omega = \langle\Omega|\cdot|\Omega\rangle$
on $\abif$, in the sense that $\omega\circ E(\msf b)
\equiv \omega(E(\msf b))
= \omega(\msf b)$ for $\msf{b}\in\abif$.  It is  
straightforward to see that $\ainfty$ is also preserved 
by the modular flow for any horizon-cut algebra $\acut_{\lambda}$,
and therefore each of these  has an associated 
conditional expectation $E_{\lambda}: \acut_{\lambda}\rightarrow
\ainfty$.

Assuming $\ainfty$ is a factor, 
the existence of the conditional expectation $E_\lambda$ implies that 
$\ainfty$ and its relative commutant ${\ainfty^c}_{,\lambda} = \ainfty'\wedge
\acut_\lambda$ embed into $\acut_\lambda$ as a tensor product, i.e.\
$\ainfty\otimes {\ainfty^c}_{,\lambda}\cong\ainfty
\vee {\ainfty^c}_{,\lambda}   \subset \acut_\lambda$
\cite{Takesaki1972}.
If in fact we have equality $\ainfty\vee {\ainfty^c}_{,\lambda}
= \acut_\lambda$,
the inclusion $\ainfty\subset \acut_\lambda$ is called {\it conormal}
\cite{Longo1984}, and along with the existence of the conditional
expectation, this implies that $\acut_\lambda$ 
tensor factorizes $\abif = \ainfty\otimes {\ainfty^c}_{,\lambda}$ into
an algebra at infinity $\ainfty$ and its relative commutant
${\ainfty^c}_{,\lambda}$.  In this case the relative commutant 
is naturally associated with a nontrivial horizon algebra 
${\ainfty^c}_{,\lambda} = \alg(\hor^+_\lambda)$, which is the setup
expected for free quantum fields.  For interacting quantum
fields, we do not expect a nontrivial horizon algebra, 
and hence may conjecture that the relative commutant
is trivial, ${\ainfty^c}_{,\lambda} = \mathbb{C} \mathbbm{1}$,
in which case the inclusion $\ainfty\subset \acut_\lambda$ 
is called {\it singular} \cite{Longo1984}.  
Since our discussion of the GSL generically involves 
interacting matter fields but free gravitons, we will 
be in an intermediate case where ${\ainfty^c}_{,\lambda}$ is nontrivial
and describes the graviton horizon algebra, but $\ainfty\vee
{\ainfty^c}_{,\lambda} \neq \acut_\lambda$ since the matter fields do 
not have a horizon algebra.

This discussion indicates that the existence of the conditional
expectation $E_\lambda$ is a generalization of the 
notion of a tensor factorization of the algebra $\acut_\lambda$. 
When $\acut_\lambda$ does factorize, there is a class 
of tensor product states $\omega = \omega_\infty \otimes
\omega_{\hor_\lambda^+}$ on $\acut_\lambda = \ainfty
\otimes {\ainfty^c}_{,\lambda}$ that 
contain no correlations between the two subfactors,
i.e.\ $\omega(\msf{bc}) = \omega(\msf{b})\omega(\msf{c})$ 
for $\msf{b}\in \ainfty$ and $\msf{c}\in{\ainfty^c}_{,\lambda}$.
The states that are preserved by the conditional expectation,
$\omega = \omega\circ E_\lambda$,
are product states which satisfy $E_\lambda(\msf{c}) = 
\omega_{\hor_\lambda^+}(\msf{c}) \mathbbm{1}$ for all $\msf{c}
\in {\ainfty^c}_{,\lambda}$.  This shows that the conditional
expectation $E_\lambda$ encodes the information 
of a state $\omega_{\hor_\lambda^+}$ on ${\ainfty^c}_{,\lambda}$,
and different choices of this state result in different conditional
expectations from $\acut_\lambda$ to $\ainfty$.  When the 
inclusion $\ainfty\subset \acut_\lambda$ is singular,
the conditional expectation $E_\lambda$ is unique 
\cite[Section IX.4]{TakesakiII}.  In this case,
there is no algebra on which to define a horizon state 
$\omega_{\hor_\lambda^+} $, but the analog of tensor 
factorized states are states fixed by the conditional 
expectation.  Such states 
can be thought of as vacuum states for fields localized 
very close to the horizon.

\begin{figure}
\centering
    \include{schwarz-ext-fig.tex} 
    \caption{The Penrose diagram of the conformally extended 
    Schwarzschild solution is shown above.  The standard 
    maximally extended Schwarzschild
    solution is the region shown in yellow, and the triangular regions 
    in the corners are the conformal extension
    of the spacetime beyond $\scri^+$ and $\scri^-$.  The algebra
    of massless or conformal fields in this spacetime naturally
    allows for smearing of the fields in the extended regions.  This 
    motivates defining the algebra at infinity for the Schwarzschild
    black hole to be associated with the region $R_\infty$ 
    shown in purple, and this region captures the notion
    of outgoing quantum field modes that never enter the black hole
    horizon.}
    \label{fig:extSch}
\end{figure}

A question that arises is why the algebra associated with the horizon
is trivial in interacting theories, while the algebra at infinity can
remain nontrivial.  For example, in the asymptotically flat Schwarzschild
black hole, both algebras appear to be associated with operators
smeared on null surfaces ($\hor_\lambda^+$ or $\scri^+$), and so
it is not immediately clear what distinguishes the two.  The resolution
is that the algebra at infinity is nontrivial in 
interacting theories whenever it is associated 
with an open region of spacetime.  This is easiest to see in
the Schwarzschild-de Sitter spacetime, where the algebra at infinity
consists of all operators smeared in the region beyond the cosmological
horizon, which is an open spacetime region 
(see figure \ref{fig:SdS}).  

\begin{figure}
    \centering
    \begin{subfigure}[t]{0.6\textwidth}
        \centering
        \includegraphics[width=0.8\textwidth]{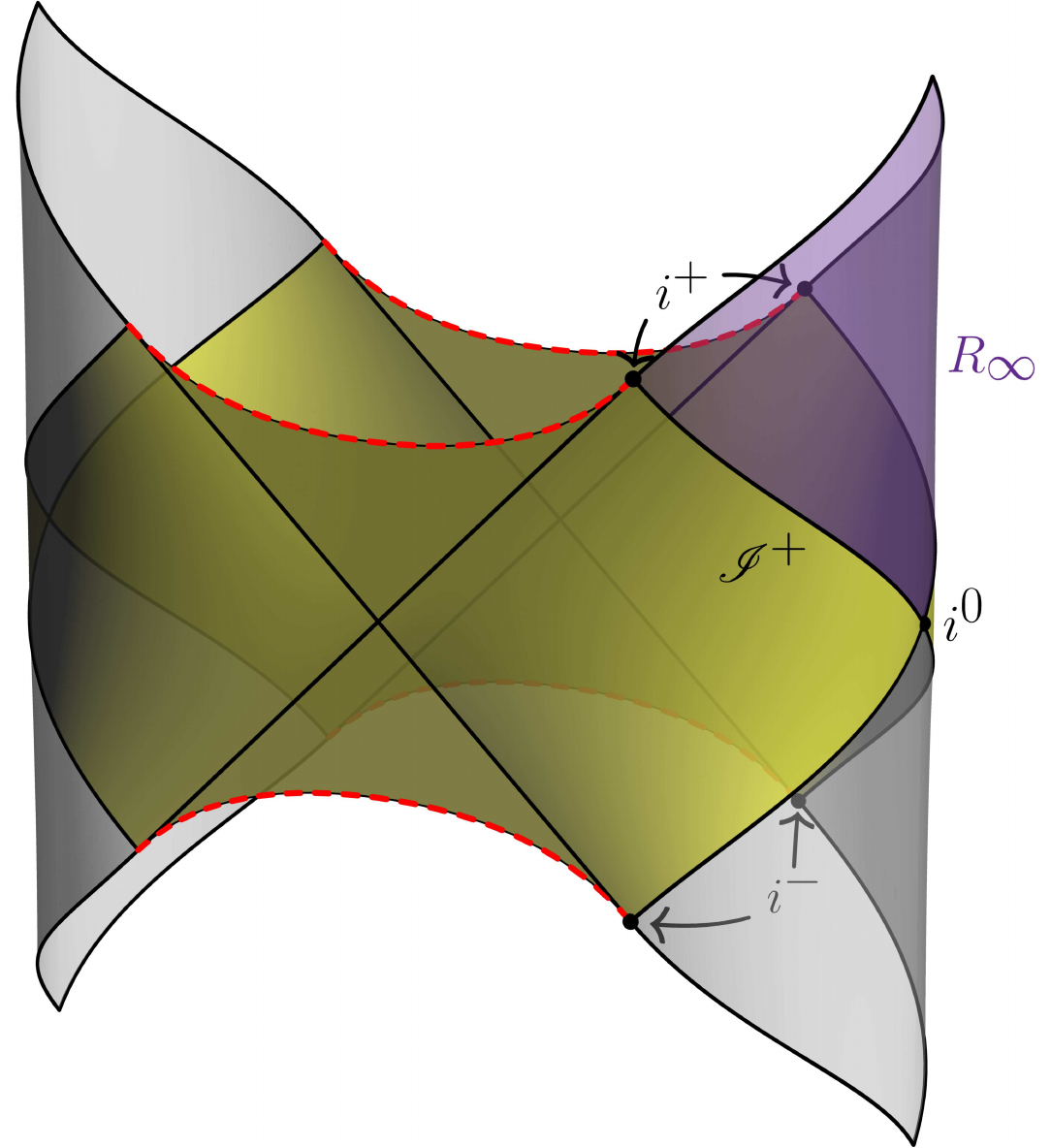}
        \caption{Conformal extension of  Schwarzschild}
    \end{subfigure}
    \begin{subfigure}[t]{0.3\textwidth}
        \centering
        \includegraphics[width=0.8\textwidth]{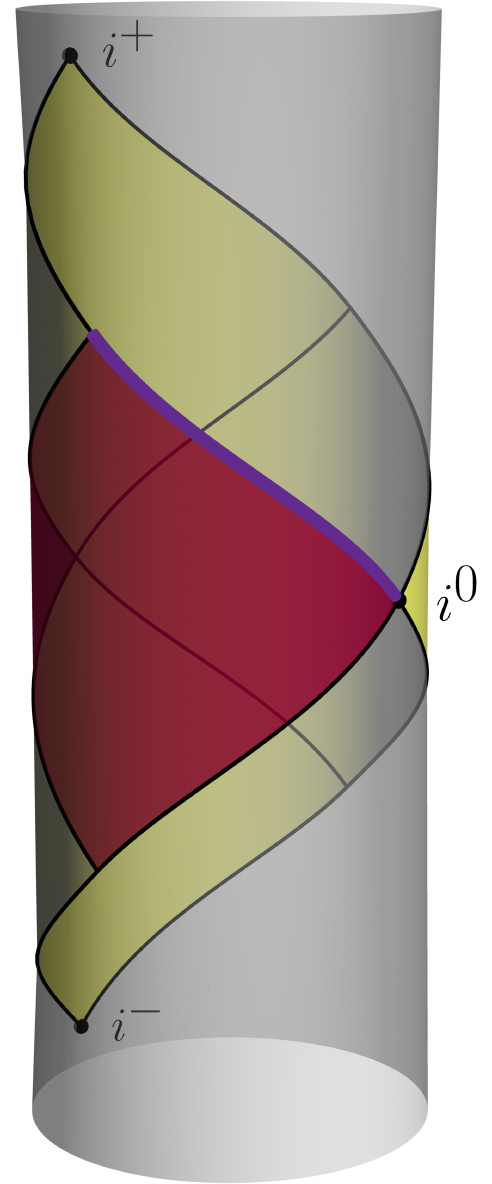}
        \caption{Conformal extension of Minkowski space}
    \end{subfigure}
    \caption{Shown above are the conformal extensions of 
    the Schwarzschild and Minkowski spacetimes which are asymptotically
    conformal to the Einstein static universe near spatial infinity
    $i^0$, which shrinks to a point in this conformal 
    compactification.  The algebra at infinity for the 
    Schwarzschild solution lives in the purple region $R_\infty$.  
    Note that the spacetime can have pathologies such as naked
    singularities in the region $R_\infty$ and singular behavior
    near $i^+$, but the figure still motivates defining a nontrivial
    algebra for these regions.  In the conformal extension of Minkowski 
    space, the maroon region indicates the Rindler wedge, 
    and we see that the 
    region at infinity for this patch is the single generator
    of $\scri^+$ shown in purple.  Because this is not an open
    region in the Einstein static universe, we do not expect an algebra
    at infinity for half-sided translations associated with Rindler
    horizons in higher than two spacetime dimensions.  }
    \label{fig:ESU}
\end{figure}

For the asymptotically flat Schwarzschild spacetime, we can give a 
similar argument  by considering 
a conformal extension of the usual spacetime to include regions
beyond $\scri^+$, $\scri^-$ into which modes propagating
out to infinity enter.  These are the triangular regions depicted in
figure \ref{fig:extSch} added on to the standard Penrose diagram of the 
Schwarzschild spacetime.  This conformal extension of 
Schwarzshild can be constructed by analytic continuation,
in which case the regions beyond $\scri^\pm$ are conformal to the negative
mass (or equivalently, negative $r$) Schwarzschild solution \cite{Halacek:2013hwa}.  This 
extended spacetime is similar to the conformal embedding of Minkowski spacetime
into the Einstein static universe \cite{Wald:1984rg}, see figure \ref{fig:ESU}.
When dealing with conformal matter fields, the net of algebras
associated with the Schwarzschild spacetime should naturally define 
a net on the extended spacetime.  In this case, we see that the algebra 
at infinity can be characterized by all fields smeared in the 
open region $R_\infty$ added beyond $\scri^+$ in the conformal
extension of the spacetime.\footnote{There may be pathologies
associated with the existence of naked singularities
in this conformal extension, since the new region has the 
geometry of the negative mass Schwarzschild spacetime.  It seems
likely that these can be handled with appropriate boundary conditions,
the details of which do no matter for the present discussion.  It is only
relevant that there is an open region beyond $\scri^+$ in which
we can smear fields in order to obtain a nontrivial algebra
at infinity.}

Figure \ref{fig:ESU} also highlights the difference between  
black hole spacetimes and a Rindler horizon.  When embedded into
the Einstein static universe, the Rindler patch covers a causal
diamond whose boundary intersects $\scri^+$ only along a single
generator.  This corresponds to the fact that only light rays propagating 
in the same direction parallel to the Rindler horizon
will remain in the Rindler patch when they intersect $\scri^+$.  In
this case, the algebra at infinity is trivial,
since the region at infinity only consists of this single null
generator along $\scri^+$.  Hence, asymptotically flat black holes 
provide an interesting arena to study  half-sided translations with 
nontrivial algebras at infinity, which is not available  
when restricting to quantum field theory in Minkowski space.

As mentioned above, any algebraic state arising from an
average null energy vacuum $|\Omega\rangle$ is fixed by the 
conditional expectation $E_\lambda$.
When the algebra $\ainfty$ is nontrivial, the vacuum $|\Omega\rangle$
is no longer unique; any other state of the form 
$\msf{b}|\Omega\rangle$ with $\msf{b}\in\ainfty$ 
is also a vacuum state, since $[P_\lambda,
\msf{b}]=0$.  
For any such vacuum, its modular flow maps $\ainfty$ 
into itself. This suggests that the modular Hamiltonian 
will decompose into a component associated with $\ainfty$ 
and a component associated with the horizon.  To describe
this decomposition, we first note that $P_\lambda$ also 
generates an inverse half-sided translation  
for the commutant algebras
$\acut_\lambda'$, and these similarly must contain a nontrivial 
translation-invariant algebra $\aminfty$ that is naturally associated
with the algebra at minus infinity on the left half of the 
black hole spacetime, containing $\scri^-_L$ and $i^-_L$.  
This algebra is equivalently 
characterized as $\aminfty = J_\Omega \ainfty J_\Omega$, where
$J_\Omega$ is the modular conjugation for a vacuum state $|\Omega\rangle$.
Modular flow in the commutant algebras again maps $\aminfty$
into itself, and there is a collection of  conditional
expectations $E_\lambda':\acut_\lambda'\rightarrow \aminfty$ 
which fix the algebraic states on $\acut_\lambda'$ defined 
by the vacuum vector $|\Omega\rangle$.   

Because states fixed by a conditional
expectation  are the analogs of tensor factorized states,  
they represent configurations in which there is no
correlation between operators in $\ainfty$ and the ingoing 
modes on the horizon.  However, $|\Omega\rangle$ is a separating
state for $\ainfty$, meaning that this algebra is
entangled with 
something.  The purifying degrees of freedom must be the 
complementary algebra at minus infinity, $\aminfty$.  
Defining the two-sided algebra at infinity as
the algebraic union of these two,
\beq
\atwoinfty = \ainfty \vee \aminfty,
\eeq
this discussion suggests
that $|\Omega\rangle$  defines a pure state on $\atwoinfty$, which,
in particular, implies that $\atwoinfty$ is type $\tone$.  Its commutant
$\ahor=\atwoinfty'$ contains operators  localized to the complete
Killing horizon.  Note that $\ahor$ is nontrivial even in the interacting
case where there is no nontrivial horizon algebra localized 
to the future of a cut.
At minimum, $\ahor$ contains the spectral projections of the 
translation generator $P_\lambda$.  

Since the modular operator $h_\Omega^\lambda$ for the 
horizon cut algebra $\acut_\lambda$ maps both $\ainfty$
and $\aminfty$ into themselves, it also maps $\atwoinfty$
into itself and similarly for the horizon algebra
$\ahor$.  As both $\atwoinfty$ and $\ahor$ are type
$\tone$, this suggests that $h_\Omega^\lambda$  
generates inner automorphisms of these algebras, meaning 
that it would be possible to find operators 
$h_\infty$ affiliated with $\atwoinfty$ 
and $h_\hor^\lambda$ affiliated with  $\ahor$ such that 
\beq \label{eqn:homfactor}
h_\Omega^\lambda = h_\infty + h_\hor^\lambda.
\eeq
At present we do not have a proof of this factorization
of the modular Hamiltonian, in part due to the fact that 
the algebras $\atwoinfty$ and $\ahor$ are not factors
when the inclusion $\ainfty \subset \acut_\lambda$ has trivial 
relative commutant.  However, we can motivate the existence
of $h_\hor^\lambda$ as a well-defined operator by appealing to the 
free field case.  In that case, the horizon cut 
algebra factorizes as $\acut_\lambda = \acut_\lambda^\hor
\otimes \ainfty$, and since there is now a quantum field 
algebra localized to the horizon to the future of the cut,
we expect that the full horizon algebra $\ahor$ becomes a factor.
In this case the modular operator factorizes between 
a horizon component and component at infinity,
$\Delta_{\Omega,\lambda} = \Delta_{\hor,\lambda}\Delta_\infty$,
implying that the modular Hamiltonian exhibits 
the decomposition (\ref{eqn:homfactor}).  
On the other hand, 
$h_\hor^\lambda$ has an expression (\ref{eqn:hclambda}) 
in terms of the integral 
of the stress tensor over the horizon, and its independent
definition in terms of modular theory means that 
this operator has finite fluctuations in a dense
class of states, despite only being
smeared on a codimension-1 null surface.  Since 
the two-point function of the stress tensor in an interacting
CFT takes the same form as in a free theory, we expect that 
$h_\hor^\lambda$ will  have finite fluctuations
for interacting theories as well.  
Hence,
it is plausible that the factorization (\ref{eqn:homfactor}) holds 
even in interacting theories where the 
inclusion $\ainfty\subset
\acut_\lambda$ is singular.  

An important aspect of this splitting is that any vacuum
state $|\Omega\rangle$ gives rise to the same horizon 
component of the modular Hamiltonian $h_\hor^\lambda$.  This
follows from the fact that any two such vacua $|\Omega_{1,2}\rangle$
induce states on $\acut_\lambda$ that are fixed by the conditional
expectation $E_\lambda$, and similarly the states $\omega_{1,2}'$
induced on $\acut_\lambda'$ are fixed by the complementary
conditional expectation $E_\lambda'$.  Hence the  
Connes cocycle  $u_{2|1}(s) = \Delta_{\Omega_2}^{is}
\Delta_{\Omega_1|\Omega_2}^{-is}$ for these two states
is an element of $\ainfty$, and the commutant cocycle
$u_{2|1}'(s) = \Delta_{\Omega_1|\Omega_2}^{-is}\Delta_{\Omega_1}^{is}$
is an element of $\aminfty$
\cite[Chapter IX, corollary 4.22(ii)]{TakesakiII},  
which then implies that $\msf{b}\defeq h_{\Omega_2}^\lambda - 
h_{\Omega_1|\Omega_2}^\lambda$ is affiliated with  
$\ainfty$ and 
$\msf{b}'\defeq h_{\Omega_1|\Omega_2}^\lambda - h_{\Omega_1}^\lambda$ is 
affiliated with $\aminfty$.  We therefore have that 
\begin{align}
h_{\Omega_2}^\lambda - h_{\Omega_1}^\lambda = \msf{b} + \msf{b}' 
\quad\implies\quad
h_{\hor,2}^\lambda - h_{\hor,1}^\lambda = 
\msf{b}+\msf{b}' -h_{\infty,2}+h_{\infty,1} \:\text{affiliated with}\; \atwoinfty.
\end{align}
Thus, the difference $h_{\hor,2}^\lambda - h_{\hor,1}^\lambda$
must be affiliated with the center of $\ahor$.  Since the 
decomposition (\ref{eqn:homfactor}) is ambiguous up to compensating
shifts of $h_\hor^\lambda$ and $h_\infty$ by elements of the center,
we can consistently choose the decomposition such that 
$h_\hor^\lambda$ is the same for all vacuum states $|\Omega\rangle$.
This point connects with the discussion of section 
\ref{sec:hstr} that the commutation relation with $P_\lambda$
uniquely fixes the modular Hamiltonian of a vacuum 
state up to terms that commute with $P_\lambda$.  Since
$h_\infty$ is affiliated with  
$\ahor'$, it commutes with $P_\lambda$, and 
hence $h_\hor^\lambda$ alone satisfies the commutation
relation (\ref{eqn:hPbrack}) associated with the half-sided 
translation.

We now apply this reasoning to the generalized second law.  
Given a vacuum state $|\Omega\rangle$, we formally
split the modular Hamiltonian into one-sided components
$h_\Omega^\lambda = K_\lambda - K_\lambda'$, defined as 
usual as sesquilinear forms respectively 
affiliated with $\acut_\lambda$ and $\acut_\lambda'$.  
This decomposition can be refined by performing a similar
split on each of $h_\infty$ and $h_\hor^\lambda$ appearing
in (\ref{eqn:homfactor}), resulting in
$K_\lambda
= K_\hor^\lambda + K_\infty$, with $K_\infty$ affiliated
with $\ainfty$ and $K_\hor^\lambda$ affiliated with $\ainfty'$.
The validity of this decomposition
only relies on the existence of the conditional
expectation $E_\lambda:\acut_\lambda\rightarrow \ainfty$,
and holds even in interacting theories where 
we cannot assume a tensor product structure 
between $\ainfty$ and a horizon algebra.  
This decomposition of $K_\lambda$ is all that is needed 
to apply Wall's argument for the GSL in setups where the algebra
at infinity is nontrivial.  The relative entropy
between an arbitrary state $|\Phi\rangle$ and a translation
vacuum $|\Omega\rangle$ is give by 
\begin{align}
S_\text{rel}^{\acut_\lambda}(\Phi||\Omega)
&= -S_{\Phi,\lambda}^\text{QFT} +\langle K_{\hor}^\lambda
+K_\infty\rangle_\Phi
+S_{\Omega,\lambda}^\text{QFT}\nonumber \\
&= -S_{\Phi,\lambda}^\text{QFT} 
-\left\langle\frac{A_\lambda}{4G_N}\right\rangle_\Phi +
\left\langle \frac{A_\infty}{4G_N}\right\rangle_\Phi +
\vev{K_\infty}_\Phi
+S_{\Omega,\lambda}^\text{QFT}
\label{eqn:Srelinfty}
\end{align}
applying the identity (\ref{eqn:horcutid}) 
and the fact that $K_{\hor}^\lambda$
continues to be given by the stress tensor integral 
$\frac{2\pi}{\kappa} H_{\zeta_\lambda}^+$ appearing 
in that relation.   
When considering relative entropy differences between
two horizon cuts, the last three terms in (\ref{eqn:Srelinfty})
will cancel since the vacuum entanglement entropy
is independent of the horizon cut and because $K_\infty$ and 
$\frac{A_\infty}{4G_N}$ are forms affiliated with the algebra 
at infinity whose expectation values are the same 
in the state $|\Phi\rangle$ for any value of $\lambda$.  
Monotonicity of relative entropy again reduces to the GSL for this 
setup.  

One interesting aspect of this discussion is that we can employ
any vacuum state $|\Omega\rangle$ when applying monotonicity
of relative entropy
in (\ref{eqn:Srelinfty}).  This independence of the choice
of vacuum can be made more explicit by invoking an identity 
satisfied by the relative entropy in the 
presence of a conditional expectation \cite{Hiai1981, 
Petz1986, Petz1991, Petz1993}.  
Using the notation $\omega = \langle\Omega|\cdot|\Omega\rangle$,
$\varphi = \langle\Phi|\cdot|\Phi\rangle$ as 
states on $\acut_\lambda$, 
and recalling $\omega = \omega\circ E_\lambda$, the identity reads
\beq
S_\text{rel}^{\acut_\lambda}(\varphi||\omega)
= S_\text{rel}^{\ainfty}\left(\varphi|_{\ainfty} \;\Big|\Big|\; \omega|_{\ainfty}\right)
+ S_\text{rel}^{\acut_\lambda}(\varphi||\varphi\circ E_\lambda).
\eeq
The first term on the right involves the relative entropy of the two
states restricted to $\ainfty$, and hence is the same when
computed for two different horizon cuts.  Monotonicity of the 
relative entropy on the left between horizon cuts therefore implies 
that $S_\text{rel}^{\acut_\lambda}(\varphi||\varphi\circ E_\lambda)$
is also monotonic.  Since this relative entropy only 
refers to the state $\varphi$ and the conditional expectation
$E_\lambda$, it is manifestly independent of the choice 
of vacuum state $\omega$.  In terms of entropies,
it can be expressed as 
\begin{align}
S_\text{rel}^{\acut_\lambda}(\varphi||\varphi\circ E_\lambda)
&= -S_{\Phi,\lambda}^\text{QFT} +\vev{K_{\hor_\lambda^+} +
K^\Phi_\infty}_\Phi + S_{\varphi\circ E_\lambda}^\text{QFT} \nonumber \\
&= -S_{\Phi,\lambda}^\text{QFT} -\left\langle \frac{A_\lambda}{4G_N}
\right\rangle_\Phi + \left\langle\frac{A_\infty}{4G_N}\right\rangle_\Phi
+S_{\varphi\circ E_\lambda}^\text{QFT}
\end{align}
using that $\vev{K_\infty^\Phi}_\Phi = 0$ and applying
(\ref{eqn:horcutid}).  Since $\varphi\circ E_\lambda$ is a 
vacuum state, the entropy $S_{\varphi\circ E_\lambda}^\text{QFT}$ 
is independent of the cut $\lambda$, and since the expectation
value of $A_\infty$ is also independent of the cut,
monotonicity of $S_\text{rel}^{\acut_\lambda}(\varphi || \varphi\circ 
E_\lambda)$ again implies the GSL.

\subsection{Gravitational algebra}
\label{sec:flatgravalg}

The discussion up to this point has been at the level of relative
entropies in quantum field theory.  We now describe how 
to incorporate the gravitational constraints to arrive at 
a semiclassical second law for a  gravitational algebra
that incorporates asymptotic charges.  
We will specialize to asymptotically flat spacetimes with massless
matter fields in this 
section, but it should be clear that similar arguments apply
to other asymptotics such as cosmological black holes.  As
emphasized by KFLS \cite{Kudler-Flam2023}, because the null Cauchy surface for 
our subregion naturally splits into a component on the black
hole horizon $\hor_{\lambda}^+$ and a component associated with $\scri^+$,
the gravitational constraints similarly decompose
into independent constraints on each component of the Cauchy surface. 
This allows for a construction of the gravitational algebra
in steps by introducing individual asymptotic charges 
associated with each constraint and constructing the 
crossed product algebras.  Following this, there is an
additional constraint arising from matching conditions 
where the two Cauchy surfaces meet (i.e.\ across $i^+$ in the 
asymptotically flat case).  The matching conditions determine
some of the gravitational charges in terms of other quantities,
so that the final algebra contains only asymptotic 
charges at spatial infinity.  We describe the details 
of this procedure below, and then demonstrate it 
explicitly for an asymptotically flat Schwarzschild black hole.

In order to keep the discussion as general as possible, it is 
convenient to assume that the constraints have already been
imposed on the algebra at infinity, so that 
$\ainfty$ already incorporates asymptotic charges associated 
with $\scri^+$ such as the ADM mass or late-time Bondi mass.
We will assume that this results in a semifinite
algebra, meaning that $\ainfty$ is type $\tone$ or type $\ttwo$ and 
possesses a normal, faithful, semifinite trace, 
which we denote $\trinfty$.  We also
expect $\ainfty$ to be properly infinite (i.e.\ type $\tone_\infty$
or type $\ttwo_\infty$), meaning that $\trinfty \mathbbm{1} = \infty$,
so $\trinfty$ does not define a normalizable state.  Instead, 
$\trinfty$ is a normal, faithful, semifinite weight, 
which still has a well-defined 
modular theory associated with it 
\cite[Chapter VIII]{TakesakiII}.  
We continue to assume that $\ainfty$ is the translation-invariant
algebra of the horizon cut algebra $\acut_\lambda$ (which
remains type $\tthr_1$), and hence 
we can define a weight on $\acut_\lambda$ by composing $\trinfty$
with the conditional expectation $E_\lambda$,
\beq
\omega_{\tr} \defeq \trinfty\circ E_\lambda.
\eeq
The modular flow for $\omega_{\tr}$ reproduces the modular 
flow associated with $\trinfty$ on operators in $\ainfty$, which
is trivial.  Hence, $\ainfty$ is the centralizer of this modular
flow, meaning that these operators commute with the modular
Hamiltonian $h_{{\tr}}^\lambda$ for this weight.\footnote{To
define $h_{\tr}^\lambda$ as an operator on the full Hilbert 
space, we must also specify a weight for the commutant
algebra $\acut_\lambda'$, which we can 
define using the trace $\tr_{-\infty}$ on $\aminfty$
and the complementary conditional expectation $E_\lambda':
\acut_\lambda\rightarrow \aminfty$.  Given such
a pair of weights, we can always find a Hamiltonian
$h_{\tr}^\lambda$ which generates modular flow
on $\acut_\lambda$ and backwards modular flow 
in $\acut_\lambda'$ \cite[Lemma 5.11]{Haagerup1979II}.} 
Using the decomposition (\ref{eqn:homfactor}), valid for
modular Hamiltonians of weights fixed by $E_\lambda$, we find
that $h_{\tr, \infty} = 0$ and hence $h_{\tr}^\lambda = h_\hor^\lambda$,
which is uniquely determined by the conditional expectation $E_\lambda$.
Hence when $\ainfty$ is semifinite, the geometric flow  generated 
by $h_\hor^\lambda$  which fixes points on $\scri^+$ and acts
like a boost around the horizon cut is the modular flow of 
a weight defined on $\acut_\lambda$.  

The procedure for imposing the gravitational constraints is now 
exactly the same as in section \ref{sec:gslsemi}.  
We introduce the 
additional asymptotic charge 
\beq
\hat{y} = -\frac{A_\infty^{(2)}}{4G_N}
\eeq
acting on a Hilbert space $\hfut = L^2(\mathbb{R}_y)$,
and determine the operators that commute with the gravitational
constraint associated with the boost diffeomorphism
which $h_\hor^\lambda$ generates.  When
represented on $\widetilde{\hs} = 
\wh{\hs} \otimes \hfut$, where $\wh{\hs}$  includes
the QFT Hilbert space as well as the Hilbert space for 
asymptotic charges in $\ainfty$, these operators
form a crossed product algebra 
\beq
\widetilde{\acut}_\lambda = \left \langle e^{i\hat{k}h_\hor^\lambda} 
\msf{a} e^{-i\hat{k} h_{\hor}^\lambda}, \hat{y}
\right\rangle,
\qquad \msf{a}\in \acut_\lambda, \quad \hat{k} = -i\frac{d}{dy}.
\eeq
Because $h_\hor^\lambda$ is the modular Hamiltonian of the 
weight $\omega_{\tr}$, $\widetilde{\acut}_\lambda$ is 
the crossed product of a type $\tthr_1$ quantum field theory
algebra by a modular automorphism group, and hence is type
$\ttwo_\infty$ and possesses a trace.

$\widetilde{\acut}_\lambda$ is not the final gravitational
algebra associated with the exterior of the cut since we 
still need to impose the constraint coming from the matching 
condition at $i^+$.  This condition arises because 
the Hilbert space $\widetilde{\hs}$ contains two operators
that measure the time shift at $i^+$ relative to $i^-_L$ 
in the complementary region.  
One of these is the operator $\hat{k}$ appearing
in the crossed product, which is conjugate to the asymptotic
area operator $\hat{y}$ which generates forward time evolution along
the horizon.  This gives $\hat{k}$ the interpretation
of a renormalized Killing
time difference between $i^-_L$ and the cut of $\hor_\lambda^+$ 
in the distant future where one matches the horizon
to a spacelike hypersurface connecting it to $\scri^+$.  
The second time operator is associated with time translations
along $\scri^+$.  We suppose that $\ainfty$ contains an
energy operator $\hat{x}$ 
generating forward time evolution along $\scri^+$, 
which, due to the semiclassical assumptions has a continuous
spectrum and is unbounded above and below.  It therefore 
possesses a conjugate time operator which we denote $\hat{p}$ 
which measures the time shift between $i_L^-$ and the cut 
of $\scri^+$ where one matches the Cauchy surface connecting 
it to the black hole horizon.  Since these two notions of time
shift must agree, we arrive at the final matching 
constraint
\beq \label{eqn:match}
\constr_\text{match}=\hat{k}-\hat{p} = 0.
\eeq

To impose this constraint on the algebra, we first 
look for operators in $\widetilde{\acut}_\lambda$ that 
commute with $\hat{k}-\hat{p}$.  This subalgebra 
\beq
\widetilde{\acut}_\lambda^0 = \widetilde{\acut}_\lambda\wedge
\left\langle \hat{k}-\hat{p}\right\rangle'
\eeq
of $\widetilde{\acut}_\lambda$ is isomorphic
to the desired gravitational algebra, but  represented on
a Hilbert space that contains other operators that do not 
preserve the constraint (\ref{eqn:match}) .  We can remedy
this by passing to a Hilbert space 
of coinvariants, following the procedure outlined by CLPW
\cite{Chandrasekaran2022a}.  This Hilbert space is a quotient 
of $\widetilde{\hs}$ obtained by identifying vectors
that differ by a state in the image of $\hat{k}-\hat{p}$,
so that $|\Omega\rangle \sim |\Omega\rangle 
+ (\hat{k}-\hat{p})|\chi\rangle$.  Passing to this 
space of coinvariants can be used to eliminate the 
factor of $\hfut$ from the Hilbert space, so that the physical
Hilbert space is identified with $\wh{\hs}$.  This identification
is achieved by defining the operator 
\beq
V \defeq \langle 0_k|e^{-i\hat{y}\hat{p}},
\eeq
which maps $\widetilde{\hs}$ to $\wh{\hs}$, and annihilates 
states of the form $(\hat{k}-\hat{p})|\chi\rangle$.
This allows us to map any operator in $\msf{c}_0\in
\widetilde{\acut}_\lambda^0$
to an operator $\ho c$ acting on $\wh{\hs}$ by the relation 
\beq\label{eqn:intertwine}
V \msf{c}_0 = \ho c V.
\eeq
The algebra in $\mathcal{B}(\wh{\hs})$ generated by all 
such $\ho c$ is the final result for the gravitational
algebra $\wh{\acut}_\lambda$.\footnote{$V$ is not bounded 
when viewed as an operator acting on $\widetilde{\hs}$, and 
is singular even  as an unbounded operator
since $VV^\dagger = \langle 0_k|e^{-i\hat{y}\hat{p}} e^{i\hat{y}\hat{p}}
|0_k\rangle = \infty$.  One interpretation of $V$ is to view it as 
determining an operator-valued weight  $T$ from operators 
in $\widetilde{\acut}_\lambda$ to operators in 
$\wh{\acut}_\lambda$ via $T(\tilde{\msf{c}}) = V\tilde{\msf{c}} V^\dagger$
(see section \ref{sec:ovw} for the definition of operator-valued weights).
Equation (\ref{eqn:intertwine}) then corresponds to the bimodule property of 
the operator-valued weight.  It is interesting that the 
BRST procedure described by CLPW \cite{Chandrasekaran2022a} 
has an interpretation
in terms of operator-valued weights.  }  

A subtlety arises at this point regarding the existence of 
a trace on $\wh{\acut}_\lambda$.  The issue is that  the 
trace defined on $\widetilde{\acut}_\lambda$ is a 
semifinite weight, meaning it assigns infinite values to 
some operators in $\widetilde{\acut}_\lambda$, although
it is finite on sufficiently many operators that they generate
the full algebra.  There is no guarantee that the trace defines 
a semifinite weight when restricted to a subalgebra, since the 
subalgebra may consist mostly of operators with infinite trace.  
Hence, we need to separately analyze whether $\wh{\acut}_\lambda$
admits a trace.  We will describe here a set of sufficient
conditions that ensure this is the case.  

Since $\hat{p}$ 
generates an automorphism of $\ainfty$, we can assume that this 
algebra takes the form of a crossed product 
$\ainfty = \overline{\ainfty^0} \rtimes \mathbb{R}$,
where $\overline{\ainfty^0}$ is the subalgebra of operators commuting 
with $\hat{p}$, $\overline{\ainfty^0} = \ainfty \wedge \left\langle\hat{p}
\right\rangle'$.  
Denoting 
$\Hinf$  as the generator of the automorphism
on $\overline{\ainfty^0}$ for this crossed product, the algebra $\ainfty$
is takes the form
\beq
\ainfty = \left\langle e^{i\hat{p}\Hinf} \msf{b} e^{-i\hat{p}\Hinf},
\hat{x}\right\rangle,\qquad \msf{b}\in {\ainfty^0}\cong
\overline{\ainfty^0}, \quad [\hat{x},\hat{p}] = i.
\eeq
The algebra ${\ainfty^0}$ simply consists 
of quantum field theory operators without the dressing factors 
$e^{i\hat{p}\Hinf}(\cdot)e^{-i\hat{p}\Hinf}$. 
The full horizon cut algebra also takes the form of a crossed product:
defining $\overline{\acut_\lambda^0} 
= \acut_\lambda\wedge \left\langle\hat{p}
\right\rangle'$, we find that 
\beq\label{eqn:Clamb}
\acut_\lambda = \left\langle e^{i\hat{p}\Hinf} \msf{c}e^{-i\hat{p}
\Hinf}, \hat{x}\right\rangle, \qquad \msf{c}\in 
{\acut_\lambda^0}\cong \overline{\acut_\lambda^0}. 
\eeq
The unconstrained gravitational algebra $\widetilde{\acut}_\lambda$
then is a crossed product of ${\acut_\lambda^0}$ 
by $\mathbb{R}^2$,
\beq \label{eqn:wtildClam}
\widetilde{\acut}_\lambda = \left\langle
e^{i\hat{k}h_\hor^\lambda}e^{i\hat{p}\Hinf}
\msf{c}e^{-i\hat{p} \Hinf} e^{-i\hat{k}h_\hor^\lambda},
\hat{x}, \hat{y}
\right\rangle, \qquad \msf{c}\in{\acut_\lambda^0},
\eeq
and when imposing the matching constraint (\ref{eqn:match}), we find
that only the combination $\hat{x}+\hat{y}$ remains in the final
gravitational algebra.  The relation  (\ref{eqn:intertwine}) then
maps $\hat{x}+\hat{y}$ to $\hat{x}$ when representing
this algebra on $\widehat{\mathcal{H}}$, so that the final
gravitational algebra is just a single crossed product by the 
combined automorphism generated by $h_\hor^\lambda + \Hinf$,
\beq
\wh{\acut}_\lambda = \left\langle
e^{i\hat{p} (h_\hor^\lambda + \Hinf)} \msf{c}
e^{-i\hat{p}(h_\hor^\lambda + \Hinf)}, \hat{x}\right\rangle,
\qquad \msf{c}\in {\acut_\lambda^0}.
\eeq

We can now characterize the status of the trace on 
$\wh{\acut}_\lambda$ in terms of properties of the automorphism
that $\Hinf$ generates on $\ainfty^0$.  
The first case to consider is when
$\hat{p}$ generates an automorphism that preserves the trace
on $\ainfty$, and  the dual automorphism generated by $\Hinf$
on $\ainfty^0$ is inner.  
An example of this case would 
be the algebra defined at $\scri^-$ when working with the Unruh
state for a black hole, 
as considered in \cite{Kudler-Flam2023},
where the analog of $\ainfty^0$ is type $\tone_\infty$. 
In this case, $\ainfty^0$ is itself
semifinite and possesses a trace, so modular flow 
on this algebra is inner.  Since  $\Hinf$
generates an inner automorphism of $\ainfty^0$, 
by the converse of the cocycle derivative theorem 
\cite[Theorem VIII.3.8]{TakesakiII},
it must be the modular Hamiltonian $h_\infty$  of some weight
defined on $\ainfty^0$.  Pulling back this weight by the 
conditional expectation $E_\lambda$, this leads to a weight 
on ${\acut_\lambda^0}$ whose modular Hamiltonian is $h_\hor^\lambda
+ h_\infty$, and therefore $\widehat{\acut}_\lambda$ is a modular
crossed product of the type $\tthr_1$ algebra $\acut_\lambda^0$,
and hence possesses a trace.  

The second case is when the automorphism generated by 
$\hat{p}$ rescales the trace on $\ainfty$. An example 
of this case, considered below, is the Schwarzschild
black hole in the Hartle-Hawking state.  In this case, we can 
apply the general structure theorem for type $\tthr$ 
von Neumann algebras \cite[Theorem XII.1.1]{TakesakiII} to conclude,
assuming both $\ainfty^0$ and $\ainfty$ are factors, 
that $\ainfty^0$ is type $\tthr_1$
and $\Hinf$ generates modular flow of some weight on $\ainfty^0$,
whose modular Hamiltonian we denote $h_\infty$.
So once again by composing this weight with the conditional
expectation $E_\lambda$, we obtain a weight on $\acut_\lambda^0$
whose modular Hamiltonian is $h_\hor^\lambda+h_\infty$.
Again, the gravitational algebra $\widehat{\acut}_\lambda$
is a modular crossed product which 
possesses a trace.  

Finally, if $\hat{p}$ preserves the trace on $\ainfty$, but 
$\Hinf$ generates an outer automorphism on $\ainfty^0$, 
then it is possible to show
that $\ainfty^0$ is already semifinite.  However, the outerness
of the automorphism generated by $\Hinf$ shows that 
it is not a modular Hamiltonian for $\ainfty^0$, 
in which case neither is 
$h_\hor^\lambda+\Hinf$ for $\acut_\lambda^0$.  This is 
enough to conclude that $\widehat{\acut}_\lambda$ is 
not semifinite.  This case might arise if $\Hinf$ were a 
different symmetry generator of $\ainfty^0$, such as rotation.  
In such situations, there are additional gravitational
charges that should be added such as the ADM angular momentum,
and by incorporating the associated constraints one may be able to 
argue that the final gravitational algebra is semifinite.  
An example where these additional constraints are important 
is for rotating black holes, such as the Kerr example
considered in \cite{Kudler-Flam2023}.  

We conclude this section by applying
the general construction to the example 
of a Schwarzschild black hole.   In this case, there is 
a natural vacuum state $|\Omega\rangle$ provided by the 
Hartle-Hawking state \cite{HH1976}.   The horizon cut
algebra $\acut_\lambda^0$ consists of all quantum fields to the 
exterior of the cut $v=\lambda(y^A)$, and possesses a subalgebra 
$\ainfty^0$ of quantum fields that propagate out through $\scri^+$.  
Since $|\Omega\rangle$ is a vacuum state for the horizon
average null energy  $P_\lambda$ which generates 
a half-sided translation and $\ainfty^0$ 
is the translation-invariant algebra, the modular Hamiltonian
for $|\Omega\rangle$ splits into a sum of two pieces,
$h_\Omega^\lambda = h_\hor^\lambda + h_\infty$, where 
$h_\hor^\lambda$ is given, as usual, as an integral of the stress 
tensor over the horizon (\ref{eqn:hclambda}).  The component $h_\infty$
affiliated with the algebra at infinity is independent of the choice
of horizon cut, and therefore can be identified by taking the cut
to lie on the bifurcation surface, $\lambda =0$. In this case, 
the modular Hamiltonian is the generator of the global time-translation
symmetry associated with the Killing vector $\xi^a$.  In general,
this generator will be given by an integral over a complete
Cauchy surface $\Sigma$ in the global spacetime,
\beq
h_\Omega = -\frac{2\pi}{\kappa}
\int_\Sigma T\indices{^a_b}\xi^b \epsilon_{a\ldots},
\eeq
and it is independent of the choice of Cauchy surface since
it is the generator of a symmetry.  Choosing the Cauchy surface 
to lie on the horizon and asymptotic infinity $\Sigma =
\hor \cup \scri^+ \cup \scri^-_L$ (and assuming massless fields 
so that contributions from $i^+, i^-_L$ can be neglected), we 
see that $h_\infty$ is a local integral of the stress tensor
over just $\scri^+$ and $\scri^-_L$,
\beq\label{eqn:hinftyint}
h_\infty = -\frac{2\pi}{\kappa}\int_{\scri^+ \cup \scri^-_L}
T\indices{^a_b}\xi^b\epsilon_{a\ldots}.
\eeq
This reflects the fact that the state
of the fields at $\scri^+$ in the Hartle-Hawking
state is thermal at the Hawking temperature $T_H = \frac{\kappa}{2\pi}$.
This modular Hamiltonian includes a contribution from 
gravitons, which is quadratic in the  Bondi
news tensor at $\scri^+$, $\scri^-_L$ 
\cite{Ashtekar1981, Wald1999, Grant2021}.

To carry out the first step in the construction of the gravitational
algebra, we must add the appropriate asymptotic charges
to $\ainfty^0$ and impose constraints to end up with a 
semifinite algebra.  The asymptotic charge is given by
the difference between the ADM mass $M_\text{ADM}$ defined 
at $i^0$ and the late time Bondi mass $M_+$ defined at $i^+$.  
The constraint that relates these operators to the quantum
fields is the Bondi mass loss formula \cite{Bondi1962, Sachs1962,
Ashtekar1979, Ashtekar1981, Wald1999, Grant2021, Prabhu2022},
\beq \label{eqn:constr+}
\constr_{\scri^+} = K_\infty +\frac{2\pi}{\kappa}(M_+-M_\text{ADM})=0,
\eeq
where $K_\infty$ involves only the component of $h_\infty$ affiliated
with $\scri^+$, 
\beq
K_\infty=
-\frac{2\pi}{\kappa}\int_{\scri^+}T\indices{^a_b}\xi^b\epsilon_{a\ldots}.
\eeq
There is a similar constraint involving operators in the complementary
algebra at infinity at $\scri^-_L$, the left ADM mass $M_\text{ADM}^L$,
and the left early time Bondi mass $M_-^L$ at $i^-_L$, which reads 
\beq \label{eqn:constr-L}
\constr_{\scri^-_L} = -K_{-\infty'} - \frac{2\pi}{\kappa}(M_-^L-
M_\text{ADM}^L) = 0, \qquad K_{-\infty'} = \frac{2\pi}{\kappa}
\int_{\scri^-_L}T\indices{^a_b}\xi^b\epsilon_{a\ldots}.
\eeq

When imposing these constraints on the algebra, it is necessary
to consider the sum of these two constraints
$\constr_\infty = \constr_{\scri^+} + \constr_{\scri^-}$, 
due to the appearance of  $K_\infty$ and 
$K_{-\infty'}$  in the individual constraints (\ref{eqn:constr+}),
(\ref{eqn:constr-L}).  
Since $\ainfty^0$ is an algebra of quantum fields in a thermal
state, the one-sided Hamiltonians $K_\infty$, $K_{-\infty'}$ are only
 defined as sesquilinear forms due to infinite 
thermal fluctuations.  However, the combination appearing 
$\constr_\infty$ is $K_\infty - K_{-\infty'}=h_\infty$,
which is a well-defined operator generating a 
strongly continuous flow and therefore amenable to constructing 
a crossed product algebra.  
Carrying out the general procedure outlined in 
sections \ref{sec:crprod} and \ref{sec:gslsemi} 
and references \cite{Chandrasekaran2022a, Jensen2023}
for imposing the constraint, the resulting 
semifinite algebra $\ainfty$ is the crossed product
\beq
\ainfty = \left \langle e^{i\hat{p} h_\infty}\msf{b}e^{-i\hat{p}h_\infty}
,\hat{x}\right\rangle, \quad \msf{b}\in\ainfty^0, \quad 
\hat{x} = \frac{2\pi}{\kappa}(M_+-M_\text{ADM}).
\eeq
Since $h_\infty$ generates modular flow on $\ainfty^0$, 
this algebra is a modular crossed product and therefore 
possesses a trace.  The algebra $\acut_\lambda$ that includes
$\ainfty$ as a subalgebra is defined exactly as in equation
(\ref{eqn:Clamb}), which is a type $\tthr_1$ algebra admitting
a half-sided translation with $\ainfty$ as the translation-invariant
subalgebra.  

We next impose the horizon constraint by including the 
asymptotic area operator
\beq
\hat{y} = -\frac{A_\infty}{4G_N},
\eeq
and forming the crossed product of $\acut_\lambda$ by the 
horizon modular flow generator $h_{\hor}^\lambda$, just as 
in section \ref{sec:gslsemi}.  This leads to the double crossed 
product algebra $\widetilde{\acut}_\lambda$ defined in 
equation (\ref{eqn:wtildClam}).  We then impose the matching 
constraint (\ref{eqn:match}) by determining the subalgebra 
that commutes with $\hat{k}-\hat{p}$.  The only asymptotic 
charge that survives this constraint is the sum 
\beq\label{eqn:x+y}
\hat{x} + \hat{y} = -\frac{2\pi}{\kappa} M_\text{ADM}+\frac{2\pi}{\kappa}
M_+ - \frac{A_\infty}{4G_N}.
\eeq
Furthermore, since we have assumed only massless fields, there 
is no stress energy flux through $i^+$, which implies a further 
condition on the gravitational charges \cite{Kudler-Flam2023},
\beq \label{eqn:AMmatch}
\frac{A_\infty}{4G_N} = \frac{2\pi}{\kappa} M_+,
\eeq
so that these terms cancel in (\ref{eqn:x+y}).
Therefore, the only asymptotic charge in the final gravitational
algebra is the ADM mass.  After representing this algebra
on a Hilbert space of coinvariants, which we can take to be 
$\mathcal{H}_\text{QFT}\otimes L^2(\mathbb{R})$ as described
above, the resulting algebra is 
\beq \label{eqn:xMADM}
\wh{\acut}_\lambda = \left\langle 
e^{i\hat{p} (h_\hor^\lambda + h_\infty)} \msf{c}
e^{-i\hat{p}(h_\hor^\lambda + h_\infty)}, \hat{x}\right\rangle,
\qquad \msf{c}\in {\acut_\lambda^0}, \qquad
\hat{x} = -\frac{2\pi}{\kappa} M_\text{ADM}.
\eeq
One confirms that this final algebra possesses a trace because it
is simply the modular crossed product
of $\acut^0_\lambda$, since $h_\hor^\lambda+h_\infty$ is the horizon-cut
modular Hamiltonian of the Hartle-Hawking state.  

The  generalized second law for these horizon
cut algebras now follows in a similar manner as in section
\ref{sec:gslsemi}.  We consider semiclassical states $|\wh\Phi\rangle = 
|\Phi, f\rangle$
with the wavefunction for $\hat{x}$ slowly varying, $\frac{f'}{f}\sim
\op(\vep)$.  Applying formula (\ref{eqn:SrhoSrel}), the semiclassical
expansion for the entropy of the algebra $\wh{\acut}_\lambda$
in this state is given by 
\beq
S(\rho_{\wh{\Phi}}^\lambda) = 
-S_\text{rel}^{\acut_\lambda^0}(\Phi||\Omega) 
-\left\langle \hat{x}\right\rangle_{\wh{\Phi}} +S_f^{M_\text{ADM}}
+\op(\vep^2).
\eeq
This can again be converted into an expression
involving the generalized entropy of the horizon
cut by writing the relative entropy as the modular
free energy $S_\text{rel}^{\acut_\lambda^0}(\Phi||\Omega)
= \langle K_\infty + K_{\hor}^\lambda\rangle_\Phi 
- S_{\Phi,\lambda}^{\text{QFT}}$ and applying the 
constraints (\ref{eqn:constr+}), (\ref{eqn:AMmatch}), 
and (\ref{eqn:horcutid}). 
Since the same formula applies for the entropy of the algebra 
for a later cut $\tilde\lambda$ 
in the same state $|\wh{\Phi}\rangle$, and 
since $\left\langle \hat{x}\right\rangle_{\wh{\Phi}}$ and 
$S_f^{M_\text{ADM}}$ remain unchanged, the entropy difference
between the two cuts reduces to a difference of relative
entropies,
\beq
S(\rho_{\tilde\lambda})-S(\rho_\lambda) = 
-S_\text{rel}^{\acut_{\tilde\lambda}^0}(\Phi||\Omega)
+ S_\text{rel}^{\acut_\lambda^0}(\Phi||\Omega) \geq 0,
\eeq
up to $\op(\vep^2)$ corrections.  

Interestingly, a stronger statement of 
the generalized second law can be proven for overall differences
in entropy between the horizon cut $\lambda$ and the 
entropy at very late times.  
As discussed above,
$\wh{\acut}_\lambda$ 
is a crossed product of a quantum 
field algebra $\acut_\lambda^0$ which admits a half-sided 
translation.  The state in which the crossed product 
is defined is fixed by the conditional expectation $E_\lambda$ 
that sends $\acut_\lambda$ to the translation-invariant
algebra $\ainfty^0$.  Modular flow in this state then 
maps $\ainfty^0$ to itself,  
and hence there is a subalgebra
of $\wh{\acut}_\lambda$ isomorphic to the crossed product 
of $\ainfty^0$,
\beq
\ainfty  = \left\langle e^{i\hat{p}h_\infty} \msf{b}
e^{-i\hat{p} h_\infty}, \hat{x}\right\rangle \subset\wh{\acut}_\lambda.
\eeq
Because modular flow commutes with the conditional expectation
$E_\lambda$, the gravitational algebra $\wh{\acut}_\lambda$
admits a conditional expectation to $\ainfty$, denoted
$\wh{E}_\lambda$, which acts on operators as 
\beq
\wh{E}_\lambda\left(e^{i\hat{p}(h_\hor^\lambda + h_\infty)}
\msf{c} e^{-i\hat{p}(h_\hor^\lambda+h_\infty)} e^{iu\hat{x}}\right)
=
e^{i\hat{p}h_\infty} E_\lambda(\msf{c}) e^{-i\hat{p}h_\infty}
e^{iu\hat{x}}.
\eeq
The formula for the trace on either algebra $\wh{\acut}_\lambda$ 
or $\ainfty$ is the same, given by 
\beq 
\tr(\cdot) 
= 2\pi \langle \Omega, 0_p| e^{-\hat{x}} (\cdot)|\Omega, 0_p\rangle,
\eeq
and hence $\tr_\infty = \tr_{\lambda|\infty}$.   Because 
the inclusion $\ainfty \subset \wh{\acut}_\lambda$ is trace-preserving, 
we are in a position to apply the results 
of \cite{Longo2022}, which imply that the entropy on $\ainfty$
is larger than the entropy on $\acut_\lambda$ for {\it any}
state $|\wh\Phi\rangle$, with no semiclassical assumption
on the wavefunction.  Hence we arrive at the global second law,
\beq \label{eqn:globalgsl}
S(\rho_{\wh{\Phi}}^\infty)\geq S(\rho_{\wh\Phi}^\lambda).
\eeq
This is similar to the global second law discussed in 
\cite{Chandrasekaran2022b}, except that it now involves 
a large algebra at infinity $\ainfty$ that consists
of more than just the asymptotic charges.  
We also note that
the conditional expectation $\wh{E}_\lambda$ 
 preserves the trace $\tr_\lambda = \tr_\infty \circ \wh{E}_\lambda$,
but we expect that the index of this inclusion is infinite.  
Hence, the upper bound on the entropy increase derived in 
\cite{Longo2022} is simply infinite, and does not yield an interesting
constraint.  

As a final comment, we mention that a  double crossed product
algebra similar to $\widetilde{\acut}_\lambda$ 
defined in (\ref{eqn:wtildClam}) appears in the construction
of the algebra for Schwarzschild-de Sitter by KFLS
\cite{Kudler-Flam2023}.  In this construction, there is an
additional  degree of freedom included in the algebra that 
describes  an observer
eternally restricted to the regions between the black hole 
and cosmological horizons.  The observer's energy therefore
contributes to the flux through $i^-$
so that the analog of the matching condition (\ref{eqn:AMmatch})
is modified by the observer's Hamiltonian, which 
is then related to the sum of the two early time 
horizon area fluctuations.
The presence of such an observer also allows the matching 
condition (\ref{eqn:match}) for the two horizon times 
to not be imposed, so that the final double crossed product
algebra is the correct description in this case.  However, 
it is also natural to ask whether the observer degree of freedom
can be eliminated in this setup.  This is done precisely by setting 
the observer Hamiltonian to zero, and imposing the matching 
constraint (\ref{eqn:match}).  The resulting algebra is 
a single crossed product, and semifiniteness follows
from the arguments discussed above.  In this 
description, the early time 
black hole horizon area replaces
the observer Hamiltonian as the asymptotic gravitational
charge, so that the black hole itself plays the 
role of the observer in de Sitter space.  
This setup is quite analogous to the model of 
an observer in holography as a small black hole,
considered in \cite{Jafferis:2020ora, deBoer:2022zps}.

\subsection{Rotating black holes and Unruh states}
\label{sec:rotUnruh}

The discussion of section \ref{sec:flatgravalg} aims to be sufficiently
general so as to provide a construction of semifinite 
gravitational algebras for generic black holes with a variety
of asymptotics and angular velocities.  However, a number of 
subtleties
arise when carrying out the details of the construction for 
more complicated cases than the Schwarzschild black hole.
In this section, we describe two such subtleties in applying 
the procedure in the physically relevant contexts of 
rotating black holes and  black holes formed from collapse.  
We conjecture how these subtleties should be resolved, but do
not perform an in-depth analysis, which we leave as an
interesting topic for future investigation.  

For rotating asymptotically flat 
black holes such as the Kerr solution or Myers-Perry
black holes in $d\geq 5$ \cite{Myers1986}, the main issue is that the 
horizon-generating Killing vector $\xi^a$ 
is not timelike throughout the 
black hole exterior.  In general, the Killing vector has the 
decomposition \cite{Hollands:2006rj}
\beq
\xi^a = t^a +\sum_i \Omega^i_{\hor} \phi_i^a,
\eeq
where $t^a$ is the Killing vector associated with asymptotic time
translations, $\phi_i^a$ are a collection of rotational Killing vectors,
and $\Omega_{\hor}^i$ are the horizon rotational velocities.  
Since the norms of $\phi_i^a$ grow at large radius while the norm
of $t^a$ remains bounded, the vector $\xi^a$ eventually becomes 
spacelike as long as there is at least one nonzero $\Omega_\hor^i$.  
There are therefore operators localized in the region that $\xi^a$ 
is spacelike that carry negative Killing energy.  The 
existence of negative energy states created by these operators
is closely tied to the phenomenon of superradiance for 
rotating black holes \cite{Hawking:1975vcx,Wald:1975kc}.

Negative Killing energy states pose a problem for 
constructing a thermal state,
since a thermal density matrix $\rho \propto e^{-\beta H_\xi}$ is not
normalizable when the spectrum of $H_\xi$ is not bounded below.  
This  
appears to preclude the existence of a global vacuum state
$|\Omega\rangle$ that satisfies 
a KMS condition with respect to the Killing flow, meaning there 
is no analog of the Hartle-Hawking state for the Kerr black hole
whose modular Hamiltonian globally generates a geometric flow
\cite{Kay:1988mu}.\footnote{Relatedly, one can show
that for any state whose modular flow is globally geometric,
the flow must be timelike throughout the spacetime region associated 
with the algebra \cite{Sorce2024}.  Interestingly,
the restriction to timelike flows does not appear to apply to 
proper weights, discussed below.}  This does not pose an immediate issue to the 
semiclassical generalized second law discussed in section \ref{sec:ainftysemi}.
There, we only required that there is a vacuum state $|\Omega\rangle$ 
for the horizon
average null energy operators $P_\lambda$, since any such state
has a modular Hamiltonian whose horizon component $h_\hor^\lambda$
is given by an integral of the stress tensor (\ref{eqn:hclambda}).
The behavior of the modular flow of $|\Omega\rangle $ on the algebra
at infinity $\ainfty$ does not affect the derivation
of the generalized second law in this context.

The actual problem arises when attempting to construct a 
semifinite gravitational algebra following the procedure 
of section \ref{sec:flatgravalg}.  In this case, we would like 
to argue that imposing the gravitational constraints on the 
algebra at infinity results in an algebra with a trace; however,
this is tantamount to assuming that there is a state $\omega_\infty$
on $\ainfty$ whose 
modular flow generates time translations along the 
vector $t^a$ at $\scri^+$. 
One possibility is that such a state  does exist, and one would 
further need to argue that it is thermal at the Hawking temperature.
Assuming this is possible, this state could be extended to a state 
on any of the horizon cut algebras by composing it with the 
conditional expectation $E_\lambda$, and then $\omega_\lambda = 
\omega_\infty\circ E_\lambda$ would be a horizon vacuum state.  For rotating
black holes, the modular flow for this state would be geometric 
only for operators on the horizon and on $\scri^+$, since it flows 
along different Killing vectors 
on these two surfaces (respectively,
$\xi^a$ and $t^a$).  Assuming that one can argue that this 
state is Hadamard throughout the two-sided Kerr geometry, this 
would provide the necessary state to construct a semifinite gravitational
algebra.  One would also need to considered modified matching 
conditions for the gravitational charges and incorporate additional
crossed products for rotations, along the lines of KFLS
\cite{Kudler-Flam2023}, due to the fact that the constraints
are associated with different Killing vectors and therefore different
asymptotic charges.  

Another possibility is to try to work directly with the 
analog of the Hartle-Hawking state, which in the case of 
Kerr is given by the Frolov-Thorne vacuum, which has 
the appropriate thermal behavior for operators near the horizon
where $\xi^a$ is timelike \cite{Frolov:1989jh}.  
However,  this vacuum cannot lead
to a well-defined state everywhere in the Kerr exterior, due to issues
related to superradiance discussed above.  This manifests in 
a failure of the Hadamard property for correlation 
functions  in the region where $\xi^a$ is 
spacelike.  
If one were to treat the Frolov-Thorne vacuum as a normalizable state,
it would produce a unitarily inequivalent GNS 
representation of the local
field algebra relative to a state that was Hadamard everywhere in the Kerr
exterior.  Such a representation would be irrelevant for investigations
of the generalized second law for the Kerr black hole.  However, it 
is possible that by reinterpreting the Frolov-Thorne vacuum as a nonnormalizable
weight as opposed to a state, it could provide the appropriate
modular operator to construct a trace on the algebra at infinity.

Recall that a weight on a von Neumman algebra is a positive linear 
function from positive operators in the algebra to the extended
positive reals $\wh{\mathbb{R}} = \mathbb{R}^+\cup \{+\infty\}$
\cite[Chapter VII]{TakesakiII}.  
The weight is said to be semifinite if the weak closure of 
the algebra generated by the operators with finite expectation value 
in the weight gives back the full von Neumann algebra.  
Since the Frolov-Thorne vacuum assigns finite expectation values
to operators in the region where $\xi^a$ is timelike, we can
interpret it as a weight on the algebra. Furthermore, the causal
completion of this region in spacetime yields back the entire exterior
wedge of the Kerr black hole, so it is plausible that the subalgebra 
of operators with finite expectation value in the Frolov-Thorne vacuum
is weakly dense in the full von Neumann algebra of the black hole exterior.  
If this is the case, the Frolov-Thorne vacuum would define a semifinite weight
on this algebra.  The modular operator for this weight would presumably 
generate the Killing flow of $\xi^a$, and hence by implementing the 
crossed product of this flow on the algebra at infinity, one would obtain
an algebra with a  trace.  While an intriguing proposal, there are many
details to check in order to establish the proposed properties
of the Frolov-Thorne vacuum as a semifinite weight.  A few of these
issues are discussed in section \ref{sec:discussion}.

A second set of questions arises when applying the arguments 
for the generalized second law to  black holes formed from
collapse,
whose natural vacuum state is the Unruh state 
\cite{Unruh:1976db, Wald:1995yp} as opposed 
to the Hartle-Hawking state.  The Unruh vacuum
is defined by initial data on the past black hole 
horizon and $\scri^-$, and imposes that there is no incoming 
radiation from past infinity.   By contrast, the Hartle-Hawking
state has incoming thermal radiation from $\scri^-$ in addition
to outgoing Hawking radiation at $\scri^+$, which is 
an unnatural idealization for astrophysical black holes; 
such a boundary condition is more appropriate for an eternal
black hole.  The question therefore arises to what extent 
the Unruh state describing an evaporating black hole is 
covered by the analysis in the present paper which relied 
on properties of the Hartle-Hawking vacuum.  

At first glance it appears that the Unruh state 
can just be considered a specific excited state 
for the algebra of quantum fields exterior to a cut 
of the horizon.  It would then describe a state with some 
amount of entanglement between modes that enter the horizon
and modes in $\ainfty$ that escape through $\scri^+$, and 
therefore satisfies a generalized second law according 
to the discussion of section \ref{sec:ainftysemi}.  
The main issue with this argument is that the Unruh vacuum
differs significantly from the Hartle-Hawking vacuum 
in the infrared, so that in standard treatments these states 
result in unitarily inequivalent representations of the field
algebras in the black hole exterior.  This results in a failure 
in the Hadamard property at the past black hole horizon in the 
maximally extended spacetime,  
leading to a divergence in the average null energy operators $P_\lambda$ 
coming from contributions near the bifurcation surface.  

As in the discussion of the Frolov-Thorne vacuum, the resolution here 
may involve reinterpreting the states under consideration instead
as weights on a single representation of the field algebra.  
If we take the perspective that the Unruh vacuum is the 
appropriate state for black holes formed from collapse, this would
then require interpreting the Hartle-Hawking vacuum as a weight 
on the von Neumann algebra constructed from the GNS representation
of the field algebra in the Unruh state.  
Fortunately, the structure
theorems of half-sided translations and modular inclusions
generalize to the case of normal semifinite weights
\cite{Araki2005}, so assuming the Hartle-Hawking vacuum
still defines a weight fixed by the translation, these
theorems would continue to allow the derivation of the 
modular Hamiltonian as a stress tensor integral 
as in sections \ref{sec:hstr} and \ref{sec:ainftysemi}.  
One possible subtlety in applying the general discussion of 
section \ref{sec:ainftysemi} is that when the translation-invariant
weight is not normalizable, there may no longer be a conditional
expectation to the translation-invariant algebra at infinity.  
A natural conjecture generalizing 
the results of \cite{Borchers1997} 
is that instead, one would find an operator-valued
weight to the algebra at infinity.  This would likely still allow
the discussion of section \ref{sec:ainftysemi} to go through, and 
would still allow the construction of  a gravitational
algebra following the procedure of section \ref{sec:flatgravalg}.  
Verifying the details of these conjectures is an important direction
for future work.

\section{Perturbative corrections to crossed product entropy}
\label{sec:pertent}

In section \ref{sec:gslsemi}, we found that the 
crossed product entropies satisfy a generalized second law 
to leading order in the semiclassical expansion for states of the 
form $|\wh{\Phi}\rangle = |\Phi,f\rangle$.  This raises the 
 question of whether these entropies continue to 
be monotonic after accounting for corrections in the semiclassical
expansion.  In this section we begin to address this question.
We derive the exact expression for the correction to the 
semiclassical entropy in the crossed product algebra, and work
out some of its main properties.  One of these properties is that 
the correction is strictly nonnegative, which 
follows from the interpretation of the classical-quantum
state as a Petz recovery of the quantum
field state $|\Phi\rangle$.  We can also obtain
an upper bound on the entropy correction using the explicit expression
for the entropy correction.  Although we are not able to 
use these bounds to conclude anything about the corrected generalized
second law, which relies on differences in entropy corrections, 
we note that  these bounds were recently
employed in \cite{Kudler-Flam:2023hkl} 
to obtain rigorous proofs of the Bekenstein
bound and quantum null energy condition in terms of vacuum-subtracted
entropies. These bounds indicate that it is difficult to violate the 
generalized second law with semiclassical corrections, although it 
remains possible that such a violation could occur over short 
time scales in regimes where the quantum field relative entropy
is nearly constant.  We leave open the question of whether the GSL
holds at all times for the crossed product entropies, but 
note that if it did hold, it would represent a novel data-processing
inequality bounding the decrease of relative entropy for quantum 
fields.  It would therefore be interesting to connect this proposed
bound to the improved data-processing inequalities employed in universal
recovery channels \cite{Junge:2015lmb, Faulkner:2020iou, Faulkner:2020kit}.

We begin by explaining the interpretation of the state $|\Phi,f\rangle$
as a Petz-recovered state, which leads to the lower bound on the 
entropy correction.  Given a state $\varphi$ for a von Neumann
algebra $\alg$, we denote by  $|\varphi^{\frac12}\rangle$ its 
canonical purification in the natural cone defined by a vacuum
state $|\Omega\rangle$ in a standard representation of $\alg$.  
We consider  a unital, completely positive (UCP) map $\alpha: 
\mathcal{B} \rightarrow 
\alg$ between von Neumann algebras, and look to construct the 
associated Petz recovery map associated with
a cyclic-separating 
state $\varphi$ on 
$\alg$.  We first define the $\varphi$-dual map $\alpha'$,
which is a UCP map
on the commutants $\alpha_\varphi':\alg'\rightarrow \mathcal{B}'$ defined
by the relation 
\beq
\left \langle(\varphi\circ \alpha)^{\frac12}
\right| \alpha_\varphi'(\msf{a}') \msf{b}
\left|(\varphi\circ\alpha)^{\frac12}\right\rangle = 
\left\langle \varphi^{\frac12}\right|\msf{a}'\alpha(\msf{b})
\left|\varphi^{\frac12}\right\rangle,
\qquad \msf{a'}\in\alg', \msf{b}\in\mathcal{B}.
\eeq
By combining $\alpha'$ with the modular conjugations $J_{\alg}$,
$J_{\mathcal{B}}$, we obtain a new UCP map $\alpha_\varphi^P:\alg\rightarrow
\mathcal{B}$
called the {\it Petz map} 
\cite{Petz1993, Accardi1982, Petz1984},
defined by
\beq
\alpha^P_\varphi(\msf{a}) = J_{\mathcal{B}}\alpha_\varphi'\big(J_{\alg}
\msf{a}J_{\alg}\big) J_{\mathcal{B}}.
\eeq

We now apply this to the crossed product algebra 
\beq \label{eqn:Ahat}
\wh{\alg} = \left\langle e^{i\hat{p}h_\Omega}\msf{a}e^{-i\hat{p}h_\Omega},
\hat{q}\right\rangle,
\eeq
where $\msf{a}\in \alg$ are operators in a QFT algebra of interest,
and $|\Omega\rangle$ is a cyclic-separating vacuum state
whose modular Hamiltonian $h_\Omega$ 
defines the crossed product.  The inclusion map
\beq\label{eqn:incA}
\inc(\msf{a}) = e^{i\hat{p} h_\Omega} \msf{a} e^{-i\hat{p} h_\Omega} 
\eeq
defines a UCP map from the QFT subalgebra $\alg$ into the crossed
product $\wh{\alg}$, and hence we would like to construct 
its Petz map.  We consider the vacuum state $|\wh\Omega\rangle
= |\Omega, f\rangle$ with wavefunction $f$ chosen to be real 
and positive.  This ensures that $|\wh\Omega\rangle$ is 
cyclic-separating for $\wh{\alg}$, and reality of $f$ implies
that $|\wh\Omega\rangle$ is in the canonical cone of the tracial 
weight $\tr = 2\pi \langle \Omega, 0_p|e^{-\hat{q}} \cdot |\Omega,0_p\rangle$,
and hence has the modular conjugation (see equation (\ref{eqn:DelhatPsi}))
\beq \label{eqn:JA}
J_{\wh{\alg}} = J_\Omega e^{-i\hat{p}h_\Omega}.
\eeq
States of this form pull back to the vacuum state $|\Omega\rangle$
under the action of the dual channel $\inc^*$, 
meaning that $(\omega_{\wh\Omega} \circ \inc)(\msf{a})
= \langle\wh\Omega|\inc(\msf{a})|\wh\Omega\rangle
= \langle\Omega|\msf{a}|\Omega\rangle$. 

We denote $\omega_{\wh\Omega}$-dual map by $\inc'_f:\wh{\alg}'
\rightarrow\alg'$, which acts as
\beq
\langle\Omega|\inc'_f(\wh{\msf{a}}')\msf{a}|\Omega\rangle
=
\langle\Omega,f|\wh{\msf{a}}'\inc(\msf{a})|\Omega,f\rangle
=
\langle \Omega,f|\wh{\msf{a}}' e^{i\hat{p}h_\Omega} \msf{a}
e^{-i\hat{p}h_\Omega}|\Omega,f\rangle
 = \langle\Omega| \langle f|e^{-i\hat{p}h_\Omega}\ho{a}'e^{i\hat{p}h_\Omega}
 |f\rangle\msf{a}|\Omega\rangle.
\eeq
This determines the expression for $\inc_f'$ to be
\beq
\inc_f'(\ho{a}') = \langle f|e^{-i\hat{p}h_\Omega} 
\ho{a}'e^{i\hat{p}h_\Omega}|f\rangle.
\eeq
Combining this with the expression (\ref{eqn:JA}) for the modular 
conjugation, we find that the Petz map $\inc_f^P:\wh{\alg}\rightarrow
\alg$ is given by 
\beq\label{eqn:incPetz}
\inc_f^P(\ho{a}) = J_\Omega\langle f|e^{-i\hat{p}h_\Omega}
J_\Omega e^{-i\hat{p} h_\Omega} \ho a e^{i\hat{p}h_\Omega}J_\Omega
e^{i\hat{p}h_\Omega}|f\rangle J_\Omega = \langle f|\ho a|f\rangle,
\eeq
where in the last equality we used that $f$ is real valued to pass 
the $J_\Omega$ past $\langle f|$ and $|f\rangle$.
Hence, we find that taking the partial expectation value 
with respect to the wavefunction $f$ defines a UCP map from $\wh{\alg}$ 
to $\alg$ that serves as a Petz map dual to the inclusion map 
(\ref{eqn:incA}).  The  quantum channel $(\inc_f^P)^*$ maps a state $\varphi$
on $\alg$ to a state $(\inc_f^P)^*\varphi = \varphi\circ \inc_f^P$ on the crossed
product $\wh\alg$. If $\varphi$ arises from the vector state $|\Phi\rangle$, it
follows  immediately 
from (\ref{eqn:incPetz}) that $|\wh\Phi\rangle = |\Phi,f\rangle$
induces the state $\varphi\circ \inc_f^P$ on $\wh\alg$.  Hence, we find that the 
classical-quantum states which were the focus of our investigations into the 
GSL have the interpretation of a Petz recovery of a state $|\Phi\rangle$ 
on the quantum field theory subalgebra $\alg$.

Because  $|\wh\Phi\rangle$ and $|\wh\Omega\rangle$ are 
images of the states $|\Phi\rangle$ and $|\Omega\rangle$ under the Petz
recovery channel $(\inc_f^P)^*$, the relative entropy computed in $\wh\alg$ 
of these states must be less than the corresponding relative entropy
computed in $\alg$, which follows from Uhlmann's general monotonicity
theorem for relative entropy under the action of a quantum channel
\cite{Uhlmann1977, Petz1993}. This monotonicity result places a lower bound
on the corrections to the semiclassical expression of the entropy
for the state $|\wh\Phi\rangle$.  To derive the bound, recall
 from equation (\ref{eqn:DelhatPsi}) that the 
density matrix for the state $|\wh\Omega\rangle$ is given by 
\beq
\rho_{\wh\Omega} = g(\hat{q})e^{\hat{q}},
\eeq
where $g(\hat{q}) = |f(\hat{q})|^2$.  We can then write the relative 
entropy in $\wh\alg$ as 
\begin{align}
S_\text{rel}^{\wh\alg}(\wh\Phi||\wh\Omega) &= \langle\wh\Phi|\log\rho_{\wh\Phi} - 
\log\rho_{\wh\Omega}|\wh\Phi\rangle
= -S(\rho_{\wh\Phi}) -\langle f|\log g(\hat{q}) + \hat{q}|f\rangle \nonumber \\
&= 
S_\text{rel}^\alg(\Phi||\Omega) - \Delta S(\rho_{\wh\Phi}),
\label{eqn:Srelexp}
\end{align}
where in the last equality we have used the semiclassical
expansion of the entropy $S(\rho_{\wh\Phi}) = -S_\text{rel}^{\alg}(\Phi||\Omega)
-\langle f|\log g(\hat{q}) + \hat{q}|f\rangle + \Delta S(\rho_{\wh{\Phi}})$
which follows from equation (\ref{eqn:SrhoSrel}),
with $\Delta S$ denoting the subleading corrections.  Monotonicity of relative entropy
under the channel $(\inc_f^P)^*$ requires that $S_\text{rel}^{\alg}(\Phi||\Omega) 
\geq S_\text{rel}^{\wh\alg}(\wh\Phi||\wh\Omega)$, which, together with
(\ref{eqn:Srelexp}), implies
\beq \label{eqn:DelSpos}
\Delta S(\rho_{\wh\Phi}) \geq 0.
\eeq
This is our first key result: the correction
to the semiclassical expression
for the crossed product entropy for the state $|\wh\Phi\rangle$ is strictly nonnegative.

We now proceed to derive an explicit expression for the entropy correction
$\Delta S(\rho_{\wh\Phi})$.  The starting point is the expression
(\ref{eqn:logrho}) for the logarithm of the density matrix $\rho_{\wh\Phi}$
for the classical-quantum state $|\wh\Phi\rangle = |\Phi,f\rangle$.  
This expression can be written
\beq
-\log \rho_{\wh\Phi} = - h -\hat{q} -\log X; \qquad X\defeq f(\hat{q})
e^{i\hat{p}h} \Delta e^{-i\hat{p}h} f^*(\hat{q}),
\eeq
where we have defined 
\beq
h = h_{\Omega|\Phi}, \quad \Delta = \Delta_\Phi,
\eeq
and now allow the wavefunction $f$ to be complex.\footnote{In fact, the phase of the 
wavefunction has no effect on the entropy,
since we can write a generic classical-quantum
state as $|\Phi,f\rangle = e^{i\phi(\hat{q})}
|\Phi,f_r\rangle$, where $f_r(q)$ is real.  
Since $e^{i\phi(\hat{q})}$ is a unitary operator
in $\wh\alg$, the states $|\Phi,f\rangle$
and $|\Phi,f_r\rangle$ have the same entropy,
since the entropy is invariant under the 
action of unitaries from the algebra.}  
The entropy is obtained by taking the expectation value of 
$-\log \rho_{\wh\Phi}$ in the state $|\Phi,f\rangle$.  The leading 
semiclassical contribution to the entropy comes from
\beq
S_0(\rho_{\wh\Phi}) = \langle \Phi,f|-h-\log \Delta - \hat{q}-\log g(\hat{q})
|\Phi,f\rangle,
\eeq
which reproduces the expression (\ref{eqn:SrhoSrel}), noting that 
$\langle \Phi |\log\Delta|\Phi\rangle = 0$.  The correction to the 
entropy is therefore given by
\beq
\Delta S(\rho_{\wh\Phi}) = \langle \Phi,f|-\log X + \log(g(\hat{q})\Delta)
|\Phi,f\rangle
\eeq

To evaluate this, we use the integral representation of the 
operator logarithm
\beq
-\log A = \int_0^\infty d\lambda \left( \frac{1}{\lambda +A} 
- \frac{1}{\lambda +1}\right)
\eeq
to write
\begin{align}
-\log X + \log(g(\hat{q})\Delta)
&=
\int_0^\infty d\lambda\left(\frac{1}{\lambda +X} 
- \frac{1}{\lambda + g(\hat{q}) \Delta}\right)
\nonumber \\
&=
\int_0^\infty d\lambda \frac{1}{f^*(\hat{q})}\left(
e^{i\hat{p}h}\frac{1}{\frac{\lambda}{g(\hat{q}-h)} + \Delta} e^{-i\hat{p}h}
-\frac{1}{\frac{\lambda}{g(\hat{q})} + \Delta}\right)\frac{1}{f(\hat{q})}
\end{align}
Since we will eventually take the expectation value of this 
expression in the state $|\Phi,f\rangle$, it is convenient to pass 
this operator through the UCP map $\langle f|\cdot|f\rangle$, which 
results in an operator acting on the QFT Hilbert space whose 
expectation value in the state $|\Phi\rangle$ yields the entropy
correction.  This simplifies the expression due to the identity
\beq
\frac{1}{f(\hat{q})}|f\rangle = \int dz |z\rangle 
= \int ds dz \frac{e^{-isz}}{\sqrt{2\pi}}|s\rangle = \sqrt{2\pi}|0_p\rangle,
\eeq
using the normalization of position and momentum eigenstates
$\langle s|z\rangle = \frac{e^{-isz}}{\sqrt{2\pi}}$.  This then
gives
\begin{align}
\langle f|-\log X + \log(g(\hat{q})\Delta)|f\rangle
&=
2\pi \langle 0_p|\int_0^\infty d\lambda 
\left(\frac{1}{\frac{\lambda}{g(\hat{q}-h)}+\Delta} - 
\frac{1}{\frac{\lambda}{g(\hat{q})} +\Delta} \right)|0_p\rangle
\nonumber \\
&=
\int_0^\infty d\lambda \int dz
\left(\frac{1}{\frac{\lambda}{g(z-h)}+\Delta} - 
\frac{1}{\frac{\lambda}{g(z)} +\Delta} \right).
\label{eqn:logXlogg}
\end{align}

In order to simplify this further, we would like to switch the order 
of integration and change variables to $\lambda = g(z)\mu$.  This 
can only be done if the $\lambda$ integral converges at fixed $z$, 
which does not hold in the present form.  We therefore manipulate
it into a form that does converge at fixed $z$, making 
use of the ability to add terms whose $z$-integral is zero.  In particular,
we define $\delta(z,h)\defeq g(z-h)-g(z)$, which satisfies $\int dz \delta(z,h)
= 0$, and then express (\ref{eqn:logXlogg}) as 
\begin{align}
\langle f|\cdots|f\rangle &=
\int_0^\infty d\lambda \int dz\frac{1}{\lambda +\Delta g(z)}
\delta(z,h)\frac{\lambda}{\lambda + \Delta g(z-h)} 
\nonumber \\
&=
\int_0^\infty d\lambda \int dz\left[
-\frac{1}{\lambda + \Delta g(z)} \delta(z,h) \Delta g(z-h)
\frac{1}{\lambda + \Delta g(z-h)}
\right.
\nonumber \\
& \hphantom{=\int_0^\infty d\lambda \int dz \;}
\left.
+\left(
\frac{1}{\lambda +g(z)\Delta} - \frac{1}{\lambda+1}
\right)\delta(z,h)
\right].
\label{eqn:fcdotsf}
\end{align}
Here we have made use of the operator identity $A^{-1}+ B^{-1}
= A^{-1}(A+B)B^{-1}$, and have added the term $-\frac{1}{\lambda+1}\delta(z,h)$
in the last line which $z$-integrates to zero.  It is now clear 
that the $\lambda$ integral of each of the two terms in 
the final expression  in (\ref{eqn:fcdotsf}) 
converges at fixed $z$, allowing the 
order of integration to be reversed.  

The integral in the second line can be performed explicitly,
resulting in
\begin{align}
\int dz \int_0^\infty d\lambda\left(\frac{1}{\lambda + g(z)\Delta}
-\frac{1}{\lambda+1}\right) \delta(z,h) 
&= 
-\int dz \log (g(z)\Delta) \delta(z,h)
\nonumber \\
&=
-\int dz \log g(z) \big[g(z-h)-g(z)\big]
\nonumber \\
&=
-\int dz g(z) \log \frac{g(z+h)}{g(z)}
\eqqcolon G(h),
\label{eqn:Ghdef}
\end{align}
where in the first line we have used that $\log (\Delta )\delta(z,h)$
integrates to zero.  

We define the first line of (\ref{eqn:fcdotsf}) to be $-R(\Delta,h)$, 
and simplify its expression by switching the order of integration
and substituting $\lambda = g(z) \mu$.  
After subtracting from the integrand 
a term $\frac{1}{\mu+\Delta} 
\delta(z,h)\frac{\Delta}{\mu+\Delta}$ which integrates to zero, 
this leads to
\begin{align}
R(\Delta,h) 
&=
\int dz \int_0^\infty d\mu \frac{1}{\mu + \Delta} \delta(z,h)\Delta
\left(\frac{1}{\frac{\mu g(z)}{g(z-h)} + \Delta} - \frac{1}{\mu +\Delta}
\right)
\nonumber \\
&=
\int dz g(z) \int_0^\infty d\mu
\frac{1}{\mu+\Delta}\frac{\delta(z,h)}{g(z)} 
\frac{\mu}{\frac{\mu}{\Delta} + \frac{g(z-h)}{g(z)}}
\frac{\delta(z,h)}{g(z)} \frac{1}{\mu+\Delta}.
\label{eqn:RDelh}
\end{align}
Hence, the final expression for the entropy correction
is given by
\beq \label{eqn:DelSexact}
\Delta S(\rho_{\wh\Phi}) = \langle\Phi|-R(\Delta,h) + G(h)|\Phi\rangle,
\eeq
with $G(h)$ defined by (\ref{eqn:Ghdef}) and $R(\Delta,h)$ by 
(\ref{eqn:RDelh}).  

The first observation to make about this expression is that 
 $R(\Delta,h)$ is a positive operator.  This follows from the fact that 
the term in the middle of the integrand $\frac{1}{\frac{\mu}{\Delta} + 
\frac{g(z-h)}{g(z)}}$ takes the form of a parallel sum $\left( 
\frac{\Delta}{\mu}:\frac{g(z)}{g(z-h)}\right)$.  The parallel 
sum of two positive, invertable operators is defined by 
$(A:B) = \frac{1}{A^{-1}+B^{-1}}$, and the resulting operator 
is also positive \cite[Chapter 36]{simon2019loewner}.  
Since both $\frac{\Delta}{\mu}$ and 
$\frac{g(z)}{g(z-h)}$ are positive for all values of $\mu$ and $z$ 
in the range of integration, the middle operator in (\ref{eqn:RDelh})
is manifestly positive.  The full integrand then takes the form
$\mu g(z) C^\dagger\left(\frac{\Delta}{\mu}:\frac{g(z)}{g(z-h)}\right)
C$, where the operator $C$ depends on $\mu$, $z$, $\Delta$, and $h$,
and hence $R(\Delta,h)$ is the integral of a collection of 
positive operators with respect to 
a positive measure.  This demonstrates that $R(\Delta,h)$ is positive.  

Positivity of $R(\Delta,h)$ then leads to an upper bound on $\Delta S
(\rho_{\wh\Phi})$ in terms of the expectation value of $G(h)$,
as is clear from (\ref{eqn:DelSexact}).  Combining with the bound
(\ref{eqn:DelSpos}), this leads to 
\beq \label{eqn:DelSbd}
0 \leq \Delta S(\rho_{\wh\Phi}) \leq \langle \Phi |G(h) |\Phi\rangle.
\eeq
Hence, the entropy variation is upper bounded by a function of the 
relative modular operator $h=h_{\Omega|\Phi}$, whose precise form
depends on the wavefunction $f(\hat{q})$.  These bounds were recently
employed in a nongravitational setting in 
\cite{Kudler-Flam:2023hkl} to obtain rigorous statements about quantum
field theory entropies regulated by the crossed product.  

The next point about the entropy correction is that it has no first
order contribution in the semiclassical expansion, so the correction
is generically $\op(\vep^2)$ with $\vep\sim \frac{g'}{g}$.  
The expression (\ref{eqn:RDelh}) for $R(\Delta,h)$ is manifestly
$\op(\vep^2)$ since $\delta(z,h) = -g'(z) h + \op(\vep^2)$, and 
two factors of $\delta(z,h)$ appear in $R(\Delta,h)$.  The contribution 
from $G(h)$ is also $\op(\vep^2)$, since the first order contribution
integrates to zero,
\begin{align}
G(h) &= 
-\int dz g(z)\left(h \frac{g'(z)}{g(z)} +\frac12 h^2 \frac{g''(z)}{g(z)}
-\frac12 h^2\left(\frac{g'(z)}{g(z)}\right)^2\right)+\op(\vep^3)
\nonumber \\
&=\frac12\int dz \frac{g'(z)^2}{g(z)} h^2 + \op(\vep^3).
\label{eqn:Gep2}
\end{align}
This demonstrates that the semiclassical expression (\ref{eqn:SrhoSrel}) 
for the crossed product
entropy is valid up to $\vep^2$ corrections.  In particular, this implies that 
it is difficult to violate the second law in perturbation theory, since generically the 
decrease in relative entropy will more than compensate the change in 
the $\op(\vep^2)$ terms in the entropy.  

We can simplify the contribution of $R(\Delta,h)$ to the entropy
correction by evaluating this operator in the state $|\Phi\rangle$.  
Because $\Delta|\Phi\rangle = |\Phi\rangle$, we find that
\begin{align}
\langle\Phi|R(\Delta,h)|\Phi\rangle
&= 
\int dz g(z) \Big\langle \Phi  \Big| \frac{\delta(z,h)}{g(z)}
\int_0^\infty d\mu \frac{\mu}{(\mu+1)^2}
\frac{1}{\frac{\mu}{\Delta}+\frac{g(z-h)}{g(z)}}
\frac{\delta(z,h)}{g(z)}
\Big|\Phi\Big\rangle
\nonumber \\
&=
\int dz g(z)
\Big\langle \Phi \Big| \frac{\delta(z,h)}{g(z)}
\Delta^{\frac12}
\int_0^\infty d\mu \frac{\mu}{(\mu+1)^2}
\frac{1}{\mu+\Delta^{\frac12}\frac{g(z-h)}{g(z)} \Delta^{\frac12}}
\Delta^{\frac12}
\frac{\delta(z,h)}{g(z)}
\Big|\Phi \Big\rangle
\nonumber \\
&=
\int dz g(z) \Big\langle \Phi \Big|\frac{\delta(z,h)}{g(z)} \Delta^{\frac12} 
 \left[
\RT\left(\Delta^{\frac12} \frac{g(z-h)}{g(z)}\Delta^{\frac12}\right)
\right]
 \Delta^{\frac12} \frac{\delta(z,h)}{g(z)}
\Big|\Phi\Big\rangle,
\label{eqn:RDelhsimp}
\end{align}
where we have defined the function
\beq \label{eqn:RTfxn}
\RT(x)\defeq \int_0^\infty d\mu \frac{\mu}{(\mu+1)^2(\mu+x)}
=
\frac{x\log x +1-x}{(1-x)^2}.
\eeq
The $\op(\vep^2)$ contribution to this expression can be obtained 
directly by replacing $\delta(z,h)$ with $-g'(z) h$ and setting
$\frac{g(z-h)}{g(z)}$ to $1$ inside the argument of $\RT$.  Combining
the resulting expression with (\ref{eqn:Gep2}) according to (\ref{eqn:DelSexact}),
we arrive at the expression for the $\op(\vep^2)$ contribution to the entropy
\beq
S^{(2)}(\rho_{\wh\Phi})
=
\int dz \frac{g'(z)^2}{g(z)}
\langle \Phi| h\left(\frac12 - \Delta \RT(\Delta)\right)h|\Phi\rangle.
\eeq
The final expression simplifies such that the only dependence
on the wavefunction $g(z)$ is through the prefactor 
\beq
k_g \defeq \int dz \frac{g'(z)^2}{g(z)},
\eeq
and the rest of the expression involves an operator constructed 
purely from the quantum field theory state, involving the 
relative modular Hamiltonian $h_{\Omega|\Phi}$ and the modular operator
$\Delta_\Phi$.  

This expression allows us to verify the bounds (\ref{eqn:DelSbd}) at 
$\op(\vep^2)$.  
First, because $\Delta \RT(\Delta)$ is a positive operator,
we immediately find that $S^{(2)}(\rho_{\wh\Phi}) 
\leq \frac12 k_g \langle\Phi| h^2|\Phi\rangle$.  To verify the 
lower bound, we first note that the operator $(1-\Delta)$ has 
vanishing expectation value in the state $h|\Phi\rangle$.  This follows
from the fact that because $h_\Phi|\Phi\rangle = 0$, we can instead
write this state as $(h_{\Omega|\Phi}-h_\Phi)|\Phi\rangle$.  The operator
$h_{\Omega|\Phi}-h_\Phi$ is affiliated with $\alg$ since it is the derivative 
of the Connes cocycle $\Delta_{\Omega|\Phi}^{is}\Delta_\Phi^{-is}$ at $s=0$.  
Because of this, $\langle\Phi|(h_{\Omega|\Phi} - h_\Phi) \Delta_\Phi(h_{\Omega|\Phi}-h_\Phi)
|\Phi\rangle$ = $\langle\Phi |(h_{\Omega|\Phi}-h_\Phi)^2|\Phi\rangle$,
which then implies $\langle\Phi|h(1-\Delta)h|\Phi\rangle = 0$.
We can therefore add any multiple of $(1-\Delta)$ to $\frac12-
\Delta \RT(\Delta)$ without changing its expectation
value in the state $h|\Phi\rangle$.  It is straightforward
to verify from (\ref{eqn:RTfxn})
that $\frac12 - \Delta\RT(\Delta) -\frac13(1-\Delta) \geq 0$,
and so
\beq
\langle \Phi|h\left(\frac12-\Delta\RT(\Delta)\right)h|\Phi\rangle
=
\langle\Phi|h\left(\frac12-\Delta\RT(\Delta) - \frac13(1-\Delta)\right)h|\Phi\rangle
\geq 0.
\eeq
This coincides with the lower bound in (\ref{eqn:DelSbd}).  Thus, the second 
order contribution to the entropy satisfies the bounds
\beq
0\leq S^{(2)}(\rho_{\wh\Phi}) \leq \frac12 k_g\langle\Phi|h^2|\Phi\rangle.
\eeq

The motivation for computing the subleading corrections to the 
crossed-product entropy was to try to verify a GSL beyond the leading
order in the semiclassical expansion.  The correction terms simplify
into a fairly compact form, and the contributions from $R(\Delta,h)$ 
can be expressed in terms of a Kubo-Ando operator mean of $\Delta_\Phi$
and $\frac{g(z)}{g(z-h)}$ 
\cite{Kubo-Ando, simon2019loewner, Furuya:2021kqx},
defined by the function $\RT$ appearing in (\ref{eqn:RDelhsimp}).
Although such operator means are useful in constructing 
functionals that satisfy data-processing inequalities, in the present
context we were unable to prove monotonicity.  The issue is that the 
operator mean appears sandwiched between the factors of $\delta(z,h)$, 
so that it is unclear whether the full entropy correction $\Delta S(\rho_{\wh\Phi})$ 
satisfies a monotonicity property
when evaluated for a crossed product associated with 
a later cut of the horizon.  It is not even clear if we should expect
$\Delta S(\rho_{\wh\Phi})$ to be separately monotonic; the GSL, if it holds 
for crossed product algebras, would allow $\Delta S(\rho_{\wh\Phi})$ to decrease
as long as it is compensated by a larger increase in leading order entropy,
determined by the decrease in relative entropy $S_\text{rel}(\Phi||\Omega)$.  
If such a monotonicity result held, it would lead to a novel data-processing
inequality for the relative entropy
\beq\label{eqn:newdpi}
S_\text{rel}^{\acut_{\tilde\lambda}}(\Phi||\Omega) - S_\text{rel}^{\acut_\lambda}(\Phi||\Omega)
\leq
\Delta S(\rho_{\wh\Phi}^{\tilde\lambda}) - \Delta S(\rho_{\wh{\Phi}}^\lambda)
\eeq
with $\Delta S(\rho_{\wh\Phi})$ determined by (\ref{eqn:DelSexact}) and an
arbitrary positive probability distribution $g(z)$.
In any situation where $\Delta S(\rho_{\wh\Phi})$ decreases,
so that the right side of (\ref{eqn:newdpi}) is negative, this 
inequality would lead to an improvement of the 
standard data-processing inequality that only requires that the relative 
entropy decreases.  

At this point, the only definitive statement we 
have is that it is difficult to violate monotonicity 
of the crossed product entropy perturbatively in the semiclassical
expansion, because the corrections are $\op(\vep^2)$.  Thus in order to violate
the second law, we would need the relative entropy for the quantum fields to 
be close to constant when moving between two cuts, so that its change is able to 
compete with the $\op(\vep^2)$ terms.  We also know that because the entropy
satisfies a global second law, as discussed in section 
\ref{sec:flatgravalg}  and in reference \cite{Chandrasekaran2022b},
any violation of the local second law for crossed product entropies must necessarily
be temporary.  We leave the question of whether the second law can be violated 
at short times scales or whether it holds at all times 
to future work.

\section{Monotonicity from crossed product subalgebras}
\label{sec:ovw}

The perturbative expressions for the crossed product entropies 
obtained in section \ref{sec:pertent} obeyed a number 
of interesting relations, but it was  ultimately inconclusive
whether these entropies satisfy a second law.  
Part of the issue is that the crossed product
obscures the subalgebra structure of 
the underlying quantum field theory algebras, as is evidenced
by the factors $e^{i\hat{p}h_\Omega}$, $e^{-i\hat{p}h_\Omega}$ 
that dress the quantum field operators appearing in the 
crossed product (\ref{eqn:Ahat}).  Since the modular 
Hamiltonians $h_\Omega^\lambda$ for subalgebras $\acut_\lambda$ 
associated with different horizon cuts are different, generically
these dressing factors will mix a subalgebra $\acut_{\tilde{\lambda}}\subset
\acut_\lambda$ throughout the entire algebra.  
Once we include the asymptotic charge $\hat{q}$ in the algebra,
this mixing will prevent the crossed product of the subalgebra,
$\wh{\acut}_{\tilde{\lambda}} = \left\langle e^{i\hat{p}h_\Omega^{\tilde{\lambda}}}
\msf{c} e^{-i\hat{p}h_\Omega^{\tilde{\lambda}}}, \hat{q}\right\rangle$ with
$\msf{c}\in\acut_{\tilde{\lambda}}$, from obviously forming
a subalgebra of the crossed product of the larger
algebra $\wh{\acut}_\lambda = 
\left\langle e^{i\hat{p}h_\Omega^\lambda}
\msf{b} e^{-i\hat{p}h_\Omega^{\lambda}}, \hat{q}\right\rangle$
with $\msf{b}\in \acut_\lambda$.  Since a second law should 
have an underlying quantum information theoretic justification
such as the existence of a quantum channel (which naturally
arises when working with subalgebras), the failure of the 
crossed product algebras to form subalgebras poses 
a possible obstruction to the existence of a generalized
second law beyond the leading semiclassical approximation.  

This suggests that one possibility for arriving at a second
law is to look for a subalgebra of $\wh{\acut}_\lambda$ 
that is naturally isomorphic to the crossed product of 
the quantum field subalgebra $\wh{\acut}_{\tilde{\lambda}}$.  
The condition for such subalgebras to occur is precisely
the existence of an {\it operator-valued weight} from 
$\acut_\lambda$ to $\acut_{\tilde{\lambda}}$ 
\cite{Haagerup1979II}.  An operator-valued
weight can be viewed as a sort of unnormalized 
conditional expectation  between von Neumann algebras
that may be finite on only a subset of operators; therefore,
it generalizes the notion of a conditional expectation 
in the same way that weights generalize 
the notion of states on the von Neumann algebra.  
Formally, an operator-valued weight for an algebra 
inclusion $\mathcal{B}\subset \alg$ is a linear map $T$
from 
$\alg_+$, the positive operators in $\alg$, into
the extended positive cone $\widetilde{\mathcal{B}}_+$ of 
$\mathcal{B}$, which roughly consists of positive 
operators in $\mathcal{B}$ along with the addition
of certain unbounded positive operators affiliated 
with $\mathcal{B}$ \cite[Section IX.4]{TakesakiII}\cite{Haagerup1979I, 
Haagerup1979II}.  Like conditional expectations,
operator-valued weights satisfy the bimodule 
property $T(\msf{b}^\dagger \msf{a} \msf{b})
= \msf{b}^\dagger T(\msf{a}) \msf{b}$ for all $\msf{b}\in \mathcal{B}$,
$\msf{a}\in \alg_+$.    

If $T$ is an operator-valued weight
satisfying the additional technical requirements of faithfulness,
normality, and semifiniteness, then we can take a weight 
$\varphi$ on $\mathcal{B}$ and compose it with $T$ to obtain
a faithful, semifinite, normal weight $\omega = \varphi\circ T$ 
on the larger algebra $\alg$.  This weight has the property
that the modular flow with respect to $\omega$ preserves the 
subalgebra $\mathcal{B}$, and agrees with the modular
flow of the weight $\varphi$ on $\mathcal{B}$.  This result
is a direct generalization of Takesaki's theorem for 
conditional expectations to the case of operator-valued weights,
and in fact the existence of the pair of weights $\omega, \varphi$
on $\alg, \mathcal{B}$ with this property is equivalent to the 
existence of the operator-valued weight $T$
\cite{Haagerup1979II}.  The main difference
from a conditional expectation is that $T$ may assign infinite
values to the entire subalgebra $\mathcal{B}$ while still
being finite on a dense subalgebra of $\alg$, whereas a 
conditional expectation assigns a finite value to 
all operators and is normalized to map the identity $\mathbbm{1}$
to itself.  

Before discussing the implications of the operator-valued 
weight on the structure of crossed products, we first consider
whether one should expect an 
operator-valued weight to exist between two horizon-cut 
algebras $\acut_{\tilde{\lambda}}\subset \acut_\lambda$.  One reason
to expect that the operator-valued weight exists
is due to the analogy with split inclusions.  
A split inclusion $\mathcal{B}\subset \alg$ is one for
which there exists a type $\tone$ factor $\mathcal{N}$
that interpolates $\mathcal{B}$ and $\alg$, 
in the sense that $\mathcal{B}\subset \mathcal{N} \subset
\alg$ \cite{Doplicher1984}.  Such inclusions often occur 
for quantum field theory algebras when $\mathcal{B}$ and 
$\alg$ are algebras associated with  causally complete nested subregions 
$R_\alg$ and $R_{\mathcal{B}}$ 
such that there is a finite corridor between
$R_\mathcal{B}$ and $R_{\alg}$; the technical requirement
on the  regions is that 
the closure of $R_{\mathcal{B}}$ is contained in the interior of 
$R_{\mathcal{A}}$, and the field
theory is said to satisfy the {\it split property} if in such 
a situation $\mathcal{B}\subset\alg$ is always a split inclusion
\cite[Section V.5]{Haag1992}.  

When both $\alg$ and $\mathcal{B}$
are type $\tthr_1$ algebras associated with local subregions 
in quantum field theory, the existence of the type $\tone$ 
interpolating algebra $\mathcal{N}$ implies that there cannot
be a (normal) conditional expectation $E$ 
from $\mathcal{A}$ to $\mathcal{B}$.  The reason is that such 
a conditional expectation would restrict to a conditional 
expectation from $\mathcal{N}$ to $\mathcal{B}$, which is not
possible since conditional expectations cannot exist
to a subalgebra with a higher type, i.e.\ there are no conditional
expectations from a type $\tone$ algebra to a type $\tthr$ subalgebra.
On the other hand, $\mathcal{N}$ at the same time guarantees the 
existence of a normal, faithful, semifinite operator-valued 
weight $T$ from $\alg$ to $\mathcal{B}$.  This is because 
there always exists such an operator-valued weight when 
either the algebra or the subalgebra is type $\tone$, so there
are operator-valued weights $T_{\mathcal{N},\alg}$ from $\alg$ to 
$\mathcal{N}$ and $T_{\mathcal{B},\mathcal{N}}$ from $\mathcal{N}$
to $\mathcal{B}$ \cite{Haagerup1979I, 
Haagerup1979II}.  Composing these results in an operator-valued 
weight $T = T_{ \mathcal{B},\mathcal{N}}\circ T_{\mathcal{N},\alg}$
from $\alg$ to $\mathcal{B}$,
which is normal, faithful, and semifinite if both 
$T_{\mathcal{N},\alg}$ and $T_{\mathcal{B}, \mathcal{N}}$ are
\cite[Proposition IX.4.16]{TakesakiII}.

Unfortunately, the existence of an operator-valued weight
in the case of a split inclusion does not give any
immediate information about the horizon-cut 
inclusion $\acut_{\tilde{\lambda}}\subset\acut_{\lambda}$,
which is instead a half-sided modular inclusion.  
Half-sided modular inclusions
are never split \cite{Wiesbrock1992}, so the above arguments do 
not apply for algebras associated with nested 
cuts of a Killing horizon.  In this case, it still 
seems unlikely that there is a conditional expectation
from $\acut_{\lambda}$ to $\acut_{\tilde{\lambda}}$; for example,
in the free field theory case where the relative commutant
$\acut_{\tilde{\lambda}}'\wedge \acut_\lambda$ forms a nontrivial
factor consisting of fields smeared between the cuts, 
the conditional expectation would imply a factorization of the algebras
across the horizon cut $\tilde{\lambda}$, which is not 
a property expected in quantum field theory.\footnote{In this
case, one can argue that the inclusion extends to a Mobius-covariant
net associated with a family of translated horizon cuts, 
which allows one to prove that no
conditional expectation from $\acut_{\lambda}$ to $\acut_{\tilde{\lambda}}$
exists
\cite{Longo:2017bae}.}  However, this leaves open 
the possibility that there exists an operator-valued weight
from $\acut_\lambda$ to $\acut_{\tilde{\lambda}}$.  
One might expect that you could argue for 
such an operator-valued weight by considering a sequence 
of algebras $\acut_{\tilde{\lambda},n}\subset
\acut_{\tilde{\lambda}} $ of regions whose entangling 
surfaces are located off of the horizon and spacelike separated
from the entangling surface of $\acut_\lambda$,
which approach $\acut_{\tilde{\lambda}}$ in the limit $n\rightarrow \infty$. 
Since each inclusion $\acut_{\tilde{\lambda},n}\subset \acut_{\lambda}$
is then split, there is a sequence of operator-valued weights $T_n$ 
to each of these subalgebras.  One might hope that 
these limit to an operator-valued weight $T$ from $\acut_\lambda$
to $\acut_{\tilde{\lambda}}$ in the limit $n\rightarrow\infty$; unfortunately,
it is difficult to make a rigorous argument to this effect.  
Nevertheless, there is no argument we know of that obviously
precludes such an operator-valued weight, and so we will 
conjecture that one indeed exists, and explore 
the implications for the crossed product algebras.  

The existence of an operator-valued weight between horizon-cut algebras
implies a subalgebra structure for the crossed product algebras.  
To simplify notation for the rest of this section, we denote the earlier 
horizon cut algebra as $\mc{B}$ and the later algebra as $\mc{C}\subset\mc{B}$,
with $T$ denoting the operator-valued weight from $\mc{B}$ to $\mc{C}$.  
To see how the crossed product $\wh{\mc{C}}$ embeds as a subalgebra of 
$\wh{\mc{B}}$, we begin by defining a weight $\chi$ 
on $\mc{B}$ by composing the vacuum state on $\mc{C}$, $\omega_{\mc{C}} = 
\langle \Omega|\cdot|\Omega\rangle$, with the operator-valued weight,
\beq
\chi = \omega_{\mc{C}}\circ T.
\eeq
We denote the unitary operator implementing modular flow for this weight on $\mc{B}$ as 
$\Delta_{\chi|\omega}^{-is} = e^{ish_{\chi|\omega}}$,
where $\omega$ refers to the vacuum weight on the commutant 
algebra $\mc{B}'$, and is needed to define how $h_{\chi|\omega}$ acts in
the global QFT Hilbert space, including its action on operators
in $\mc{B}'$.
Since $\chi$ is obtained as a pullback via  $T$, Haagerup's 
theorem for operator-valued weights \cite{Haagerup1979II}
guarantees that modular flow with respect to 
$\Delta_{\chi|\omega}^{-is}$ on the subaglebra $\mc{C}$ agrees with modular flow 
for the state $\omega_{\mc{C}}$, generated by $\Delta_{\omega_{\mc{C}}}^{-is}
= e^{ish_{\omega_\mc{C}}}$,
\beq
\Delta_{\chi|\omega}^{-is} \msf{c} \Delta_{\chi|\omega}^{is} = 
\Delta_{\omega_{\mc{C}}}^{-is}\msf{c}\Delta_{\omega_{\mc{C}}}^{is},
\quad \forall \msf{c}\in\mc{C}.
\eeq

This relation immediately demonstrates that the crossed product $\widetilde{\mc{B}}$
constructed using  modular flow for the weight $\chi$ contains the crossed product
$\wh{\mc{C}}$  as a subalgebra, since $\wh{\mc{C}} = 
\left\langle \Delta_{\omega_{\mc{C}}}^{-i\hat{p}}\msf{c} \Delta_{\omega_{\mc{C}}}^{i\hat{p}}
e^{it\hat{q}}\right\rangle
=
\left\langle \Delta_{\chi|\omega}^{-i\hat{p}}\msf{c} \Delta_{\chi|\omega}^{i\hat{p}}
e^{it\hat{q}}\right\rangle \subset 
\left\langle \Delta_{\chi|\omega}^{-i\hat{p}}\msf{b} \Delta_{\chi|\omega}^{i\hat{p}}
e^{it\hat{q}}\right\rangle = \widetilde{\mc{B}}.$
We can then apply Connes's result 
\cite{Connes1973}\cite[Section VIII.3]{TakesakiII}
that modular flows with respect to different weights
are related to each other by a unitary in the algebra,
\beq
\Delta_{\chi|\omega}^{-is} \msf{b}\Delta_{\chi|\omega}^{is} = 
u_{\chi|\omega}(-s)\Delta_\omega^{-is} \msf{b}\Delta_{\omega}^{is} u_{\chi|\omega}(-s)^\dagger,
\eeq
where the Connes cocycle $u_{\chi|\omega}(-s)$ is defined by
\beq
u_{\chi|\omega}(-s) = \Delta_{\chi|\omega}^{-is}\Delta_\omega^{is}\in\mc{B}.
\eeq
Following the arguments from \cite{Witten2021}, we can conjugate the crossed product
algebra $\widetilde{\mc{B}}$ by a unitary constructed from the cocycle
\beq \label{eqn:U}
U \equiv \msf{u}(-\hat{p}) \defeq u_{\chi|\omega}(-\hat{p}) = \Delta_{\chi|\omega}^{-i\hat{p}}
\Delta_\omega^{i\hat{p}}
\eeq
in order to map it into the 
original crossed product algebra
$\wh{\mc{B}} = \left \langle \Delta_\omega^{-i\hat{p}}
\msf{b}\Delta_{\omega}^{i\hat{p}}e^{it\hat{q}}\right\rangle$.  
To show this, we need the identity
\begin{align}
U^\dagger e^{it\hat{q}} U 
= 
\Delta_\omega^{-i\hat{p}} e^{it(\hat{q}-h_{\chi|\omega})}
\Delta_\omega^{i\hat{p}}
=
\Delta_\omega^{-i\hat{p}} e^{-ith_{\chi|\omega}}\Delta_\omega^{i\hat{p}} e^{it h_\omega} 
e^{it\hat{q}}
=
\Delta_\omega^{-i\hat{p}} u_{\chi|\omega}(t)\Delta_\omega^{i\hat{p}}e^{it\hat{q}}
\end{align}
which is an element of $\wh{\mc{B}}$ since $u_{\chi|\omega}(t)\in\mc{B}$. 
Hence, a generic element $\tilde{\msf{b}} = \Delta_{\chi|\omega}^{-i\hat{p}}
\msf{b} \Delta_{\chi|\omega}^{i\hat{p}} e^{it\hat{q}}$ in $\widetilde{\mc{B}}$ 
maps to
\begin{align}
U^\dagger \tilde{\msf{b}} U = \Delta_\omega^{-i\hat{p}}\msf{b} u_{\chi|\omega}(t)
\Delta_{\omega}^{i\hat{p}} e^{it \hat{q}} \in \wh{\mc{B}}.
\end{align}
This shows that $U^\dagger\widetilde{\mc{B}} U = \wh{\mc{B}}$.

Now because $\wh{\mc{C}} \subset \widetilde{\mc{B}}$, the conjugated algebra 
$\bar{\mc{C}} = U^\dagger \wh{\mc{C}} U$ is a subalgebra of $\wh{\mc{B}}$.  
By conjugating a generic element $\Delta_{\chi|\omega}^{-i\hat{p}} \msf{c} 
\Delta_{\chi|\omega}^{i\hat{p}} e^{it\hat{q}}$ in $\wh{\mc{C}}$ by $U^\dagger (\cdot) U$,
we find a generating set of operators for $\bar{\mc{C}}$:
\beq
\bar{\mc{C}} = \left\langle \Delta_{\omega}^{-i\hat{p}} \msf{c}\, u_{\chi|\omega}(t) 
 \Delta_{\omega}^{i\hat{p}}e^{it\hat{q}}\right\rangle
\eeq
Thus, we have constructed
a subalgebra of $\wh{\mc{B}}$ unitarily equivalent to $\wh{\mc{C}}$.
We can therefore define an inclusion map $\inc_{\wh{\mc{B}}, \wh{\mc{C}}}$ from 
$\wh{\mc{C}}$ into $\wh{\mc{B}}$ by 
\beq \label{eqn:inchBhC}
\inc_{\wh{\mc{B}}, \wh{\mc{C}}}(\ho{c}) = U^\dagger \ho c U.
\eeq
Since this is a UCP map, its dual is a quantum channel from states on $\wh{\mc{B}}$ 
to states on $\wh{\mc{C}}$.  We would therefore like to leverage this channel and 
the monotonicity of relative entropy to arrive at some version of a second law for the crossed
product algebras.  

The most straightforward way to do this is to examine the 
relative entropy of a generic state $|\wh{\Phi}\rangle$ with respect to 
a specific vacuum weight on the algebra.  In general, it is possible to define
the relative entropy $S_\text{rel}(\varphi|| \psi)$ where $\varphi$ is a state 
while $\psi$ is only a weight which may not be normalized, and could even
be unbounded in the sense that $\psi(\mathbbm{1}) = \infty$
\cite[Chapter 7]{Petz1993}.
In fact, the type $\ttwo_\infty$ entropies that we have computed 
throughout this paper are naturally interpreted as minus the 
relative entropy with respect to the tracial weight on the algebra
\cite{Petz1993, Segal1960, Longo2022}.  
For algebras possessing a trace, one can always 
find a density matrix $\rho_\psi$ affiliated with the algebra
that reproduces expectation values for the weight
via
\beq \label{eqn:rhodef}
\psi(\msf{b}) = \tr(\rho_\psi \msf{b}).
\eeq
Because the weight $\psi$ need not be normalized, this density matrix 
no longer necessarily satisfies $\tr\rho_\psi = 1$. 

For the crossed product, choosing the trace as the reference weight is problematic
because the trace on $\wh{\mc{B}}$ does not restrict to a semifinite weight 
on the subalgebra $\bar{\mc{C}}$.  Instead, there is an operator-valued weight
$\wh{T}$ from $\wh{\mc{B}}$ to $\bar{\mc{C}}$ constructed 
from $T$ 
such that $\tr_{\wh{\mc B}} = \tr_{\bar{\mc{C}}}
\circ\wh{T}$, where $\tr_{\bar{\mc C}}$ is the semifinite trace on $\bar{\mc C}$.  However,
unless $\wh{T}$ is bounded so that it can be normalized to be a conditional expectation
(which we do not expect for nested horizon cuts),
$\wh{T}$ will blow up on the subalgebra $\bar{\mc{C}}$, showing that $\tr_{\wh{\mc B}}$ 
is not semifinite on this algebra.  A more convenient reference is the dual weight $\wh\omega$
arising from the vacuum state $|\Omega\rangle$ in the crossed product
construction.  This dual weight can be formally represented as arising from
an infinite norm vector $|\wh{\Omega}\rangle = \sqrt{2\pi} |\Omega,0_p\rangle$,
where $|0_p\rangle$ is the zero momentum eigenstate used in the expression for the 
trace on the crossed product.  This 
vector also defines the dual weight for the $\wh{\mc C}$ crossed product algebra,
and furthermore  has the property that it is invariant under
the unitary $U = \msf{u}(-\hat{p})$ in equation
(\ref{eqn:U}) defining the map from $\wh{\mc C}$ to $\bar{\mc{C}}$.  
Defining the weight $\wh\omega_{\mc B} = 2\pi \langle\Omega,0_p| (\cdot)|\Omega,0_p\rangle$
and similarly for $\wh{\omega}_{\mc C}$, we find that $\wh\omega_{\mc B}$ restricts 
to a semifinite weight on $\bar{\mc C}$, and further more pulls back to the 
weight $\wh\omega_{\mc C}$ under the inclusion map (\ref{eqn:inchBhC}),
\beq (\label{eqn:omBincBC}
\wh\omega_{\mc B} \circ \inc_{\wh{\mc B}, \wh{\mc C}}) (\ho c)
= 2\pi \langle\Omega,0_p|U^\dagger \,\ho c \,U|\Omega,0_p\rangle = \wh\omega_{\mc C}(\ho c).
\eeq

We now consider another state $\wh{\vphi}$ arising 
from a vector $|\wh\Phi\rangle$.  Monotonicity of relative entropy for the 
UCP map $\inc_{\wh{\mc B}, \wh{\mc C}}$ then states that 
\beq \label{eqn:crmono}
S_\text{rel}^{\wh{\mc B}}(\wh \vphi||\wh\omega_{\mc B})
\geq S_\text{rel}^{\wh{\mc C}}(\wh\vphi\circ \inc_{\wh{\mc B},\wh{\mc C}}
|| \wh\omega_{\mc B}\circ \inc_{\wh{\mc{B}}, \wh{\mc C}})
=
S_\text{rel}^{\wh{\mc C}}(\wh\vphi\circ \inc_{\wh{\mc B},\wh{\mc C}}
|| \wh\omega_{\mc C})
\eeq
where the second equality applies (\ref{eqn:omBincBC}).  The state $\wh{\vphi}
\circ\inc_{\wh{\mc B},\wh{\mc C}}$ on $\wh{\mc C}$ is the expectation
value in the vector state $U|\wh\Phi\rangle$.  To convert this into 
a statement about entropies, we note that the density matrices for the weights
$\wh\omega_{\mc B}$ and $\wh\omega_{\mc C}$ are given in both cases by
\beq
\rho_{\wh\omega} = e^{\hat{q}},
\eeq
which follows immediately from the definition (\ref{eqn:rhodef})
of the density matrix and the expression (\ref{eqn:Tr}) for the trace on the 
two algebras.  

In terms of the density matrices,
the relative entropy is given by the usual expression
\beq
S_\text{rel}(\wh\vphi||\wh\omega) = \tr(\rho_{\wh{\vphi}} (\log\rho_{\wh\vphi}-
\log\rho_{\wh\omega}))
\eeq
and hence
\beq \label{eqn:crfreeen}
S_\text{rel}^{\wh{\mc B}}(\wh\vphi||\wh\omega_{\mc B})
= - S(\rho_{\wh\Phi}^{\wh{\mc B}}) -\langle\wh\Phi|\hat{q}|\wh\Phi\rangle,
\eeq
showing that this relative entropy can be interpreted as a free energy
with respect to the asymptotic charge $-\hat{q}$, which is given by 
either the asymptotic area 
 or the ADM mass in the crossed product 
constructions discussed in sections
\ref{sec:gslsemi} and \ref{sec:flatgravalg}
(see equations (\ref{eqn:qArea}) and (\ref{eqn:xMADM})).  
The right hand side of 
(\ref{eqn:crmono}) evaluates to
\beq
S_\text{rel}^{\wh{\mc C}}(\wh\vphi\circ\inc_{\wh{\mc B}, \wh{\mc C}}
|| \wh\omega_{\mc C}) = 
- S(\rho_{U\wh{\Phi}}^{\wh{\mc C}}) - \langle\wh\Phi| U^\dagger \hat{q} U|\wh\Phi\rangle.
\eeq
Hence, the monotonicity result (\ref{eqn:crmono}) implies the entropy
inequality
\beq \label{eqn:entineq}
S(\rho_{U\wh\Phi}^{\wh{\mc C}}) + \langle\wh\Phi| U^\dagger \hat{q} U - \hat{q}|\wh\Phi\rangle
\geq
S(\rho_{\wh\Phi}^{\wh{\mc B}}).
\eeq

This is not quite the statement of the second law we were seeking.  First, the entropy
on the left hand side is computed for a different state $U|\wh{\Phi}\rangle$ for the 
algebra $\wh{\mc C}$ instead of the original state $|\wh\Phi\rangle$.  Second, there 
is an additional term involving the difference in expectation 
value of $\hat{q}$ between the state $U|\wh\Phi\rangle$ and $|\wh\Phi\rangle$.  
As it turns out, at leading order in the semiclassical expansion,
these two differences cancel out, and the inequality (\ref{eqn:entineq}) 
reduces to the semiclassical second law discussed in sections
\ref{sec:gslsemi} and \ref{sec:flatgravalg}.  To see this, we set $|\wh\Phi\rangle = |\Phi,f\rangle$
with the wavefunction $f$ slowly varying, $\frac{f'}{f}\sim \op(\vep)$.
As discussed in \cite{Chandrasekaran2022b}, 
because the unitary $U$ is a cocycle flow
with parameter $\hat{p}$, it leaves the state $|\Phi,f\rangle$ invariant up
to corrections of order $\vep$.  This means that at this 
order, we can approximate the density matrix $\rho_{U\wh\Phi}$ by $\rho_{\wh\Phi}$,
and hence
\beq
-\log \rho_{U\wh\Phi}\approx-\log\rho_{\wh\Phi}
\approx -\hat{q} - h_{\Omega|\Phi}+ h_\Phi - \log |f(\hat{q})|^2 + \op(\vep).
\eeq

The entropy $S(\rho_{U\wh\Phi})$ is obtained as the expectation
value of $-\log\rho_{U\wh\Phi}$ in the state $U|\wh\Phi\rangle$.  
Again because $U$ leaves the state $|\wh\Phi\rangle$ nearly invariant, 
most terms in this expectation value are the same as in the state 
$|\wh\Phi\rangle$ at leading order in the semiclassical expansion.  
The exception is the operator $\hat{q}$, which shifts when conjugated by the cocycle,
\beq
U^\dagger \hat{q} U = \hat{q} + h_\omega 
- e^{i\hat{p} h_\omega} h_{\chi|\omega} e^{-i\hat{p} h_\omega}.
\eeq
However, we see that this shift is compensated by the additional terms
appearing in the inequality (\ref{eqn:entineq}):  the entropy of
$\rho_{U\wh\Phi}$ admits the approximation,
\begin{align}
S(\rho_{U\wh\Phi}^{\wh{\mc{C}}})
&\approx
\langle\wh\Phi| U^\dagger(-\hat{q}
-h_{\Omega|\Phi} + h_\Phi - \log |f(\hat{q})|^2)U|\wh\Phi\rangle
\nonumber \\
&\approx
\langle\wh\Phi| - \hat{q}- h_{\Omega|\Phi}+ h_\Phi -\log|f(\hat{q})|^2|\wh\Phi\rangle
+\langle\wh\Phi|\hat{q} - U^\dagger \hat{q} U|\wh\Phi\rangle
\nonumber \\
&\approx
S(\rho_{\wh{\Phi}}^{\wh{\mc C}}) + \langle\wh\Phi|\hat{q} - U^\dagger \hat{q} U|\wh\Phi\rangle
\end{align}
and  plugging this into the inequality (\ref{eqn:entineq})  reproduces the second law
for crossed product algebras at leading order in the semiclassical expansion,
\beq
S(\rho_{\wh\Phi}^{\wh{\mc C}}) +\op(\vep) \geq S(\rho_{\wh\Phi}^{\wh{\mc B}}).
\eeq
Hence, assuming the existence of the operator-valued weight between
horizon cuts, we are able to establish a subalgebra explanation
for the semiclassical generalized second law derived 
in sections \ref{sec:gslsemi} and \ref{sec:flatgravalg}.

Finally, it is interesting to note that the terms in (\ref{eqn:entineq})
that could spoil the proposed second law can be written as a relative entropy
difference.  Noting that $S_\text{rel}^{\wh{\mc{C}}}
(\wh\Phi||\wh{\omega}_{\mc{C}}) = -S(\rho_{\wh\Phi}^{\wh{\mc{C}}}) - 
\vev{\hat{q}}_{\wh\Phi}$, we can rewrite the inequality (\ref{eqn:entineq})
as
\beq \label{eqn:xgslsrel}
S(\rho_{\wh\Phi}^{\wh{\mc C}}) + S_\text{rel}^{\wh{\mc C}}(\wh{\Phi}||
\wh\omega_{\mc C})
- S_\text{rel}^{\wh{\mc C}}(U\wh\Phi || \wh\omega_{\mc C})
\geq
S(\rho_{\wh\Phi}^{\wh{\mc B}}).
\eeq
Hence, depending on the sign of the relative entropy difference
$S_\text{rel}^{\wh{\mc C}}(\wh{\Phi}||
\wh\omega_{\mc C})
- S_\text{rel}^{\wh{\mc C}}(U\wh\Phi || \wh\omega_{\mc C})$
between the states $|\wh\Phi\rangle$ and $U|\wh\Phi\rangle$ on $\wh{\mc C}$,
we can either guarantee the second law holds (when the difference is 
negative or zero),
or derive a necessary condition on the state $|\wh\Phi\rangle$ for a violation 
to occur (the difference is positive).  At present, there does not 
seem to be any constraint on the sign of this relative entropy difference, but 
examining it more closely may result in further insights into the 
crossed product second law.  One statement that seems to hold is that 
this difference should be $\op(\vep^2)$
in the semiclassical expansion, since the perturbative
calculations of section \ref{sec:pertent} demonstrated that the 
second law for semiclassical states holds up to linear order in $\vep$.

\section{Discussion}
\label{sec:discussion}

The overarching goal of this paper has been to explore properties 
of subsystem gravitational algebras for regions defined by arbitrary
cuts of a Killing  horizon. These regions provide nontrivial
examples where the modular Hamiltonian of a vacuum state $|\Omega\rangle$
is known and generates a geometric flow along the horizon, 
which facilitates the construction of type $\ttwo_\infty$ 
gravitationally dressed crossed-product algebras following the general
procedure of JSS \cite{Jensen2023}.  
As reviewed in section \ref{sec:hstr}, the geometric flow of this 
state follows from properties of half-sided translations and the 
ANEC, and  holds even for interacting quantum field theories.  
We found that entropies of semiclassical states in the crossed product
algebra reproduce the generalized entropy of the horizon cut, and 
for these states 
we were able to reduce the second law for crossed product algebras 
to Wall's derivation of the generalized second law \cite{Wall2011}.  
The crossed-product second law improves on Wall's result by identifying
a gravitational von Neumann algebra whose entropy is computed by 
the generalized entropy, with the further advantage that crossed-product
entropies are
manifestly UV finite, so that one does not need to worry about subtleties
involving the renormalization of the quantum field entanglement entropy.  

After discussing the case of AdS black holes, we presented 
in section \ref{sec:ainftysemi} a novel approach to treating
 asymptotically flat and de Sitter black holes 
using the concept of an algebra at infinity for the half-sided translation.  
This algebra naturally characterizes the degrees of freedom that never enter
the black hole.  Using a theorem of Borchers which guarantees the 
existence of a conditional expectation to this algebra at infinity
\cite{Borchers1997},
we discussed how to handle the generalized second law in this context,
even in the case of an interacting theory where 
one expects the algebra of fields strictly localized to the horizon
to the future of a cut
to be trivial.  We also described the construction of gravitational
algebras for these alternative asymptotics, which lead to nontrivial
matching conditions at asymptotic infinity.  

Finally, we explored the question of whether the crossed product algebras
satisfy a generalized second law beyond the leading semiclassical approximation
of the states.  We derived in section
\ref{sec:pertent} formulas for the corrections to the 
crossed-product entropy, and showed that the first correction appears 
at second order in the semiclassical expansion.  Although we obtained explicit
expressions for the entropy corrections, we were unable to determine 
whether these corrections can be used to violate the second law
at short time scales.  However, we noted that because the gravitational 
algebra 
at infinity defines a trace-preserving inclusion inside the dressed
horizon-cut algebras, the entropy at infinity must be larger than that
of the horizon cut algebra.  This gives a global second law 
which generalizes a similar result appearing in \cite{Chandrasekaran2022b}, 
and 
necessarily constrains
any possible violation of the second law between horizon cuts to 
 be temporary.  After this, we explored in
 section \ref{sec:ovw} the possibility 
that nested horizon-cut gravitational algebras could form a subalgebra
structure using the conjectured existence of an operator-valued weight
between horizon cuts.  This leads to a monotonicity constraint on entropies
between two horizon cuts, but involving different states at different 
cuts.  We showed that this monotonicity result still reduces to the 
second law at leading order in the semiclassical expansion, and left
open the question of whether it can be leveraged to produce a stronger
result.

The investigations carried out in this work have lead to a number of 
intriguing questions and directions for future work, so we conclude
by describing some of these.

\paragraph{Singular vacua as semifinite weights.}

In the discussion of rotating black holes 
in section \ref{sec:rotUnruh}, we confronted the issue
that there does not appear to be a natural vacuum state
whose modular flow is geometric on the algebra at infinity.  
The closest analog is the Frolov-Thorne vacuum, which is 
well known to become singular in the region far from the black hole
where the horizon-generating Killing vector is spacelike.  Ordinarily,
quantization about such a singular state results in a unitarily 
inequivalent representation of the quantum field algebra,
which is likely not the correct description 
of quantum fields on a smooth background geometry.  However, 
we raised the possibility that one could reinterpret this vacuum
as a nonnormalizable weight on the physical von Neumann algebra 
of fields in the black hole exterior.  Having a weight, as opposed to 
a state, whose modular flow is geometric is sufficient to arrive 
at a gravitational algebra with a trace following the JSS construction.  
Similarly, the Unruh vacuum, which provides the more accurate description 
of an astrophysical black hole formed from collapse, yields 
a unitarily inequivalent representation of the quantum field 
algebra from the Hartle-Hawking vacuum.  However, the latter 
vacuum provides the state whose modular flow is 
geometric on the horizon, and hence one would like a description
where both states can be treated in a single representation of the field 
algebra.  We suggested the possibility that when quantizing around the 
Unruh vacuum, the Hartle-Hawking vacuum could be treated as a semifinite 
weight.  

These points motivate the broader question of determining when
a singular state of the quantum field algebra can be reinterpreted
as a semifinite weight.  One important aspect of such a procedure is 
identifying the correct definition subalgebra for the weight
(see e.g. \cite[Section VII.1]{TakesakiII}).  Given
a weight $\varphi$ on a von Neumann algebra $\alg$, we can 
identify a subcone of the set of positive operators $\alg_+$
with finite weight,
\beq
\mathfrak{p}_\varphi = \{\msf{a}\in \alg_+| \varphi(\msf{a})<\infty\}.
\eeq
The linear span of $\mathfrak{p}_\varphi$ forms the {\it definition
subalgebra}
$\mathfrak{m}_\varphi \subset \alg$ of the weight $\varphi$.  In
general, $\mathfrak{m}_\varphi$ is not a $C^*$-algebra since it need
not contain uniform limits of operators, and therefore is 
not a von Neumann algebra either.  
The definition subalgebra is also not unital for nonnormalizable
weights, since if $\varphi(\mathbbm{1}) < \infty$, one could rescale 
$\varphi$ to make it a state.  
If the weight $\varphi$ is 
semifinite, the weak closure of $\mathfrak{m}_\varphi$ yields the full
von Neumann algebra $\alg$.  Because of this last fact, one can
still perform an analog of the GNS representation for this algebra
with respect to the weight $\varphi$ called the {\it semicyclic representation},
and such a representation with respect to a faithful, semifinite, normal
weight is unitarily equivalent to the GNS representation with 
respect to a faithful normal state.  This fact allows many of the results
of Tomita-Takesaki theory to apply to the case of semifinite weights.  

The re-interpretation of singular states in quantum field theory as 
semifinite weights thus requires a careful analysis of the associated
definition
subalgebra $\mathfrak{m}_\varphi$.  Since in the case of the Frolov-Thorne
vacuum, the region in the black hole exterior where the state is regular
has a causal completion consisting of the entire exterior, it is possible
this region can be used to identify an appropriate weakly dense 
definition subalgebra for the Frolov-Thorne vacuum, viewed
as a weight.  Similar comments apply to the Hartle-Hawking vacuum
for the algebra constructed from the Unruh state.  We believe that 
working out the details of this weight interpretation of singular vacua
may provide important insights into the relevance of these states and 
further inform the construction of gravitational algebras.

\paragraph{GSL violation or improved monotonicity.}

Upon computing the corrections to the crossed product
entropy for states of the form $|\Phi,f\rangle$
in section \ref{sec:pertent}, we were ultimately unable to determine
whether these corrections can be used to violate the leading 
order second law found in sections \ref{sec:gslsemi}
and \ref{sec:flatgravalg}.  The leading 
order statement follows from monotonicity of relative entropy
for the quantum field subalgebra of the crossed product,
and so if the entropy still satisfies a second law upon
including the subleading corrections, it would require
a modified bound on the amount the quantum field relative entropy decreases
between horizon cuts.  In some cases, this bound would improve the 
standard data-processing inequality that relative entropy must 
decrease between horizon cuts.  It would therefore be interesting
to relate the proposed bound to other recent works on improving 
the data-processing inequality and its connection to universal
recovery channels
\cite{Junge:2015lmb, Faulkner:2020iou, Faulkner:2020kit}.

One possible avenue at arrive at a nontrivial bound is the conjectured
subalgebra structure for crossed products discussed in section
\ref{sec:ovw}.  This structure relies on the existence of an operator-valued
weight between algebras for nested cuts of the Killing horizon, whose
existence has not been proven.  This leads to  the following purely
mathematical questions in the theory of operator algebras: given a
half-sided modular inclusion $\mathcal{N}\subset \mathcal{M}$, can
there exist an operator-valued weight $T$ from $\mathcal{M}$ to 
$\mathcal{N}$?  Does this always occur, and if one exists,
what are its properties?  In the context of half-sided modular 
inclusions arising from horizon cuts, if such an operator-valued
weight existed, it should possess a natural construction using 
tools from quantum field theory.  In the interacting case where 
we expect the relative commutant $\mathcal{N}'\wedge \mathcal{M}$ 
to be trivial, the operator-valued weight would have to be unique,
since if another operator-valued weight existed, the cocycle between
the two weights would define a nontrivial element of the relative commutant.  
Even in the free case where the relative commutant is nontrivial, 
there should be a preferred operator-valued weight that 
restricts to the horizon vacuum state on the relative commutant.  
Hence, a tangible goal for the future is to explicitly
construct the operator-valued weight, or prove that one 
does not exist.  

Assuming the existence of such an operator-valued weight, we found a version
of the second law for crossed products, equation (\ref{eqn:entineq}), 
that does not rely on any semiclassical approximation.  
Unfortunately, we do not at present have a simple interpretation of 
the derived inequality, since it involves the entropy of a different 
state at a later cut of the horizon, as 
well as the change in the expectation value of the asymptotic 
charge between the two states.  It would be interesting to 
better understand this inequality, since at the very least it constrains
any violation of the crossed product second law to be bounded 
by a relative entropy difference of two states, as evidenced 
by equation (\ref{eqn:xgslsrel}).

\paragraph{Connections to dynamical horizon entropy.}

One output of the construction of gravitational algebras 
for horizon cuts is that it yields an algebra whose entropy 
directly coincides with the generalized entropy of the horizon
cut.  This supports the idea that in the limit of weak gravitational
coupling, this entropy can be interpreted as an entanglement entropy,
subject to the caveat that entropies in type $\ttwo$ algebras 
are most naturally thought of as vacuum-subtracted entropies
\cite{Witten2021, Longo2022}.  Furthermore,
it reinforces the interpretation of the horizon area
as capturing the classical component of the black hole entropy.  
Recently, an alternative proposal 
for the entropy of dynamical black holes has appeared 
that modifies this classical component by 
a term involving the perturbative expansion of the horizon
\cite{Hollands:2024vbe, Visser:2024pwz, Rignon-Bret:2023fjq}.  
This raises the question of whether this entropy has a quantum
generalization, and whether one can account for it in the framework
of semiclassical gravitational algebras.  

We note that one aspect of the construction of the gravitational
algebras in the present work is the use of the event horizon boundary
condition in section \ref{sec:geometric}
in order to interpret the asymptotic gravitational 
charge as the late time area of the horizon.  
Interestingly, the new dynamical entropy proposal
can be interpreted as the area of an apparent horizon inside the black
hole, 
as opposed to the area of the event horizon  
\cite{Hollands:2024vbe, Visser:2024pwz}.  This 
is strongly reminiscent of the Engelhardt-Wall proposal for the 
coarse-grained outer entropy in the context of AdS/CFT
\cite{Engelhardt:2017aux, Engelhardt:2018kcs}.  
In our setup, we could arrive at a null surface that 
intersects the apparent horizon by relaxing the event horizon
boundary condition that the asymptotic expansion $\Theta$ goes to zero. 
This also suggests that we should include
an asymptotic  gravitational charge associated with this freedom to 
choose this boundary condition, since the late time expansion naturally
arises as an asymptotic charge associated to the null translation
along the horizon.  The asymptotic past and future expansions
are related to the average null energy along the horizon via a gravitational
constraint (\ref{eqn:intRay}), 
so it raises the possibility that one should include this asymptotic
charge and gravitational dressing 
for this constraint as well.  It may 
be that including this translational gravitational charge would 
lead to a natural connection to the new dynamical entropy, and 
may also help in a more complete formulation of the generalized
second law, without the possibility of temporary entropy decrease
beyond the leading semiclassical approximation.

\paragraph{Applications to quantum focusing.}

A natural generalization of the present work is to investigate
the crossed product construction for more general
subregions.  Ultimately, we would like to be able to formulate the 
quantum focusing conjecture (QFC) \cite{Bousso:2015mna}
as a statement about crossed product
entropies for subregions bounded by a causal horizon.  One benefit 
of such a description is that it would handle the problem
of UV divergences in the entanglement entropy, which are one 
of the main technical obstructions to obtaining rigorous 
statements about quantum focusing.  On the other hand, there is a 
question of what more we could learn from a crossed-product 
formulation of the QFC.  The gravitational algebras considered here
appear in the $G_N\rightarrow 0$ limit of quantum gravity,
and in this limit the quantum focusing conjecture in most cases
reduces to the classical focusing theorem of general relativity.  
In order to arrive at a nontrivial quantum result, one needs to 
consider background geometries that classically saturate
the focusing inequality, so that perturbative corrections to the 
geometry can compete with contributions coming from the change in
entanglement entropy as one evolves along a light sheet.  
In this limit, quantum focusing generally reduces to the 
quantum null energy condition (QNEC), which is a statement
purely about quantum field theory in a fixed background geometry,
and there now exist proofs of the QNEC independent of any
assumption about quantum focusing \cite{Bousso:2015wca,
Balakrishnan:2017bjg, Ceyhan:2018zfg}.  
However, we found even when considering the generalized 
second law using crossed products, proving that the entropy
increases beyond the leading semiclassical approximation
is nontrivial, and could result in improved data-processing inequalities
in quantum field theory.  Therefore, it would be interesting to determine
if a similar improvement could arise from the crossed product formulation
of the QFC.  

One of the main advantages of working with a Killing horizon
in the present work was the ability to leverage the ANEC to prove 
the existence of a state with a geometric modular flow along the 
horizon.  This suggests that a strategy for finding the necessary states
conjectured in \cite{Jensen2023}
to carry out the crossed product for generic subregions is to look
for related energy conditions adapted to the subregion under consideration.  
One could look for positive operators involving a smeared
stress tensor over the subregion horizon to try to establish 
the existence of a half-sided translation, which then would determine
a vacuum 
state and modular operator whose flow is boost-like near the subregion
entangling surface.  It is possible that the novel energy conditions 
recently investigated in \cite{Freivogel:2018gxj, Fliss:2021gdz,
Fliss:2021phs} could be helpful
in establishing these positivity results.  This may provide 
a crucial clue for how to rigorously establish that local gravitational
algebras can be realized as a crossed product, and ultimately help
in formulating the QFC in this language.

\subsection*{Acknowledgments}
We thank Daine Danielson, Ben Freivogel, Stefan Hollands, 
Ted Jacobson, Nima Lashkari, Roberto Longo, Geoff Penington,
Pranav Pulakkat, 
Gautam Satishchandran, Jon Sorce, Erik Verlinde, Bob Wald, Aron Wall, and
Victor Zhang for 
helpful discussions.
 A.J.S. thanks the Isaac Newton Institute for Mathematical Sciences for support and hospitality during the programme
“Black holes: bridges between number theory and holographic quantum information” where work on this paper
was undertaken. 
This work was supported by EPSRC grant EP/R014604/1.
This research is supported in part by 
the Air Force Office of Scientific Research under award number FA9550-19-1-036 and by the DOE award number DE-SC0015655. 

\appendix

\section{Tomita operators for crossed product algebra} \label{app:modth}

The Tomita operators considered in section
\ref{sec:crprod} for states of the form 
$|\wh{\Phi}\rangle = |\Phi,f\rangle$ were
crucial for obtaining density matrices and computing the perturbative 
expansion of entropies for the crossed product algebra in 
section \ref{sec:pertent}.  Here, we derive
the formula for these Tomita operators, and additionally obtain
expressions for the Tomita operators for a more general class of states
obtained from arbitrary superpositions of the $|\Phi,f\rangle$ states.  
This  simplifies the derivation of the density matrices for these 
states from \cite{Jensen2023}, and additionally 
yields new results in the explicit expressions for the modular conjugations 
$J_{\wh\Phi}$.
This appendix assumes familiarity with Tomita-Takesaki
theory; see \cite{Witten:2018zxz}, 
\cite[Appendix C]{Jensen2023} for an 
introduction.

The crossed product of a QFT algebra $\alg$ by the modular automorphism
group associated with the state $|\Omega\rangle \in \hqft$ 
is generated by the operators
\beq
\wh{\alg} = \left\langle 
e^{i\hat{p} h} \msf{a}e^{-i\hat{p}h}, \hat{q}\right\rangle
\eeq
acting on the Hilbert space $\wh{\hs} = \hqft\otimes L^2(\mathbb{R})$,
where $h = h_\Omega$ is the modular Hamiltonian 
for the state $|\Omega\rangle$. For a given vector 
state $|\wh{\Phi}\rangle \in \wh{\hs}$
for the crossed product algebra, the Tomita operator
is defined to act as 
\beq\label{eqn:tomitadefn}
S_{\wh{\Phi}} \ho a|\wh\Phi\rangle 
= \ho a^\dagger |\wh\Phi\rangle.
\eeq
Hence, we can solve for $S_{\wh{\Phi}}$ 
by determining the action of a linearly spanning 
set of operators $\ho a\in\wh{\alg}$ along with their 
conjugates $\ho a^\dagger$ 
on the state $|\wh{\Phi}\rangle$. Considering the 
additive basis of operators
\beq
\ho a = e^{i\hat{p} h} \msf a e^{-i\hat{p} h} e^{iu\hat q},
\eeq
we can compute their action on the classical-quantum states
$|\wh{\Phi}\rangle = |\Phi,f\rangle$ which were used 
in the investigations of the GSL in this paper.  

Before doing this, it is helpful to note how the QFT Tomita 
operators act in the larger Hilbert space $\hqft\otimes 
L^2(\mathbb{R})$.  This is somewhat subtle because the Tomita
operators are antilinear, so there is a choice to make
when defining its action on the larger 
Hilbert space $\wh{\hs}$.  We will define any QFT Tomita
operator $S$ to act trivially on the position basis $|y\rangle$
of $L^2(\mathbb{R})$, i.e.
\beq
S|\Phi,y\rangle = |S\Phi,y\rangle
\eeq
This implies that $S$ commutes with $\hat{q}$:
\begin{align}
S\hat{q}|\Phi,y\rangle 
= S y|\Phi,y\rangle
=y S|\Phi,y\rangle
=y|S\Phi,y\rangle
=\hat{q}|S\Phi,y\rangle
= \hat{q} S|\Phi,y\rangle.
\end{align}
On the other hand, $S$ must anticommute with $\hat{p}$:
\begin{align}
S\hat{p}|\Phi,y\rangle 
&= S\int dy' \langle y'|\hat{p}|y\rangle|\Phi,y'\rangle
=\int dy'\langle y|\hat{p}|y'\rangle S|\Phi,y'\rangle
=-\int dy' \langle y'|\hat{p} |y\rangle |S\Phi,y'\rangle
\nonumber \\
&=
-\hat{p}|S\Phi,y\rangle = -\hat{p} S|\Phi,y\rangle
\end{align}
where we have used that $\langle y|\hat{p}|y'\rangle
= -\langle y'|\hat{p}|y\rangle$.  We can remember these 
commutation relations by noting that $\hat{q}$ acts as 
multiplication by a real number in the $|y\rangle$ basis,
while $\hat{p}$ is a derivative times an imaginary number,
$\hat{p} = -i\frac{d}{dq}$.  These relations imply that $S$ 
acts as a standard time-reversal transformation on the 
$L^2(\mathbb{R})$ Hilbert space, and hence sends 
a momentum eigenstate to its negative counterpart, 
$S|\Phi, s_p\rangle = |S \Phi, -s_p\rangle$.  Finally, we note the 
action on the state $|\Phi,f\rangle$,
\beq
S|\Phi,f\rangle = \int dy Sf(y)|\Phi,y\rangle = 
\int dy f^*(y) |S\Phi,y\rangle = |S\Phi,f^*\rangle
\eeq
These relations also hold for the antiunitary modular conjugations
$J$.

We now compute\footnote{We follow the conventions 
of \cite[Appendix E]{Jensen2023} for Fourier transforms.
In particular, we have $|f\rangle = \int dy f(y)|y\rangle
= \int ds \tilde{f}(s)|s\rangle$ and $\langle y|s\rangle = 
\frac{e^{isy}}{\sqrt{2\pi}}$. }
\begin{align}
\ho a^\dagger |\wh\Phi\rangle
&=
e^{-iu\hat{q}}e^{i\hat{p} h}\msf{a}^\dagger e^{-i\hat{p} h}
|\Phi,f\rangle 
\nonumber \\
&=
e^{-iu\hat{q}} e^{i\hat{p}h}\msf{a}^\dagger e^{-i\hat{p}h}
S_{\Phi|\Omega}S_\Omega |\Omega,f\rangle
\nonumber \\
&=
S_{\Phi|\Omega} S_\Omega e^{-iu\hat{q}} e^{i\hat{p} h}
\msf{a}^\dagger|\Omega,f\rangle
\nonumber \\
&=
S_{\Phi|\Omega}S_\Omega e^{-iu\hat{q}} e^{i\hat{p} h}S_\Omega 
\msf{a}|\Omega,f^*\rangle
\nonumber \\
&=
S_{\Phi|\Omega} e^{iu\hat{q}}e^{-i\hat{p}h}
\msf{a}|\Omega,f^*\rangle
\nonumber \\
&=
S_{\Phi|\Omega}\int ds \widetilde{f^*\,}(s)
e^{-ish}\msf{a}e^{iu\hat{q}}|\Omega,s\rangle
\nonumber \\
&=
S_{\Phi|\Omega}\int ds\widetilde{f^*\,}(s)e^{is(\hat{q}-h)}\msf{a}
|\Omega,u\rangle
\nonumber \\
&=
\sqrt{2\pi} S_{\Phi|\Omega}f^*(\hat{q}-h) \msf{a}|\Omega,u\rangle
\end{align}
In the third line we have used that $S_{\Phi|\Omega}S_\Omega$
is affiliated with $\alg_\text{QFT}'$ to commute it to the left,
and in the fifth line we used that $S_\Omega^2 = \mathbbm{1}$ and 
$S_\Omega e^{i\hat{p} h}S_\Omega = e^{-i\hat{p}h}$.
On the other hand, we have that 
\begin{align}
\ho a|\wh\Phi\rangle
&=
e^{i\hat{p} h}\msf{a} e^{-i\hat{p}h} e^{iu\hat{q}}|\Phi,f\rangle
\nonumber \\
&=
e^{i\hat{p}h}\msf{a}e^{-i\hat{p}h}e^{iu\hat{q}}S_{\Phi|\Omega}
S_\Omega|\Omega,f\rangle
\nonumber \\
&=
S_{\Phi|\Omega}S_\Omega \int ds \tilde{f}(s)e^{i\hat{p}h}
\msf{a}e^{-i\hat{p} h} e^{iu\hat{q}}|\Omega,s\rangle
\nonumber \\
&=
S_{\Phi|\Omega} S_\Omega \int ds \tilde f(s) e^{i\hat{p} h}
\msf{a} e^{is\hat{q}}|\Omega,u\rangle
\nonumber \\
&=
\sqrt{2\pi}S_{\Phi|\Omega}S_\Omega e^{i\hat{p}h}f(\hat{q})
\msf{a}|\Omega,u\rangle.
\end{align}
These produce an equation for the Tomita operator
via (\ref{eqn:tomitadefn}), 
\beq
S_{\wh{\Phi}} S_{\Phi|\Omega}S_\Omega e^{i\hat{p}h}f(\hat{q})
= S_{\Phi|\Omega}f^*(\hat{q}-h)
\eeq
whose solution is 
\beq
S_{\wh{\Phi}} = S_{\Phi|\Omega} f^*(\hat{q}-h)\frac{1}{f(\hat{q})}
e^{-i\hat{p} h} S_\Omega S_{\Omega|\Phi},
\eeq
noting that $S_\Omega^{-1} = S_\Omega$, $S_{\Phi|\Omega}^{-1}
 = S_{\Omega|\Phi}$.
 We can simplify this to the form
\begin{align}
S_{\wh{\Phi}} 
&=
J_{\Phi|\Omega} \Delta_{\Phi|\Omega}^{\frac12}
f^*(\hat{q}-h)e^{\frac{\hat{q}}{2}}e^{-i\hat{p}h}
\frac{1}{e^{\frac{\hat{q}}{2}}f(\hat{q}+h)}\Delta_\Omega^{\frac12}
S_\Omega S_{\Omega|\Phi}
\\
&= 
J_{\Phi|\Omega}e^{-i\hat{p}h} \cdot \left[e^{i\hat{p}h}
\Delta_{\Phi|\Omega}^{\frac12} f^*(\hat{q}-h) 
e^{\frac{\hat{q}}{2}} e^{-i\hat{p} h} \right]
\cdot\left[ \frac{1}{e^{\frac{\hat{q}}{2}} f(\hat{q}+h)}
J_\Omega J_{\Omega|\Phi}\Delta_{\Omega|\Phi}^{\frac12}\right]
\end{align}
where the first term in brackets is affiliated with $\wh{\alg}$,
and the second term is affiliated with $\wh{\alg}'$.
Note by computing $S_{\wh{\Phi}}^\dagger S_{\wh{\Phi}}$,
we immediately arrive at the expression for the density
matrices derived in \cite{Jensen2023},
\begin{align}
\rho_{\wh\Phi} 
&= 
e^{i\hat{p} h} f(\hat{q}-h) e^{\hat{q}}
\Delta_{\Phi|\Omega} f^*(\hat{q}-h) e^{-i\hat{p}h}
\\
\rho_{\wh{\Phi}}'
&=\Delta_{\Omega|\Phi}^{-\frac12} J_{\Phi|\Omega} J_\Omega e^{\hat{q}}
\big|f(\hat{q}+h)\big|^2 J_\Omega J_{\Omega|\Phi} \Delta_{\Omega|\Phi}^{-\frac12}.
\end{align}

It is clear from
this construction that we can obtain expressions for 
the Tomita operator of arbitrary superpositions
$|\wh{\Phi}\rangle = \int d\lambda |\Phi_\lambda, 
f_\lambda\rangle$ as well, 
where $d\lambda$ is some measure for the wavefunctions 
indexed by $\lambda$, which could in principle include discrete 
$\delta-$function components. In this case we have
\begin{align}
\ho a^\dagger \int d\lambda |\Phi_\lambda, f_\lambda\rangle
&=
\sqrt{2\pi}\int d\lambda S_{\Phi_\lambda|\Omega} 
f_\lambda^*(\hat{q}-h) \msf{a}|\Omega,u\rangle
\\
\ho a\int d\lambda |\Phi_\lambda, f_\lambda\rangle
&=
\sqrt{2\pi}\int d\lambda S_{\Phi_\lambda|\Omega}S_\Omega e^{i\hat{p}h}
f_\lambda(\hat{q}) \msf{a}|\Omega,u\rangle
\end{align}
and hence
\begin{align}
S_{\wh{\Phi}} &= \left[\int d\lambda S_{\Phi_\lambda|\Omega}
f^*_\lambda(\hat{q}-h)\right] 
\left[\int d\mu S_{\Phi_\mu|\Omega}S_\Omega e^{i\hat{p}h}f_\mu(\hat{q})
\right]^{-1}
\nonumber \\
&=
J_\Omega e^{-i\hat{p}h}\left[\int d\lambda e^{i\hat{p} h}
J_\Omega J_{\Phi_\lambda|\Omega} \Delta_{\Phi_\lambda|\Omega}^{\frac12}
f_\lambda^*(\hat{q}-h) e^{-i\hat{p}h}
\right]
\left[
\int d\mu \Delta_{\Omega|\Phi_\mu}^{-\frac12}J_{\Phi_\mu|\Omega}J_\Omega
\Delta_\Omega^{\frac12} f(\hat{q}+h)\right]^{-1}
\nonumber \\
&=
J_\Omega e^{-i\hat{p}h}
\left[\int d\lambda e^{i\hat{p}h} J_\Omega J_{\Phi_\lambda|\Omega}
\Delta_{\Phi_\lambda|\Omega}^{\frac12} f_\lambda^*(\hat{q}-h) 
e^{\frac{\hat{q}}{2}}
e^{-i\hat{p}h}\right]
\left[\int d\mu \Delta_{\Omega|\Phi_\mu}^{-\frac12}
J_{\Phi_\mu|\Omega}J_\Omega f(\hat{q}+h)e^{\frac{\hat{q}}{2}}
\right]^{-1}
\end{align}
Writing it in this way makes it clear that the Tomita operator
can always be expressed in the form
\beq
S_{\wh{\Phi}} = J_\Omega e^{-i\hat{p}h} \,\ho s\; \ho s',\qquad 
\ho s \in \wh{\alg},\; \ho s'\in\wh{\alg}'.
\eeq
Using the polar decomposition 
$\ho s = \ho u\,\rho_{\wh{\Phi}}^{\frac12}$,
with $\ho u$ a unitary element of  $\wh{\alg}$ and $\rho_{\wh{\Phi}}$
the density matrix for $\wh{\alg}$, and similarly writing
$\ho{s}' = \ho{u}'\, (\rho_{\wh{\Phi}}')^{-\frac12}$, 
we arrive at the expression for the
modular operator and modular conjugation:
\beq
J_{\wh{\Phi}} = J_\Omega e^{-i\hat{p}h}\ho{u}\,\ho{u}',
\qquad
\Delta_{\wh{\Phi}} = \rho_{\wh{\Phi}} (\rho_{\wh{\Phi}}')^{-1}
\eeq

It is useful to consider this expression for the vacuum
classical-quantum states $|\wh{\Omega}\rangle = |\Omega,f\rangle$.
In that case, the Tomita operator simplifies to
\beq
S_{\wh{\Omega}} = 
J_\Omega e^{-i\hat{p} h} \frac{f^*(\hat{q})}
{f(\hat{q}+h)}\Delta_\Omega^{\frac12}
=
J_\Omega e^{-i\hat{p}h}
\left(\frac{f^*(\hat{q})f^*(\hat{q}+h)}
{f(\hat{q}) f(\hat{q}+h)}\right)^{\frac12}
\frac{|f(\hat{q})|}{|f(\hat{q}+h)|}\Delta_\Omega^{\frac12}
\eeq
which is now in the form of a polar decomposition,
allowing the modular conjugation to be read off
\beq
J_{\wh{\Omega}} = J_\Omega e^{-i\hat{p} h}
\left(\frac{f^*(\hat{q})f^*(\hat{q}+h)}{f(\hat{q})f(\hat{q}+h)}
\right)^{\frac12},
\qquad \Delta_{\wh{\Omega}} 
= \frac{g(\hat{q})}{g(\hat{q}+h)}\Delta_\Omega.
\eeq
This is particularly simple when the wavefunction is real,
in which case $J_{\wh{\Omega}} = J_\Omega e^{-i\hat{p}h}$.

It is also helpful to have the Tomita operator for the twirled
state $|\tilde{\Phi}\rangle = e^{i\hat{p} h}|\Phi,f\rangle$.
We proceed as before, computing
\begin{align}
\ho a |\tilde{\Phi}\rangle
&= e^{i\hat{p}h}\msf{a}e^{-i\hat{p}h} e^{iu\hat{q}} 
e^{i\hat{p}h}|\Phi,f\rangle
\nonumber \\
&=\int ds \tilde{f}(s) e^{i\hat{p}h}\msf{a}e^{-i\hat{p}h}e^{iu\hat{q}}
e^{ish}|\Phi,s\rangle
\nonumber \\
&=\int ds \tilde{f}(s) e^{i\hat{p}h}\msf{a}e^{-i(\hat{p}-s)h} 
e^{is\hat{q}}|\Phi,u\rangle
\nonumber \\
&=
\int ds\tilde{f}(s) e^{is(\hat{q}+h)} e^{i\hat{p}h}\msf{a}e^{-i\hat{p}h}
S_{\Phi|\Omega} S_\Omega|\Omega,u\rangle
\nonumber \\
&=
\sqrt{2\pi}f(\hat{q}+h) S_{\Phi|\Omega} S_\Omega e^{i\hat{p}h} \msf{a}|\Omega,u\rangle
\end{align}
and
\begin{align}
\ho a^\dagger|\tilde\Phi\rangle
&=
e^{-iu\hat{q}} e^{i\hat{p}h}\msf{a}^\dagger e^{-i\hat{p}h} e^{i\hat{p}h}
|\Phi,f\rangle
\nonumber \\
&=
\int ds \tilde{f}(s) e^{ish}\msf{a}^\dagger e^{-iu\hat{q}}|\Phi,s\rangle
\nonumber \\
&=
\int ds \tilde{f}(s)e^{ish}\msf{a}^\dagger e^{is\hat{q}}|\Phi,-u\rangle
\nonumber \\
&=
\sqrt{2\pi} f(\hat{q}+h) \msf{a}^\dagger |\Phi,-u\rangle
\nonumber \\
&=
\sqrt{2\pi}f(\hat{q}+h)S_{\Phi|\Omega}\msf{a}|\Omega,u\rangle.
\end{align}
Then using that $S_{\tilde\Phi} \ho a|\tilde\Phi\rangle = \ho{a}^\dagger
|\tilde\Phi\rangle$, we find that 
\begin{align}
S_{\tilde\Phi}
&=
f(\hat{q}+h) S_{\Phi|\Omega} e^{-i\hat{p}h} S_\Omega S_{\Omega|\Phi}
\frac{1}{f(\hat{q}+h)}
\end{align}
We can then simplify this into a factorized form,
\begin{align}
S_{\tilde{\Phi}}
&=
J_\Omega f^*(\hat{q}-h)J_\Omega J_{\Phi|\Omega}\Delta_{\Phi|\Omega}^{\frac12}
e^{-i\hat{p}h} \Delta_{\Omega}^{-\frac12}J_\Omega J_{\Omega|\Phi}
\Delta_{\Omega|\Phi}^{\frac12}\frac{1}{f(\hat{q}+h)}
\nonumber \\
&=
J_\Omega e^{-i\hat{p}h} \left[f^*(\hat{q}) e^{-i\hat{p}h}J_\Omega J_{\Phi|\Omega}
\Delta_{\Phi|\Omega}^{\frac12} \Delta_\Omega^{-\frac12} e^{-i\hat{p}h}
e^{\frac{\hat{q}}{2}}\right]J_\Omega J_{\Omega|\Phi}\Delta_{\Omega|\Phi}^\frac12 
\frac{e^{-\frac{\hat{q}}{2}}}{f(\hat{q}+h)}
\nonumber \\
&=
J_\Omega e^{-i\hat{p}h}J_\Omega J_{\Omega|\Phi}
\left[f^*(\hat{q}) e^{-i\hat{p}h}J_\Omega J_{\Phi|\Omega}
\Delta_{\Phi|\Omega}^{\frac12} e^{\frac{\hat{q}}{2}} e^{-i\hat{p}h}\right]
\Delta_{\Omega|\Phi}^{\frac12} 
\frac{e^{-\frac{\hat{q}}{2}}}{f(\hat{q}+h)}
\nonumber \\
&=
e^{i\hat{p}h}J_{\Omega|\Phi}
\left[f^*(\hat{q}) e^{-i\hat{p}h}J_\Omega J_{\Phi|\Omega}
\Delta_{\Phi|\Omega}^{\frac12} e^{\frac{\hat{q}}{2}} e^{-i\hat{p}h}\right]
\cdot \left[
\Delta_{\Omega|\Phi}^{\frac12} 
\frac{e^{-\frac{\hat{q}}{2}}}{f(\hat{q}+h)}
\right]
\end{align}
so that in particular the density matrices are given by,
\begin{align}
\rho_{\tilde{\Phi}} 
&= 
e^{i\hat{p}h} \Delta_{\Phi|\Omega}^{\frac12}
J_{\Omega|\Phi}J_\Omega e^{\hat{q}}g(\hat{q}-h) J_\Omega J_{\Phi|\Omega}
\Delta_{\Phi|\Omega}^\frac12 e^{-i\hat{p}h}
\label{eqn:twirlrho}\\
\rho_{\tilde{\Phi}}'
&=
f(\hat{q}+h)e^{\frac{\hat{q}}{2}} \Delta_{\Omega|\Phi}^{-1} e^{\frac{\hat{q}}{2}}
f^*(\hat{q}+h)
\end{align}
again reproducing the expressions from \cite{Jensen2023}.

\section{Null surfaces}
The identities considered in section 
\ref{sec:geometric} that are crucial in obtaining the 
crossed product gravitational algebras and relating the 
crossed product entropy to generalized entropy make 
extensive use of geometric relations for null surfaces.
This appendix summarizes the necessary properties of 
null surfaces needed to arrive at these identities by 
way of the canonical analysis on null surfaces, discussed 
in appendix \ref{app:canon}.  

\subsection{Null surface geometry}
We begin by briefly reviewing
the geometry of null surfaces, see \cite{Chandrasekaran2020, 
Chandrasekaran:2021hxc, Gourgoulhon:2005ng} for additional details.  
Let $\ns$ be a null hypersurface with a null normal $l_a$ whose norm
vanishes on $\ns$, $l\cdot l \overset{\ns}{=}0$.  We can define a null
rigging vector $n^a$ that satisfies $n\cdot l = -1$ and $n\cdot n = 0$, and 
with it define a projector onto the null surface
\beq
\Pi\indices{^a_b} = \delta^a_b +n^a l_b.
\eeq 
The metric at $\ns$ then naturally decomposes as 
\beq
g_{ab} = q_{ab} + n_a l_b + l_a n_b,
\eeq
where $q_{ab}$ is the degenerate induced metric on $\ns$.  Additionally, the 
spacetime volume form $\epsilon$ decomposes as 
\beq
\epsilon = - l \wedge \eta 
\eeq
where 
\beq
\eta = -n\wedge \mu
\eeq
is the volume form on $\ns$, and $\mu=i_l\eta$ is the degenerate area 
form that induces a volume for on horizon cuts.
Note that the choice of rigging vector $n^a$ also allows for an inverse
degenerate metric $q^{ab}$ to be defined, which satisfies $q^{ab}n_b = q^{ab}l_b
= 0$, $q^{ab} q_{ac} q_{bd} = q_{cd}$. 

The shape operator $W\indices{^a_b}$
for the null surface is defined as the projection
of $\nabla_a l^b$ onto the null surface, and decomposes as
\beq
W\indices{^a_b} = \Pi\indices{^c_b}\nabla_c l^a = -k n_b l^a+\varpi_b l^a +
q^{ac}(\sigma_{cb} +\frac{1}{d-2}\Theta q_{cb})
\eeq
where $k$ is the inaffinity, $\varpi_b$ is the \Hajicek one-form, $\sigma_{cb}$ is the 
shear, and $\Theta$ is the expansion, defined by
\begin{align}
k &= -n_bl^a\nabla_a l^b\\
\varpi_b &=- q\indices{^c_b}n_a\nabla_c l^a \\
\Theta &= \nabla_a l^a -k \\
\sigma_{cb} &= \frac12 \lie_l q_{cb} - \frac{1}{d-2}\Theta q_{cb}
\end{align}

\subsection{Gaussian null coordinates}
\label{app:gnuc}

Gaussian null coordinates (GNuC) \cite{Penrose1972, 
Hollands:2006rj, Hollands:2012sf}
provide a convenient coordinate
system for describing the metric near a null hypersurface $\ns$.  
On such a surface, we take $l^a$ to be an affinely parameterized
null generator satisfying $l^a\nabla_a l^b \overset{\ns}{=} 0$,
and let $v$ denote the affine parameter so that $l^a\nabla_a v = 1$.
Surfaces of constant $v$ then define a foliation of the null
hypersurface, and we can define a future-directed
transverse null rigging vector
$n^a$ orthogonal to these surfaces and normalized such that 
$n \cdot l = -1$.  Extending $n^a$ off of $\ns$ as an affinely
parameterized null geodesic generator, it can be chosen 
to satisfy $n\cdot n = 0$, $n^a\nabla_a n^b = 0$, and 
$\nabla_{[a}n_{b]} = 0$ throughout the
open coordinate patch near $\ns$.  We can furthermore extend
$l^a$ off of $\ns$ by requiring $[n,l]^a = 0$.  By taking $u$ to 
be an affine parameter for $n^a$ with $\ns$ at $u=0$, coordinates
$(u,v,y^A)$ can be chosen such that 
\beq
l^a =\left(\frac{\partial}{\partial v}\right)^a, \qquad
n^a = \left(\frac{\partial}{\partial u}\right)^a.
\eeq
The metric in this coordinate system takes the form
\beq \label{eqn:GNuCexpr}
ds^2 = -2 du dv - u^2 A dv^2 + 2u B_A dv d y^A + 
q_{AB}dy^A dy^B
\eeq
where $A$, $B_A$ and $q_{AB}$ are functions 
of the spacetime coordinates. Note in this coordinate system
the relations 
\beq \label{eqn:covlnrelns}
n_a = -\nabla_a v, \quad l_a = -\nabla_a u + uB_a
-u^2 A \nabla_a v, \quad l\cdot l = -u^2 A
\eeq
hold,
where $B_a =B_A \nabla_a y^A$.

When $\ns$ is a Killing horizon with bifurcation
surface located at $u=v=0$, the Killing vector in these coordinates
takes the form \cite{Hollands:2012sf}
\beq
\xi^a = \kappa v l^a - \kappa u n^a.
\eeq
Additionally, requiring the metric to be invariant under the flow
of $\xi^a$ leads to the condition that $A$, $B_A$, and $q_{AB}$ depend
on the coordinates $u$ and $v$ only via their product $uv$.

\subsection{Quasi-Killing vectors} \label{app:quasi-Killing}
In addition to the global Killing vector $\xi^a$, 
the analysis of section \ref{sec:geometric} deals extensively with 
quasi-Killing vectors, which satisfy Killing's equation
only on the null surface $\ns$.  A generic such vector 
parallel to the horizon generator $l^a$ can be expressed
in terms of a function $\tau(v, y^A)$ as 
\beq \label{eqn:quasi-Killing}
\zeta_\tau^a = \tau  l^a + u \nabla^a \tau.
\eeq
The Lie derivative of the metric on $\ns$ can then 
be written as 
\begin{align}
\lie_{\zeta_\tau} g_{ab} \overset{\ns}{=}
\tau \lie_l g_{ab} + 2\nabla_{(a} \tau l_{b)} +
2 \nabla_{(a}u \nabla_{b)} \tau \overset{\ns}{=}
\tau \lie_l g_{ab},
\end{align}
where we have dropped terms proportional to $u$.  
Hence we need only verify that $\lie_l g_{ab} = 0$.  
Note using the GNuC expressions (\ref{eqn:GNuCexpr}), one
finds
\begin{align}
\lie_l l_a =  (i_l dl)_a + \nabla_a(l\cdot l)  
\overset{\ns}{=} 0\\
\lie_l n_a = (i_l dn)_a + \nabla_a( l\cdot n) = 0.
\end{align}
Then from the decomposition of the metric $g_{ab} = q_{ab}
+ n_a l_b + l_a n_b$, we find that 
\beq
\lie_l g_{ab} = \lie_l q_{ab} = \sigma_{ab} + \frac{1}{d-2}
\Theta q_{ab} =0,
\eeq
where the last equality uses that a Killing horizon has 
vanishing expansion and shear.  This then
verifies that $\zeta_\tau^a$ is a quasi-Killing vector.

An additional restriction on the quasi-Killing vector 
is that it preserve the gauge conditions (\ref{eqn:CFPgauge}) employed 
in the canonical analysis for the null surface.
The third condition there is nontrivial, since it 
involves derivatives transverse to the null surface.  
To check whether the vector $\zeta^a_\tau$ preserves
this condition, we evaluate
\begin{align}
\frac12 n^a\nabla_a(l^b l^c\lie_{\zeta_\tau} g_{bc})
&=
n^a\nabla_a\Big(l^b l^c\nabla_b(\tau l_c + u \nabla_c \tau)\Big)
\nonumber
\\
&=
n^a\nabla_a\Big(l^b\nabla_b \tau (l\cdot l) 
+ \tau l^b l^c\nabla_b l_c
+ l^b\nabla_b u l^c\nabla_c \tau
+ u l^b l^c\nabla_b \nabla_c\tau \Big)
\nonumber
\\
&=
l^b l^c \nabla_b \nabla_c\tau + \op(u) 
=
\lie_l \lie_l \tau + \op(u).
\label{eqn:preservegauge}
\end{align}
For this to vanish, the function $\tau$ must be at most 
linear in the coordinate $v$.  
This agrees with the set of null surface symmetries
derived in \cite{CFP2018}.  

It is also useful to derive expressions involving the 
derivatives of the vectors $\zeta_\tau^a$, which can 
be applied when evaluating gravitational charges in appendix
\ref{app:canon}.  These expressions only rely on the 
form of the vector (\ref{eqn:quasi-Killing}) in GNuC, 
and do not rely on $\ns$ being a Killing horizon.  
We have
\begin{align}
l^a\nabla_a \zeta_\tau^b 
&= 
l^b l^a\nabla_a \tau +
\tau  l^a\nabla_a l^b +
l^a\nabla_a u \nabla^b\tau
+u l^a\nabla_a \nabla^b \tau
\overset{\ns}{=} l^b l^a\nabla_a \tau
\label{eqn:ldelzeta}
\\
l_b\nabla_a \zeta_\tau^b
&=
(l\cdot l) \nabla_a \tau + \frac12 \tau \nabla_a( l\cdot l)
+ \nabla_a u l^b\nabla_b \tau + u l_b\nabla_a \nabla^b\tau
\overset{\ns}{=}
- l_a l^b\nabla_b \tau
\label{eqn:lgradzeta}
\\
\nabla_a \zeta_\tau^a
&=
l^a\nabla_a \tau + \tau \Theta + \nabla_a u \nabla^a\tau +
u \nabla_a \nabla^a\tau
\overset{\ns}{=} \tau \Theta.
\label{eqn:divzeta}
\end{align}

\section{Canonical formalism and gravitational charges}
\label{app:canon}

In this appendix, we provide additional details on the canonical formulation
of general relativity on null hypersurfaces needed to arrive 
at the identities discussed in section \ref{sec:geometric}.  
We will largely follow the conventions and notation
of \cite{Chandrasekaran2020}, see also 
\cite{CFP2018, 
Chandrasekaran2021gen, Ciambelli:2023mir, Odak:2023pga, 
Chandrasekaran:2023vzb, Adami:2021nnf}
for related recent treatments of the canonical analysis
of general relativity on null hypersurfaces.  
This canonical analysis is done in the framework
of the covariant phase space, see \cite{Crnkovic1987,
Lee:1990nz, Iyer:1994ys, Harlow2019}
for overviews of this formalism.

The Lagrangian for the theory consists of a sum of a gravitational and matter 
contributions, $L = L^g + L^\psi$, with the gravitational Lagrangian comprised
of an Einstein-Hilbert term and cosmological constant,
\beq
L^g = \frac{1}{16\pi G_N} \epsilon  (R-2\Lambda) .
\eeq
The covariant symplectic potential $\theta = \theta^g + \theta^\psi$ 
arises from the boundary term appearing 
in the variation of the Lagrangian, $\delta L = E_\phi\cdot \delta \phi + d\theta$, 
where $\phi = (g_{ab},\psi)$ collectively denotes the full set 
of dynamical fields, and $E_\phi = 0$ are the equations of motion.  
The gravitational piece is given by
\beq
\theta^g = \frac1{16\pi G_N} \epsilon_{a\ldots}\Big(g^{bc}\delta\Gamma^a_{bc} - 
g^{ac}\delta \Gamma^b_{bc}\Big).
\eeq

When pulled back to a null Cauchy surface $\ns$, $\theta^g$ may 
be decomposed as \cite{Chandrasekaran2020}
\beq \label{eqn:thetag}
\underline{\theta}^g \overset{\ns}{=} -\delta \ell + d\beta + \beom^g
\eeq
where the underline denotes the pullback, and 
\begin{align}
\ell &= -\frac{1}{16\pi G_N}\eta(\Theta + 2 k) \\
\beta &= \frac{1}{32\pi G_N} g^{ab}\delta g_{ab} \mu \\
\beom^g &= \frac{1}{16\pi G_N}\eta \left[\sigma^{ab}\delta q_{ab}-
\left(k +\frac{d-3}{d-2}\Theta\right) q^{ab}\delta q_{ab}-2(\Theta n_a +\varpi_a)\delta l^a 
\right].
\end{align}
For the present applications, it is helpful
to perform a partial gauge-fixing with respect to
diffeomorphisms tangent to $\ns$.  The relevant
gauge-conditions were considered 
by Chandrasekaran, Flanagan, and Prabhu (CFP)
\cite{CFP2018}, who showed that one
can always impose the conditions that $l_a$ and 
$l^a$ are fixed, $l^a$ remains null, and 
the inaffinity of $l^a$ vanishes, $k=0$.  These
translate to the following conditions on the variations:
\begin{align}\label{eqn:CFPgauge}
\delta l_a \overset{\ns}{=} 0;  \qquad 
g_{ac}\delta l^c = - l^b\delta g_{ab} \overset{\ns}{=} 0;
\qquad \delta k = 
\frac12 n^b\nabla_b(l^a l^c\delta g_{ac}) 
\overset{\ns}{=} 0.
\end{align}
Note that we can satisfy these conditions by gauge fixing 
to Gaussian null coordinates and imposing that the only variations 
of the metric come from variations of $A$, $B_A$ and $q_{AB}$ appearing
in (\ref{eqn:GNuCexpr}). 
With these gauge conditions, the terms appearing in the decomposition
(\ref{eqn:thetag}) simplifty to
\begin{align}
\ell &= -\frac{1}{16\pi G_N} \eta \Theta 
\label{eqn:ell}\\
\beta &= \frac{1}{32\pi G_N} q^{ab}\delta g_{ab} \mu 
\label{eqn:beta}\\
\beom^g &=\frac{1}{16\pi G_N} \eta
\left[ 
\sigma^{ab}\delta q_{ab} - \frac{d-3}{d-2}\Theta q^{ab}
\delta q_{ab}
\right]. \label{eqn:beomg}
\end{align}
 An additional consequence of the above  gauge
fixing is that it allows us to define the location of a horizon
cut in a diffeomorphism invariant manner: in Gaussian null
coordinates, a surface at $v= \lambda(y^A)$ can simply be interpreted
as a cut located a given affine distance away from the bifurcation
surface.  

The quantity $\beom = \beom^g + \beom^\psi$ serves as a 
symplectic potential current on the null surface, and hence
the integral of its variation defines the symplectic form
\beq
\Omega = \int_{\ns} \delta\beom,
\eeq
where we are treating $\delta$ as an exterior
derivative on field space, so that in terms 
of individual variations, 
$\delta\beom[\delta_1\phi, \delta_2\phi] 
= \delta_1\beom[\delta_2\phi] - \delta_2 \beom[\delta_1\phi]$.
Note that $\delta \beom$ 
differs from the symplectic form constructed
from the spacetime-covariant current $\omega = \delta \theta$ 
by a total derivative $-d\delta\beta$.  We 
can determine the charge associated with a diffeomorphism
generated by a vector $\xi^a$ parallel to $\ns$ by
evaluating \cite{Chandrasekaran2020, Chandrasekaran2021gen}
\begin{align} \label{eqn:Omliexiphi}
\Omega[\delta\phi, \lie_\xi \phi]
=
\int_{\ns} d(\delta M_\xi - M_{\delta \xi} - i_\xi\beom)
+
\int_{\ns}\Big(\delta C_\xi - C_{\delta\xi}\Big)
\end{align}
where 
\beq \label{eqn:Mxi}
M_\xi = Q_\xi + i_\xi \ell - \beta[\lie_\xi\phi],
\eeq
$Q_\xi$ is the covariant Noether potential,
which for general relativity evaluates to \cite{Iyer:1994ys}
\beq \label{eqn:Q}
Q_\xi = -\frac{1}{16\pi G_N} \epsilon\indices{^a_b_{\ldots}} 
\nabla_a\xi^b,
\eeq
and $C_\xi$ are combinations of the equations of motion 
comprising the constraints of the theory.  

To satisfy Hamilton's equation, the right hand side 
side of (\ref{eqn:Omliexiphi}) must take the form of a total
variation.  The possible obstructions arise either
from field-dependence of the generator $\xi^a$ in the terms 
$M_{\delta\xi}$ and $C_{\delta \xi}$ or from the flux
term $i_\xi \beom$ evaluated at the asymptotic boundaries
of the null surface $\partial\ns$.  We verify below that
the field dependence for quasi-Killing vectors on a null
surface has vanishing contributions to the fluxes.  The 
remaining obstruction to integrability comes from $i_\xi\beom$.
To handle this term, we impose boundary conditions on the 
future horizon that the matter field fluxes go to zero and 
that the shear is asymptotically vanishing.  Additionally,
the requirement that $\ns$ is an event horizon imposes
that the expansion vanishes in the far future, so examining
the form of $\beom^g$ in (\ref{eqn:beomg}) shows that
it vanishes on the future boundary, and hence the full
obstruction to integrability vanishes here.  

Integrability at the past boundary of $\ns$ is more subtle
since $\Theta$ is constrained to satisfy (\ref{eqn:semianec}),
and 
hence cannot vanish at both the past and future boundaries
of $\ns$ unless there is zero null energy flux.  However,
when working perturbatively in $\gc = \sqrt{32\pi G_N}$,
the expansion itself is $\op(\gc^2)$, and since the metric
variation $\delta q_{ab}$ is $\op(\gc)$, we find that 
the flux term coming from the expansion in $\beom^g$ is 
suppressed.  Thus, in the $G_N\rightarrow 0$ limit, the 
expansion term does not obstruct integrability of the charges,
and assuming the shear and matter flux fall off at the 
past boundary of $\ns$, the full contribution from $i_\xi \beom$
vanishes there.  

Hence, we can define the Hamiltonian
\beq
H_\xi = \int_{\partial\ns} M_\xi + \int_\ns C_\xi,
\eeq
which, according to the discussion of integrability above,
satisfies
\beq
\delta H_\xi = \Omega[\delta\phi, \lie_\xi\phi].
\eeq
The Hamiltonian can also be usefully expressed 
in terms of a local integral on $\ns$ by noting the relation
\beq
dM_\xi + C_\xi \overset{\ns}{=} \beom[\lie_\xi\phi],
\eeq
which holds when all forms are pulled back to $\ns$.  
In general this equation contains corrections if the vector
$\xi^a$ is not tangent to $\ns$ or does not act covariantly
on $\ell$, as discussed in detail in \cite{Chandrasekaran2020,
Chandrasekaran:2021hxc}, but one can show that there is 
no contribution
from the anomalous transformation of $\ell$ when
the vector field $\xi^a$ preserves the gauge-fixing
conditions (\ref{eqn:CFPgauge}).  According to 
equation (\ref{eqn:preservegauge}), this will hold 
in particular for quasi-Killing vectors $\zeta^a_\tau$, with 
$\tau(v,y^A)$ containing only linear and constant terms in $v$.  This
results in the equivalent expression for the Hamiltonian
\beq \label{eqn:Hxi}
H_\xi = \int_{\ns} \beom[\lie_\xi\phi].
\eeq

The gravitational contribution to $H_\xi$ follows
directly from the form of $\beom^g$ given in 
(\ref{eqn:beomg}).  The matter contribution in
the simplest cases is directly related to the matter
stress tensor.  To arrive at this conclusion, we assume
that the matter symplectic potential $\theta^\psi$ does not
contain nontrivial terms involving the metric
in its decomposition on $\ns$, and 
hence $\theta^\psi = \beom^\psi$.  We can then consider 
the matter contribution to the Noether current 
\beq \label{eqn:Jxipsi}
J_\xi^\psi = \theta^\psi[\lie_\xi\phi] - i_\xi L 
\overset{\ns}{=}\beom^\psi[\lie_\xi\phi]
\eeq
where the second equality holds upon pulling back to $\ns$
and using that $\xi^a$ is tangent to $\ns$.  The divergence
of the Noether current satisfies
\beq \label{eqn:dJpsi}
dJ_\xi^\psi = -E_\psi\cdot \lie_\xi \psi 
-\epsilon T_{(\psi)}^{ab} \nabla_a \xi_b = d C^\psi_\xi + N^\psi_\xi
\eeq
where both $C_\xi^\psi$ and $N_\xi^\psi$ depend 
algebraically on $\xi^a$.  The Noether identities 
$N_\xi^\psi = 0$ hold identically due to 
diffeomorphism invariance, and hence $J_\xi^\psi = 
C_\xi^\psi + dQ_\xi^\psi$.  Again, for most standard 
matter theories, $Q_\xi^\psi = 0$, and hence the 
Noether current agrees with the contribution of the matter
terms to the constraint.  Examining (\ref{eqn:dJpsi}), 
we see that 
\beq
C_\xi^\psi = -\epsilon_{a\ldots}(T_{(\psi)})\indices{^a_b}\,\xi^b 
+ (E_\psi \;\text{terms}).
\eeq
Together with (\ref{eqn:Jxipsi}) and (\ref{eqn:Hxi}), 
this shows that the matter contribution to the full
Hamiltonian is given simply by the integral of the matter 
stress tensor, up to terms that vanish on shell.  
Note that for the most general matter theories,
one must allow for nonzero values of $\delta \ell^\psi$ and 
$\beta^\psi$ in the decomposition of $\theta^\psi$, and 
also allow for a nonzero value of $Q_\xi^\psi$.  
These can be straightforwardly incorporated into the above
discussion, but for the remainder of this work we do not consider
these generalizations.  

We now would like to evaluate the gravitational charge 
potential $M_{\zeta_\tau}$ for the vectors defined 
by (\ref{eqn:quasi-Killing}), which are quasi-Killing
vectors of the background spacetime.   
Using that $\epsilon = l\wedge n \wedge \mu$ and 
the identities (\ref{eqn:ldelzeta}) and (\ref{eqn:lgradzeta}),
 the Noether potential (\ref{eqn:Q}) pulled back to $\ns$ 
 evaluates to
\begin{align}
Q_{\zeta_\tau}
&\overset{\ns}{=}
-\frac{1}{16\pi G_N}
(l^a n_b- n^a l_b)\nabla_a \zeta_\tau^b \mu
=
\frac{1}{8\pi G_N} \mu l^a\nabla_a \tau.
\end{align}
Additionally, from the definitions (\ref{eqn:ell}) and 
(\ref{eqn:beta}) and the identity (\ref{eqn:divzeta}),
the remaining terms in $M_{\zeta_\tau}$ are given by
\begin{align}
i_{\zeta_\tau}\ell 
&\overset{\ns}{=} 
-\frac{1}{16\pi G_N} 
i_{\zeta_\tau} \eta \Theta = -\frac{1}{16\pi G_N}\mu \tau \Theta
\\
-\beta[\lie_{\zeta_\tau} g_{ab}]
&\overset{\ns}{=}
-\frac{1}{16\pi G_N}\nabla_a \zeta_\tau^a \mu 
= -\frac{1}{16\pi G_N} \mu \tau \Theta.
\end{align}
This results in the expression
\beq \label{eqn:CFPpot}
M_{\zeta_\tau}\overset{\ns}{=} \frac{1}{8\pi G_N}\mu
\Big(l^a\nabla_a \tau - \tau \Theta\Big),
\eeq
consistent with the expressions in CFP \cite{CFP2018}.

Finally, we must also verify that the field-dependence in the 
vector $\zeta_\tau^a$ does not spoil integrability of the charge.  Given
the expression (\ref{eqn:quasi-Killing}) for the quasi-Killing vector,
we require that $\delta\tau = 0$, and by gauge fixing to Gaussian
null coordinates we can ensure
that $\delta l^a = 0$ and $\delta u = 0$ in a neighborhood
of $\ns$.  This leads 
to 
\beq
\delta \zeta_\tau^a = -u \delta g_{bc}g^{ba} \nabla^c \tau.
\eeq
Hence $\delta\zeta_\tau^a$ vanishes on $\ns$, and its covariant 
derivative there does as well, since
\beq
\nabla_a \delta \zeta_\tau^b \overset{\ns}{=}-\nabla_a u \delta g_{bc}g^{ab}
\nabla^c\tau=l^b\delta g_{bc}\nabla^c\tau = 0,
\eeq
where we have used (\ref{eqn:covlnrelns}) and (\ref{eqn:CFPgauge})
 evaluated at $\ns$.  Since the possible obstructions to integrability
$M_{\delta\zeta_\tau}$ and $C_{\delta\zeta_\tau}$ 
appearing in (\ref{eqn:Omliexiphi}) involve only $\delta\zeta_\tau^a$
and $\nabla_a\delta\zeta_\tau^b$ at $\ns$, we see that these terms 
vanish.

\bibliography{gsl-refs}

\bibliographystyle{JHEPthesis}

\end{document}